\documentclass[12pt, preprint]{aastex}

\shorttitle{Light Neutron-Capture Element Abundances in PNe II}
\shortauthors{Sterling \& Dinerstein}

\begin{document}

\title{The Abundances of Light Neutron-Capture Elements in Planetary Nebulae -- II.\ \emph{s}-process Enrichments and Interpretation\protect \altaffilmark{1}}

\author{N. C. Sterling\altaffilmark{2, 3, 4} and Harriet L. Dinerstein\altaffilmark{2}}

\altaffiltext{1}{This paper includes data taken at the McDonald Observatory of the University of Texas at Austin.}
\altaffiltext{2}{The University of Texas, Department of Astronomy, 1 University Station, C1400,
    Austin, TX 78712-0259; harriet@astro.as.utexas.edu}
\altaffiltext{3}{Presently a NASA Postdoctoral Fellow at the Goddard Space Flight Center.  The NASA Postdoctoral Program is administered by Oak Ridge Associated Universities through a contract with NASA.}
\altaffiltext{4}{Current address: NASA Goddard Space Flight Center, Code 662, Greenbelt, MD 20771; sterling@milkyway.gsfc.nasa.gov}

\begin{abstract}

We present the results of a large-scale survey of neutron(\emph{n})-capture elements in Galactic planetary nebulae (PNe), undertaken to study enrichments from \emph{s}-process nucleosynthesis in their progenitor stars.  From new $K$~band observations of over 100 PNe supplemented by data from the literature, we have detected the emission lines [\ion{Kr}{3}]~2.199 and/or [\ion{Se}{4}]~2.287~$\mu$m in 81 of 120 objects.  We determine Se and Kr elemental abundances, employing ionization correction formulae derived in the first paper of this series.  We find a significant range in Se and Kr abundances, from near solar (no enrichment) to enhanced by $>$1.0~dex relative to solar, which we interpret as self-enrichment due to \textit{in situ} \emph{s}-process nucleosynthesis.  Kr tends to be more strongly enriched than Se; in 18 objects exhibiting both Se and Kr emission, we find that [Kr/Se]~=~0.5$\pm$0.2.

Our survey has increased the number of PNe with \emph{n}-capture element abundance determinations by a factor of ten, enabling us for the first time to search for correlations with other nebular properties.  As expected, we find a positive correlation between \emph{s}-process enrichments and the C/O ratio.  Type~I and bipolar PNe, which arise from intermediate-mass progenitors ($>3$--4~M$_{\odot}$), exhibit little to no \emph{s}-process enrichments.   Finally, PNe with H-deficient Wolf-Rayet central stars do not exhibit systematically larger \emph{s}-process enrichments than objects with H-rich nuclei.  Overall, 44\% of the PNe in our sample display significant \emph{s}-process enrichments ($>0.3$~dex).  Using an empirical PN luminosity function to correct for incompleteness, we estimate that the true fraction of \emph{s}-process enriched Galactic PNe is at least 20\%.

\end{abstract}


\section{INTRODUCTION}

\subsection{Nucleosynthesis in Planetary Nebula Progenitor Stars}

\setcounter{footnote}{4}

Low- and intermediate-mass stars (1--8~M$_{\odot}$), the progenitors of planetary nebulae (PNe), are important sources of He, C, N, and neutron(\emph{n})-capture elements (atomic number $Z>30$) in the Universe (Busso et al.\ 1999, hereafter BGW99).  These elements are produced by nucleosynthesis in PN progenitor stars and can be brought to the stellar surface via convective mixing, or ``dredge-up'' (Iben \& Renzini 1983; BGW99).  The enriched material is expelled into the ambient interstellar medium (ISM) via stellar winds and PN ejection.

PN progenitors may experience up to three stages of dredge-up after evolving off the main sequence.  The first dredge-up occurs during the red giant branch phase when the convective envelope penetrates regions that underwent CN-processing, resulting in enhancements of $^{13}$C and $^{14}$N, and a decrease in $^{12}$C at the stellar surface (Iben \& Renzini 1983; Sweigart et al.\ 1989; El~Eid 1994; BGW99).  The second dredge-up occurs during the early asymptotic giant branch (AGB) phase for stars more massive than $\sim$3.5~M$_{\odot}$ (hereafter denoted ``intermediate-mass stars,'' or IMS) when the convective envelope again descends into regions that experienced nuclear processing.  The second dredge-up enriches the stellar envelope with $^4$He and $^{14}$N, while $^{12}$C is depleted (Becker \& Iben 1979).

The third dredge-up (TDU) is the most relevant to our study, and occurs during the thermally-pulsing AGB (TP-AGB) phase of stars with initial masses higher than $\sim$1.5~M$_{\odot}$ (Straniero et al.\ 1997, 2006; BGW99).  During the TP-AGB, the H-burning shell is the main source of energy, while the He shell is primarily inactive.  Periodically, enough mass builds up on the He shell that it violently ignites, an event called a He shell flash, or thermal pulse.  This causes regions exterior to the He-burning layer to expand and cool, deactivating the H-burning shell and allowing the convective envelope to descend into the intershell zone where partially He-burnt material resides (Iben \& Renzini 1983; BGW99; Mowlavi 1999; Herwig 2005).  TDU is a recurrent process that operates after each thermal pulse until mass-loss reduces the envelope mass to less than 0.3--0.5~M$_{\odot}$ (Straniero et al.\ 1997, 2006).

TDU generates the conditions that lead to slow \emph{n}-capture nucleosynthesis (the ``\emph{s}-process''), by leaving a sharp discontinuity between the H-rich envelope and H-poor, C-rich intershell material after each thermal pulse.  Protons are mixed across this discontinuity by a poorly understood mechanism likely to involve convective overshoot (Herwig 2000), rotational shear (Herwig et al.\ 2003; Siess et al.\ 2004), and/or internal gravity waves (Denissenkov \& Tout 2003).  The protons are captured by $^{12}$C nuclei to form a layer rich in $^{13}$C, called the ``$^{13}$C-pocket.''  During the time intervals between thermal pulses, free neutrons are produced in this layer by the reaction $^{13}$C($\alpha,n$)$^{16}$O, and are captured by iron-peak ``seed'' nuclei.  These seed nuclei undergo a series of \emph{n}-captures interlaced with $\beta$-decays that transform them into isotopes of heavier elements, and are conveyed to the stellar envelope via TDU (K\"{a}ppeler et al.\ 1989; BGW99; Goriely \& Mowlavi 2000, hereafter GM00; Lugaro et al.\ 2003; Herwig 2005).

An additional neutron source, $^{22}$Ne($\alpha$,\emph{n})$^{25}$Mg, can be activated if the intershell layer reaches sufficiently high temperatures ($\gtrsim3.5\times10^8$~K).  This reaction plays a minor role in the \emph{s}-process nucleosynthesis of low-mass stars, which do not attain such high intershell temperatures (BGW99; GM00; Herwig 2005), but can be important in IMS (BGW99; Goriely \& Siess 2005; Lattanzio \& Lugaro 2005).  If $^{22}$Ne is the primary neutron source, the element-by-element distribution of \emph{s}-process enrichments is expected to be distinct from that of the $^{13}$C source, with larger enhancements of light \emph{n}-capture elements ($Z=30$--40) relative to heavier ones (Busso et al.\ 1988; Goriely \& Siess 2005).  However, the small intershell mass of an IMS relative to less massive AGB stars, and the significant dilution of processed material into its massive envelope can suppress \emph{s}-process enrichments, regardless of the neutron source (Lattanzio \& Lugaro 2005).

TDU thus conveys material enriched in $^4$He, $^{12}$C, and \emph{s}-process nuclei to the surfaces of TP-AGB stars, and can lead to such large C enrichments that the chemistry of the stellar envelopes change from O-rich to C-rich.  This mechanism is widely believed to cause TP-AGB stars to evolve along the sequence M$\rightarrow$MS$\rightarrow$S$\rightarrow$SC$\rightarrow$C, based on the increasing abundances of C and \emph{n}-capture elements (Smith \& Lambert 1990; Mowlavi 1999; Abia et al.\ 2002).  However, the formation of a C-rich envelope may be prevented or delayed in IMS, which can experience ``hot bottom burning'' during the TP-AGB (Boothroyd et al.\ 1993; Frost et al.\ 1998).  In this process, the temperature at the base of the convective envelope becomes large enough for the CNO-cycle to activate, leading to enhancements of $^4$He, $^{14}$N, and $^{13}$C, and depletion of $^{12}$C and possibly $^{16}$O at the stellar surface.

\subsection{Planetary Nebulae as Tracers of AGB Nucleosynthesis}

TDU and the \emph{s}-process have been widely studied in AGB stars (e.g., Smith \& Lambert 1990; Wallerstein \& Knapp 1998; BGW99; Abia et al.\ 2002; Herwig 2005, and references therein), and more recently in post-AGB stars (Van~Winckel 2003 and references therein).  However, investigations of \emph{s}-process enrichments in PNe contribute information complementary to stellar abundance determinations.

First, the most easily detected \emph{n}-capture elements in PNe are the lightest ones ($Z=30$--36), due to their relatively large abundances compared to heavier \emph{n}-capture elements.  These elements are generally detectable in stellar spectra only in the UV (e.g., Cowan et al.\ 2005; Chayer et al.\ 2005), where AGB stars produce little flux.  In the context of Galactic chemical evolution, light \emph{n}-capture elements are thought to be produced predominantly in the ``weak \emph{s}-process'' during core He- and shell C-burning in massive stars (Prantzos et al.\ 1990; The et al.\ 2000), in addition to the ``main \emph{s}-process'' in AGB stars (described above) and rapid \emph{n}-capture nucleosynthesis (the ``\emph{r}-process'') in supernovae or neutron star mergers (e.g., Truran et al.\ 2002).  However, light \emph{n}-capture elements have been poorly studied in their proposed sites of formation, due to the difficulty in detecting these elements in stellar spectra or absorption line analyses of supernova remnants (e.g., Wallerstein et al.\ 1995).  Therefore, current ideas regarding their origins are based almost exclusively on theoretical considerations.  Our survey provides some of the first empirical measurements of light \emph{n}-capture elements near one of their sites of production.

Furthermore, in PNe the \emph{s}-process can be studied in classes of stars which are not easily observed during the AGB or post-AGB stages of evolution.  For example, IMS are often obscured by dusty, optically thick circumstellar envelopes during the AGB and post-AGB, and are difficult to study with optical and UV spectroscopy (Habing 1996; Garc\'{i}a-Lario 2006).  Very little is known about the \emph{s}-process in Galactic IMS, and abundance analyses of these stars during their AGB phase have been performed only very recently (Garc\'{i}a-Hern\'{a}ndez et al.\ 2007).  In contrast, \emph{s}-process enrichments can be readily studied in Peimbert Type~I PNe, which are believed to be descendants of IMS (Peimbert 1978; Kingsburgh \& Barlow 1994; Torres-Peimbert \& Peimbert 1997).

Finally, elemental yields derived from AGB stars can be uncertain due to the unknown number of remaining TDU episodes the star will undergo (and extent of subsequent enrichment) before exiting the AGB phase.  PNe are the final evolutionary stage of low- and intermediate-mass stars, as nucleosynthesis has ceased and enriched material is directly fed into the ISM.  If the total nebular masses can be determined, then abundances of PNe can be used to directly determine elemental yields from their progenitor stars.  These are essential ingredients to Galactic chemical evolution models that aim to study the role of low- and intermediate-mass stars in the production of various elements in the Universe.

There are inherent difficulties in using nebular abundance analyses to investigate TDU and the \emph{s}-process in PN progenitors.  Specifically, it is difficult to reliably determine the C abundance in ionized nebulae (Kaler 1983; Rola \& Stasi\'{n}ska 1994), because most of its strong collisionally-excited lines are in the UV and are very sensitive to uncertainties in the extinction and gas temperature.  These uncertainties have led to disparate C/O determinations even from the same datasets.  For example, using the same \emph{International Ultraviolet Explorer} dataset for the bright PN NGC~6572, Hyung et al.\ (1994a) found C/O~=~0.6, Rola \& Stasi\'{n}ska (1994) computed C/O~=~0.7--1.1, and Liu et al.\ (2004b) derived C/O~=~1.6.  Similar discrepancies exist for the C determinations of other PNe, and hence C abundances are likely to be uncertain by a factor of two or three in most PNe.

The low initial abundances of \emph{n}-capture elements ($\lesssim10^{-9}$ relative to H in the Solar System; Asplund et al.\ 2005) make them more sensitive tracers of enrichments than C, but also cause their emission lines to be weak.  These elements were not seriously considered to be detectable in PNe until P\'{e}quignot \& Baluteau (1994, hereafter PB94) identified emission lines from Kr ($Z=36$), Xe ($Z=54$), and possibly other trans-iron species in a deep optical spectrum of the bright PN NGC~7027.  These authors approximated the unknown collision strengths of these lines and estimated the abundances of unobserved ions to derive elemental Kr and Xe abundances.  PB94 concluded that there are large enrichments of Kr and Xe in NGC~7027, providing evidence for \emph{s}-process nucleosynthesis and TDU in its progenitor star.

The study of PB94 led Dinerstein (2001) to realize that two anonymous emission lines observed in the near-infrared (NIR) spectra of several PNe are in fact fine-structure transitions of [\ion{Kr}{3}] and [\ion{Se}{4}].  She found that the strengths of the Se ($Z=34$) and Kr lines in IC~5117 and NGC~7027 are consistent with self-enrichment from the \emph{s}-process, and furthermore postulated that the presence of these NIR lines in some PNe but not others implies a spread in \emph{s}-process enrichments among Galactic PNe.

At the onset of our survey, the only other investigations of \emph{n}-capture element abundances in PNe were those of Sterling et al.\ (2002) and Sterling \& Dinerstein (2003a), who detected Ge ($Z=32$) in absorption against the central star continua of six PNe with the \emph{Far Ultraviolet Spectroscopic Explorer} (\emph{FUSE}).  They determined Ge abundances in five of the PNe, four of which are enriched in Ge by factors of $\geq3$--10 relative to solar, depending on the level of Ge depletion into dust.  Subsequently, Sterling et al.\ (2005) determined the gaseous Fe abundance from the \emph{FUSE} spectrum of one of these objects (SwSt~1), and found it to be only slightly depleted ([Fe/S]~=~$-0.35\pm0.12$).  If the elemental depletion pattern of SwSt~1 is similar to that of the diffuse ISM (Savage \& Sembach 1996; Welty et al.\ 1999), this result indicates that Ge is negligibly depleted in the absorption line of sight, and hence is enriched by a factor of five relative to solar in SwSt~1.

Recently, Sharpee et al.\ (2007) identified optical emission lines of Br, Kr, Xe, Rb, and possibly Ba, Pb, Te, and I in five PNe.  They derived abundances for Kr and Xe in each object (and Br in one), and found evidence for \emph{s}-process enrichments in three of the five PNe.  While their correction for the abundances of unobserved ionization stages were only approximate for Br, Kr, and Xe (they assumed the fractional abundances of the observed ions were the same as similar charge states of Ar), they used the same methods to find that the Kr abundance of the Orion Nebulae is approximately solar, as would be expected for an \ion{H}{2} region.

In all, these studies determined \emph{n}-capture element abundances in only eleven Galactic PNe.  Such a small sample divulges little information about \emph{s}-process enrichments and TDU in PNe as a population, and their overall role in the Galactic chemical evolution of trans-iron species.

\subsection{A Large-Scale Survey of \emph{n}-capture Elements in PNe}

We have conducted the first large-scale survey of \emph{n}-capture elements in PNe, by searching for the NIR Se and Kr emission lines first identified by Dinerstein (2001).  We observed 103 Galactic PNe in the $K$~band, and use literature data to expand our sample to 120 objects.  Overall, we have detected [\ion{Kr}{3}]~2.199 and/or [\ion{Se}{4}]~2.287~$\mu$m in 81 objects, for a detection rate of nearly 70\%.  Our study has increased the number of PNe with known \emph{n}-capture element abundances by nearly a factor of ten.  Preliminary results from our survey have been presented in Sterling \& Dinerstein (2003b; 2004; 2005a,b; 2006) and Sterling (2006).

Se and Kr are particularly useful tracers of \emph{s}-process enrichments in PNe, since they are not depleted into dust (Kr is a noble gas, and Se has not been found to be significantly depleted in the diffuse ISM; Cardelli et al.\ 1993).  Furthermore, although the Solar System's Se and Kr are believed to have been formed primarily by the \emph{r}-process and weak \emph{s}-process (Arlandini et al.\ 1999), theoretical models of the main \emph{s}-process indicate that Se and Kr can be significantly enriched in AGB stars and their descendants (Gallino et al.\ 1998; GM00; Busso et al.\ 2001).

The primary challenge in deriving elemental Se and Kr abundances from our observational data lies in correcting for the abundances of unobserved ionization stages.  Sterling et al.\ (2007, hereafter Paper~I) used the photoionization codes Cloudy (Ferland et al.\ 1998) and XSTAR (Kallman \& Bautista 2001; Bautista \& Kallman 2001) to derive formulae for computing Se and Kr ionization correction factors (ICFs).  Unfortunately, the atomic data governing the ionization balance of Se and Kr (photoionization cross-sections and rate coefficients for various recombination processes) are poorly known, and we were forced to use a number of approximations to calculate the cross-sections and rate coefficients for these processes.  We empirically adjusted the Kr atomic data by modeling ten PNe exhibiting emission lines from multiple Kr ions in their optical and NIR spectra, and optimizing the photoionization cross-sections so that our models satisfactorily reproduced the line intensities of the observed Kr ions in these nebulae.  No such correction is possible for the Se atomic data, since to our knowledge no transitions of other Se ions have been clearly identified in PNe.  Therefore, our derived Se abundances will likely be more uncertain than those of Kr.  We conducted Monte Carlo simulations to determine the effect of the atomic data uncertainties on our abundance determinations, and found that these can result in errors approaching 0.3~dex (a factor of two) in the derived Se abundances, and up to 0.2--0.25~dex for Kr.

In this paper, we apply the ICF formulae determined in Paper~I to derive elemental Se and Kr abundances for our full sample of 120 PNe.  In \S2, we discuss our observations and data reduction procedure, and provide an overview of features detected in the spectra.  In \S3, we describe the Se and Kr abundance determinations, and review the literature data we use to compute the ICFs.  The derived abundances are discussed and compared to predictions of theoretical \emph{s}-process models in \S4, and correlations with other nebular abundances and properties are explored in \S5.  We estimate the fraction of all Galactic PNe whose progenitors experienced the \emph{s}-process and TDU in \S6.  Finally, in \S7 we summarize the main conclusions of our study and discuss potential improvements to \emph{n}-capture element abundance determinations in PNe.

\section{OBSERVATIONS AND DATA REDUCTION}

We observed 103 PNe in the $K$~band with the CoolSpec spectrometer (Lester et al.\ 2000) on the 2.7-m Harlan J.\ Smith Telescope at McDonald Observatory.  Each PN was observed from 2.14--2.30~$\mu$m with our survey setting, which consists of a $2\farcs 7 \times90\arcsec$ slit and 75~l/mm grating.  The resolution is estimated to be $R=500$ from the measured widths of calibration lamp lines.  Each PN was observed in adjacent ``on-off'' pairs to correct for sky and instrumental background; the observed PNe are sufficiently compact that we nodded along the length of the slit (by $20$--40$\arcsec$, depending on the nebular diameter) to maximize observing efficiency.  In Table~1, we provide an observing log, including observing dates, exposure times, and resolution (see \S2.2).  We also list relevant properties of each observed object, including whether it is Type~I, its central star type and temperature, morphology, and dust composition (C-rich, O-rich, or both).

The data reduction was carried out with IRAF\footnote{IRAF is distributed by the National Optical Astronomy Observatories, which are operated by the Association of Universities for Research in Astronomy, Inc., under cooperative agreement with the National Science Foundation.}.  We removed cosmic rays and bad pixels with the \textit{crutil} and \textit{crmed} tasks, and corrected for dark current in the detector by applying an additive constant to the two-dimensional spectra.  Flat-fields were obtained by imaging a diffusive flat surface illuminated by an incandescent lamp.  The one-dimensional spectra were extracted using the \textit{apall} task, wavelength calibrated using an Ar lamp source, and dispersion corrected.  The on and off spectra were extracted and reduced separately, and coadded after response correction (see below).

For each PN, we observed at least one A0 standard star at a similar airmass.  The standard star spectra were reduced in the same manner as the PNe, and used to response correct the spectra and remove telluric features.  We also used these stars to perform flux calibrations.  However, most of these objects do not have known NIR photometric fluxes, and we assumed that their $K$ and $V$ magnitudes are equal.  Therefore, while the relative fluxes within each individual spectrum are well-calibrated, the absolute fluxes reported in this paper are only approximate.

We measured emission line fluxes with the \textit{splot} task, assuming Gaussian line profiles.  Flux uncertainties\footnote{All uncertainties cited in this paper are 1-$\sigma$ estimates, and all reported upper limits are 3-$\sigma$ limits.} were estimated by varying the continuum placements of the Gaussian fits.  We did not attempt to deredden our spectra, because of the small wavelength separation of the observed lines and low extinction in the $K$~band.  To illustrate that extinction corrections are not necessary for our data, we dereddened the fluxes for the most highly reddened object in our sample, K~3-17 ($c_{\mathrm{H} \beta}=4.29$; Kaler et al.\ 1996), using the Seaton (1979) interstellar extinction law.  Relative to Br$\gamma$, the (dereddened) intensities of [\ion{Kr}{3}]~2.199 and [\ion{Se}{4}]~2.287~$\mu$m are smaller than the measured fluxes by only 6 and 19\%, respectively.  Reddening corrections are much smaller for other objects in our sample (all other observed PNe have $c_{\mathrm{H} \beta} \leq 3.0$, except K~3-55 and M~1-51, with $c_{\mathrm{H} \beta}=3.82$ and 3.02, respectively).  Therefore, ignoring extinction has a negligible effect on our results.

In Table~2, we present the fluxes of all measured emission lines in our data, in units of 10$^{-13}$~erg~cm$^{-2}$~s$^{-1}$ for \ion{H}{1}~Br$\gamma$, and relative to $F$(Br$\gamma$) for other lines.  When [\ion{Kr}{3}]~2.199 or [\ion{Se}{4}]~2.287~$\mu$m were not detected, we provide upper limits to their fluxes, estimated from the RMS noise in the adjacent continuum.  Figure~1 displays the $K$~band spectra of four representative objects from our sample.

\subsection{Literature Data}

We utilize $K$~band PN spectra from the literature with reported [\ion{Kr}{3}] and/or [\ion{Se}{4}] fluxes or upper limits to expand our sample to 120 objects.  In Table~3, we list these objects along with the reference, [\ion{Kr}{3}] and [\ion{Se}{4}] fluxes, and nebular properties as in Table~1.  Flux uncertainties are taken from the literature, except for the objects from Lumsden et al.\ (2001), who did not cite flux errors.  For these objects, we assume flux uncertainties of 25\%.

For sixteen of the PNe we observed, $K$~band line fluxes have been reported elsewhere (Table~3), which provide useful comparisons to our measurements.  Our [\ion{Kr}{3}] and [\ion{Se}{4}] line fluxes agree within the errors with those from the literature in all but one case.  The lone exception is NGC~6537, in which we detected [\ion{Se}{4}], but the upper limit of Geballe et al.\ (1991) is below our detected flux.

\subsection{Corrections to $F$([\ion{Kr}{3}]) and $F$([\ion{Se}{4}]) in PNe With H$_2$ Emission}

The [\ion{Kr}{3}] and [\ion{Se}{4}] lines are well-resolved from nearby features, except in PNe exhibiting H$_2$ emission (roughly 30\% of our targets).  In these objects, [\ion{Kr}{3}] and [\ion{Se}{4}] may be blended with H$_2$~3-2~$S$(3)~2.201 and H$_2$~3-2~$S$(2)~2.287~$\mu$m, respectively.

To resolve the [\ion{Kr}{3}]/H$_2$~3-2~$S$(3) blend at 2.20~$\mu$m, we re-observed many of the PNe exhibiting H$_2$ emission with a high resolution setting ($1\farcs 0\times90\arcsec$ slit, 240~l/mm grating, with $R=4400$).  We performed these observations for PNe that exhibit the H$_2$~1-0~$S$(0)~2.224~$\mu$m line in the survey resolution data, although poor weather conditions prevented us from obtaining these spectra for a small number of objects, and we did not attempt high resolution measurements for any PNe with spectra reported only in the literature.  The high resolution data span the wavelength range from 2.155--2.205~$\mu$m, and were reduced in the same manner as the survey resolution data, except that a Kr lamp source was used for wavelength calibration.  The much smaller wavelength separation of [\ion{Se}{4}] and H$_2$~3-2~$S$(2) renders this blend unresolvable even at the highest resolution possible with CoolSpec, although it is usually possible to remove the H$_2$ contribution by using the ratios of other H$_2$ lines (see below).

The line fluxes from our high resolution data are listed in Table~4.  In many of these spectra, we detected [\ion{Kr}{3}] but not H$_2$~3-2~$S$(3)~2.201~$\mu$m (see Figure~2), indicating that H$_2$ contributes a negligible amount to the flux of the 2.199~$\mu$m line in the survey resolution spectra of these objects.  The H$_2$~3-2~$S$(2)~2.287~$\mu$m line arises from the same vibrational level and displays a flux less than 1.3 times that of its companion line at 2.201~$\mu$m in all of the H$_2$ excitation models of Black \& van~Dishoeck (1987, hereafter BvD87).  Consequently, we can be confident that, in these cases, the H$_2$ contribution to the 2.287~$\mu$m line is also negligible.

For H$_2$-emitting objects in which the [\ion{Kr}{3}] and H$_2$ lines were not detected in the high resolution spectrum, or no high resolution observations were performed, another method is needed to correct the [\ion{Kr}{3}] and [\ion{Se}{4}] fluxes for possible contamination.  The contribution of H$_2$ to the 2.199 and 2.287~$\mu$m lines depends on the H$_2$ excitation mechanism.  Models of BvD87 show that the strengths of the H$_2$~3-2 lines in the $K$~band are negligible if the H$_2$ is collisionally excited, but may reach 30--40\% of the H$_2$~1-0~$S$(0)~2.224 flux if fluorescently excited.  These two excitation mechanisms can be distinguished by comparing the observed flux ratio $F$(2.248/2.224)~=~$F$(H$_2$~2-1~$S$(1)~2.248)/$F$(H$_2$~1-0~$S$(0)~2.224) with theoretical predictions.  The canonical model of pure fluorescent H$_2$ excitation (in which the $v=3$--2 line strengths are maximal), Model~14 of BvD87, predicts that $F$(2.248/2.224)~=~1.22, while their thermal excitation ($T=2000$~K) model S2 predicts $F$(2.248/2.224)~=~0.38.

In Table~5, we list the fluxes of H$_2$ lines relative to H$_2$~1-0~$S$(0)~2.224~$\mu$m in all of the PN with detected H$_2$ emission, as well as the H$_2$ line ratios predicted by BvD87's Models~14 and S2.  We use our high resolution observations to remove the H$_2$ contributions to the [\ion{Kr}{3}] and [\ion{Se}{4}] fluxes when possible.  If the high resolution spectrum displays [\ion{Kr}{3}] emission but not H$_2$~3-2~$S$(3), we assume that the contribution of H$_2$ to the [\ion{Kr}{3}] and [\ion{Se}{4}] fluxes is negligible.  We detected H$_2$~3-2~$S$(3) in the high resolution spectra of only two PNe, M~1-40 and M~1-57.  The marginal detections of both [\ion{Kr}{3}] and H$_2$~3-2~$S$(3) in the high resolution spectrum of M~1-40 lead to highly uncertain flux determinations, and hence these lines are not useful for correcting the H$_2$ contribution to the blend.  The $F$(2.248/2.224) line ratio indicates that H$_2$ is collisionally-excited in this object, and that corrections to the [\ion{Kr}{3}] or [\ion{Se}{4}] fluxes in the survey resolution data are not necessary.  For this reason, we do not correct the [\ion{Kr}{3}] and [\ion{Se}{4}] fluxes for H$_2$ contamination in M~1-40.  In the case of M~1-57, we use the high resolution data to determine the fractional contribution of H$_2$~3-2~$S$(3) to the feature at 2.199~$\mu$m in the survey resolution data.  We assume that H$_2$~3-2~$S$(2)~2.287~$\mu$m has a flux of 1.3 times that of the 2.201~$\mu$m line in the survey resolution data (as in Model~14 of BvD87) to subtract its contribution to the [\ion{Se}{4}] flux.

For the other objects, we have chosen a cutoff of $F$(2.248/2.224)~=~0.75 to characterize the H$_2$ excitation mechanism in each PN.  If $F$(2.248/2.224)~$>0.75$, we assume that the H$_2$ is fluorescently excited, and subtract the H$_2$~3-2~$S$(3) and $S$(2) fluxes determined from Model~14 and the observed H$_2$~1-0~$S$(0)~2.224~$\mu$m flux.  Alternatively, if $F$(2.248/2.224)~$<0.75$, we consider the H$_2$ to be collisionally excited, in which case no correction is needed for the [\ion{Kr}{3}] and [\ion{Se}{4}] fluxes (BvD87).  In Table~5, we note the model that best describes the H$_2$ excitation mechanism for each PN, the corresponding fluxes of the H$_2$~3-2 lines at 2.201 and 2.287~$\mu$m, and the corrected [\ion{Kr}{3}] and [\ion{Se}{4}] fluxes.

It should be noted that it is not necessarily straightforward to distinguish between fluorescent and thermal H$_2$ excitation.  When the density of the H$_2$-emitting region is sufficiently high, collisions can modify the low-energy vibrational level populations of radiatively-excited H$_2$ (e.g., Sternberg \& Dalgarno 1989).  Thus, it is possible for a PN to have a low $F$(2.248/2.224) while its higher vibrational states exhibit a fluorescently-excited distribution.  This process may be at work in some of the objects of our sample, particularly the high density PNe IC~5117, M~3-25, NGC~6537, NGC~6886, NGC~7027, SwSt~1, and Vy~2-2, in which high resolution observations were not performed or [\ion{Kr}{3}]/H$_2$~3-2~$S$(3) was not detected.  Our values for the [\ion{Kr}{3}] and/or [\ion{Se}{4}] fluxes may be overestimated in some of these objects.  However, Likkel et al.\ (2006), whose resolution was sufficiently high to resolve the 2.20~$\mu$m blend, did not detect H$_2$~3-2~$S$(3) in IC~5117, SwSt~1, or Vy~2-2, indicating that [\ion{Kr}{3}] and [\ion{Se}{4}] are negligibly contaminated in these objects.

\subsection{Overview of the Observed Spectra}

We have detected lines from a number of species in the observed PNe, including \ion{H}{1}~Br$\gamma$, \ion{He}{2}~2.189~$\mu$m, [\ion{Fe}{3}]~2.218 and 2.243~$\mu$m, [\ion{Kr}{3}]~2.199~$\mu$m, [\ion{Se}{4}]~2.287~$\mu$m, and a number of vibrationally-excited H$_2$ lines.  These are the first $K$ band spectra reported for several objects.

Notably, we have detected [\ion{Kr}{3}] and/or [\ion{Se}{4}] in 81 of 120 objects.  This corresponds to a remarkably high detection rate of 67.5\%, considering the low initial abundances of Se and Kr ($\sim$$2 \times 10^{-9}$ relative to H in the Solar System; Asplund et al.\ 2005).  The high detection rate illustrates the utility of the NIR Se and Kr lines for studying \emph{s}-process enrichments in a large number of Galactic PNe.  [\ion{Se}{4}] is more easily detected than [\ion{Kr}{3}] (70 versus 36 detections), owing to the fact that the ionization potential range of Se$^{3+}$ (30.8--42.9~eV) causes it to have a large fractional abundance in many PNe, whereas Kr$^{++}$ exists at lower energies (24.4--37.0~eV) and Kr$^{3+}$ is often the dominant ion.

H$_2$ is detected in several PNe for the first time (He~2-459, K~3-17, M~1-17, M~1-32, M~1-40, M~1-51, M~1-61, M~1-72, M~2-43, M~3-25, M~3-35, NGC~6741, and NGC~6778), while in others we provide the first spectroscopic measurements of H$_2$ (M~1-57, M~1-75, and NGC~6881).  As previously noted by several authors, H$_2$ emission tends to be most prevalent in bipolar PNe (e.g., Zuckerman \& Gatley 1988; Kastner et al.\ 1996; Guerrero et al.\ 2000); 18 of the 28 bipolar PNe (64\%) in our sample exhibit H$_2$ emission.  This is a much higher H$_2$ detection rate than for other morphological types in our sample (12 of 46, or 26\%).

We have also detected [\ion{Fe}{3}]~2.218 and 2.243~$\mu$m in 14 objects, for the first time in all but two PNe (Hb~5 and Vy~2-2; Davis et al.\ 2003; Likkel et al.\ 2006).  These lines are often weak and their fluxes are uncertain (especially since they can be mildly blended with H$_2$~1-0~$S$(0)~2.224 and H$_2$~2-1~$S$(1)~2.248~$\mu$m).  As discussed by Likkel et al.\ (2006), the $K$ band [\ion{Fe}{3}] lines arise from high energy levels, despite their wavelengths, and are thus strongly temperature sensitive.  They can be used to determine gaseous Fe abundances (and hence the level of Fe depletion into dust), and their relative fluxes may be used as density diagnostics.  We do not consider these lines further in the present study.

\section{ABUNDANCE DETERMINATIONS}

In Paper~I, we derived formulae that can be used to correct for the abundances of unobserved Se and Kr ions, and hence derive their elemental abundances.  We use Equations~(1)--(3) of Paper~I to determine the Se and Kr abundances of PNe in our sample.  These ICF formulae are:
\begin{equation}
\mathrm{ICF}(\mathrm{Kr}) = \mathrm{Kr}/ \mathrm{Kr}^{++} = (-0.009205 + 0.3098x + 0.0007978e^{6.297x})^{-1},
\end{equation}
\begin{displaymath}
x = \mathrm{Ar}^{++}/\mathrm{Ar} \geq 0.027;
\end{displaymath}
\begin{equation}
\mathrm{ICF}(\mathrm{Kr}) = \mathrm{Kr}/ \mathrm{Kr}^{++} = (-0.3817 + 0.3796e^{1.083y})^{-1},
\end{equation}
\begin{displaymath}
y = \mathrm{S}^{++}/\mathrm{S} \geq 0.0051;
\end{displaymath}
and
\begin{equation}
\mathrm{ICF}(\mathrm{Se}) = \mathrm{Se}/ \mathrm{Se}^{3+} = (-0.1572 - 0.3532z^{17.56} + 0.153e^{1.666z})^{-1},
\end{equation}
\begin{displaymath}
z = \mathrm{O}^{++}/\mathrm{O} \geq 0.01626.
\end{displaymath}
As discussed in Paper~I, Equation~(1) is likely to be more reliable than Equation~(2), since more Ar ions are detectable in the optical spectra of PNe than S ions, leading to a more robust abundance determination for Ar.  Henry et al.\ (2004) have found that S abundances derived from optical data using model-derived ICFs are systematically lower than \ion{H}{2} region and stellar abundance determinations.  They attribute this ``S~anomaly'' to the inability of photoionization models to correctly determine the ionic fraction of S$^{3+}$ in PNe, which can only be observed in the infrared.  Therefore, we preferentially use Equation~(1) to determine elemental Kr abundances, although in some cases it is necessary to utilize Equation~(2) due to uncertain Ar$^{++}$/Ar values or the lack of a reported Ar abundance.

Equations~(1)--(3) require ionic and elemental abundances of O, S, and Ar.  Furthermore, the electron temperature $T_{\rm e}$ and density $n_{\rm e}$ of each PN are needed to determine the Se$^{3+}$ and Kr$^{++}$ ionic abundances.  We have taken this information from the literature (in some cases, it was necessary for us to compute O$^{++}$, S$^{++}$, and Ar$^{++}$ ionic abundances from the reported line fluxes), and discuss this data in \S3.1.  In \S3.2, we present the Se and Kr abundance determinations for our full sample of objects.

\subsection{Physical Parameters and Abundances from the Literature}

We have conducted an extensive literature search to obtain abundance information for the PNe in our sample.  We consider only abundance determinations from the last 25 years (with the exceptions of M~3-35 and Vy~1-1), which use more recently determined atomic data.  Furthermore, we consider only abundance determinations using collisionally-excited lines.  There is a well-known discrepancy between abundances derived from recombination and collisionally-excited lines, with recombination lines generally indicating larger ionic abundances (e.g., Rola \& Stasi\'{n}ska 1994; Peimbert et al.\ 1995a,b; Liu et al.\ 2004b; Tsamis et al.\ 2003, 2004; Wesson et al.\ 2005).  The cause for this discrepancy is not well understood at this time (see Liu et al.\ 2004b for a discussion).

We have selected up to five abundance references for each object in our sample, which we use to derive ICFs for Se and Kr.  The full set of references is given in Table~6, along with references for other nebular properties, including the NIR line fluxes, central star temperatures, morphologies, and dust compositions reported in Tables~1 and 3.  The indices assigned to the references in Table~6 are used in other tables throughout this paper to indicate the source of the adopted parameters.

For each PN, we have chosen a ``primary'' abundance reference, based on our judgment of the reliability of the abundance analysis compared to other sources.  Our main criteria for selecting a primary reference are as follows: (1) We preferentially use abundance determinations derived from spectra covering multiple wavelength regimes (e.g., UV and IR in addition to optical).  Several ions that are not detectable in the optical can be observed in the UV and IR, allowing for more robust determinations of C, N, O, Ne, and S abundances.  (2) Abundances derived from deep, high resolution optical data are preferred.  This allows weak transitions to be detected, and blended lines used in abundance analyses to be resolved.

In Table~7 (available in the electronic version of this article), we list the $T_{\rm e}$, $n_{\rm e}$, and ionic abundances needed for our ICFs reported in the primary abundance references.  We adopt the cited $T_{\rm e}$ and $n_{\rm e}$ uncertainties when given, and otherwise assume errors of $\pm1000$~K in $T_{\rm e}$ and 20\% in $n_{\rm e}$.  In some sources, line fluxes were reported but ionic abundances were not provided.  In these cases, we derived the Ar$^{++}$/H$^+$, S$^{++}$/H$^+$, and O$^{++}$/H$^+$ ionic abundances using the dereddened line intensities, [\ion{O}{3}] temperatures, and $n_{\rm e}$ from these references (these values are marked in Table~7) with the aid of the IRAF task \textit{nebular.ionic} (Shaw \& Dufour 1995).

Table~8 (available in the electronic version of this article) lists the elemental abundances of the PNe in our sample, taken from the primary abundance references.  We were unable to find any abundance determinations for K~3-62, and only very limited and uncertain abundances are available for K~3-17, K~3-55, M~3-28, M~3-35, and Vy~1-1.  New abundance determinations for these objects are needed.  We did not attempt to ``update'' any of the abundance determinations from older references by re-deriving them with newer atomic data (e.g., transition probabilities and effective collision strengths).  The abundances were determined in diverse manners, using photoionization models or various ICFs, and we feel that using an arbitrary method to homogenize the abundance determinations is not warranted.

Unfortunately, many of the references in Table~6 do not explicitly state the uncertainties in their ionic and elemental abundance determinations.  The uncertainties for the abundances we use in our ICFs are likely to be major sources of error in our Se and Kr abundance determinations.  For objects in which we derived O$^{++}$, Ar$^{++}$, and S$^{++}$ ionic abundances from the published line intensities, we estimated the uncertainties by calculating them at the minimum and maximum $T_{\rm e}$ and $n_{\rm e}$ allowed within the 1-$\sigma$ uncertainties.  On average, we find that the ionic abundance uncertainties for objects in the Aller \& Keyes (1987) sample are about 20\%.  We take this value to be representative of the uncertainties in the ionic abundances in Table~7, except when error estimates are reported in the references, or we explicitly computed the ionic abundances (and their uncertainties).  For elemental abundances, when uncertainties were not stated in the source paper we assume them to be 20\% for He (50\% if the abundance is marked as uncertain), and 30\% for other elements (75\% if marked as uncertain).  However, for references using high quality data (including UV and/or IR spectra, or high resolution optical spectra), we use lower abundance uncertainties of 20\% (60\% if marked as uncertain).

While this method of error analysis is crude, it is not possible to provide more robust uncertainty estimates when they are not explicitly reported in the source papers.  We feel that using the error estimates described above for the ionic and elemental abundances taken from the literature is preferable to ignoring the uncertainties altogether, since these are needed to determine the accuracy of our ICFs.

\subsection{Se and Kr Abundance Determinations}

\subsubsection{Ionic and Elemental Abundances Relative to H}

We compute Se$^{3+}$ and Kr$^{++}$ ionic abundances or upper limits for all objects in our sample from the observed [\ion{Kr}{3}]~2.199, [\ion{Se}{4}]~2.287~$\mu$m, and \ion{H}{1}~Br$\gamma$ fluxes.  To compute the emissivities of [\ion{Kr}{3}]~2.199 and [\ion{Se}{4}]~2.287~$\mu$m, we use the $T_{\rm e}$ (from [\ion{O}{3}] when available) and $n_{\rm e}$ values in Table~7.  We employ a five-level model atom to solve for the Kr$^{++}$ level populations, with transition probabilities from Bi\'{e}mont \& Hansen (1986) and collision strengths calculated by Sch\"{o}ning (1997).  Se$^{3+}$ has a $4p$ ground electronic configuration, so a two-level atom is sufficient to solve for the [\ion{Se}{4}]~2.287~$\mu$m emissivity; we utilize transition probabilities from Bi\'{e}mont \& Hansen (1987) and collision strengths calculated by Butler (2007, in preparation).  Energy levels for these two ions are taken from the NIST Atomic Spectra Database.\footnote{National Institute of Standards and Technology Atomic Spectra Database v3.0; see http://physics.nist.gov/PhysRefData/ASD/index.html}  We determine the Br$\gamma$ emissivity for each PN by interpolating upon Tables~B.5 and B.9 of Dopita \& Sutherland (2003).

The Se$^{3+}$ and Kr$^{++}$ ionic abundances are given in the second and third columns of Table~9.  The error bars account for uncertainties in the line fluxes, $T_{\rm e}$, and $n_{\rm e}$.  In general, flux uncertainties dominate the error bars to the Se$^{3+}$ and Kr$^{++}$ abundances, except in PNe with low electron temperatures, where uncertainties in $T_{\rm e}$ can be equally important.  The effects of uncertainties in $n_{\rm e}$ are negligible, owing to the large critical densities of [\ion{Kr}{3}]~2.199 ($2.1\times10^7$~cm$^{-3}$) and [\ion{Se}{4}]~2.287~$\mu$m ($4.6\times10^6$~cm$^{-3}$).

We compute ICFs for Se and Kr using Equations~(1)--(3) and the ionic and elemental O, S, and Ar abundances listed in Tables~7 and 8.  The abundance uncertainties are propagated into the ICFs, as is the dispersion about the fits to the ICFs (see Paper~I).  As discussed above, we use Equation~(1) to calculate the Kr ICF, except when [\ion{Ar}{3}] was not detected in the nebula, or the ICF was very large (and hence uncertain) compared to that from Equation~(2).

For comparison, we plot the Kr ICFs derived from Equation~(1) against those from Equation~(2) in Figure~3.  We do not plot the error bars for reasons of clarity (see Table~9).  It can be seen that the ICFs generally agree with each other within the (significant) scatter, although there is a slight tendency for the ICFs from Equation~(1) to be larger than those from Equation~(2).  This is likely due to the underestimated S abundances derived from optical data (Henry et al.\ 2004); the lower S abundances (and hence larger S$^{++}$ ionic fractions) cause the Kr ICFs from Equation~(2) to be smaller than those derived using Equation~(1).  The most discrepant objects in Figure~3 are generally very high or low excitation PNe, and have very low Ar$^{++}$ or S$^{++}$ fractional abundances, causing the ICFs to be large and uncertain.

The computed Se and Kr ICFs for each object are listed in Table~9.  We use these to compute Se and Kr elemental abundances,\footnote{We report elemental abundances in the notation [X/H] = log(X/H)~--~log(X/H)$_{\odot}$, where solar abundances are taken from Asplund et al.\ (2005).} with error bars accounting for uncertainties in the ICFs and the Se$^{3+}$ and Kr$^{++}$ abundances.  In Table~9, we also list Se and Kr abundances derived using NIR line fluxes from the literature.  As discussed in \S2.1, the Se and Kr fluxes we measured are in excellent agreement with those from the literature, leading to abundance determinations that are consistent within the errors in all cases (with the exception of NGC~6537; see \S2.1).

The Se and Kr abundances have been determined to within a factor of 2--3 (0.3--0.5~dex) for most objects.\footnote{Note that this does not include uncertainties in the Se and Kr atomic data used to derive the ICF formulae, which we discussed in Paper~I.  The atomic data uncertainties add a systematic error of up to 0.3~dex in the derived Se abundances, and up to 0.2--0.25~dex for Kr.}  The largest source of error stems from the derived ICFs, whose uncertainties in many cases are larger than 50\%, and sometimes exceed 100\%.  The only way to reduce these uncertainties is to observe additional Kr ions (no other Se ions have been clearly detected in PNe), which requires optical observations.  We considered ten PNe from our sample with optical [\ion{Kr}{4}] detections in Paper~I, and derived their Kr abundances with photoionization models.  In general, we found good agreement between our model-derived Kr abundances and those derived from [\ion{Kr}{3}] lines with Equations~(1)--(3).

\subsubsection{\emph{s}-process Enrichments: Choice of a Reference Element}

In order to determine whether Se and Kr are enriched in a PN, it is necessary to scale their abundances to a reference element whose abundance is indicative of the object's metallicity.  It is not possible to use the usual stellar metallicity indicator Fe, which can be depleted into dust in PNe (Perinotto et al.\ 1999; Sterling et al.\ 2005), and therefore we consider O for this purpose.  O is not expected to be processed by most PN progenitor stars (Kaler~1980; Henry~1989, 1990), with two possible exceptions.

First, O may be enriched during TDU in low-metallicity objects (Garnett \& Lacy 1993; P\'{e}quignot et al.\ 2000; Dinerstein et al.\ 2003; Leisy \& Dennefeld 2006).  Since our sample consists almost exclusively of Galactic disk objects with near-solar metallicities, O should not be enriched in objects in our sample.  Second, Type~I PNe, which are descendants of IMS, may exhibit O depletion from ON-cycling during hot bottom burning\footnote{Theoretical predictions indicate that ON-processing during the second dredge-up results in negligible O destruction in solar-metallicity stars (e.g., $\lesssim 10$\% reduction in the surface O abundance; Becker \& Iben 1979; Boothroyd \& Sackmann 1999; Karakas 2003).  Therefore, any significant O destruction in Type~I PN progenitors is likely to occur during HBB.} (HBB; Peimbert 1985; Henry 1990).  Kingsburgh \& Barlow (1994) found no evidence for ON-processing in their sample of Type~I PNe, and questioned the evidence presented in previous studies.  However, recent studies have found more compelling evidence for O depletion in Type~I PNe in both the Galaxy (Perinotto \& Corradi 1998; Pottasch \& Bernard-Salas 2006) and Magellanic Clouds (Leisy \& Dennefeld 2006).  To examine the possibility that O is depleted in Type~I PNe in our sample, we compare the O abundances to those of Ar, S, and Cl, which are unaffected by nucleosynthesis in PN progenitors.  We focus our discussion on Ar/O, since Ar abundances are generally better determined than those of S and Cl.

We find that Ar/O is systematically higher in Type~I PNe in our sample than in non-Type~I objects by a factor of 2.1.  This suggests that Type~I PNe do indeed suffer oxygen depletion during HBB.  This claim is supported by the fact that O/H is on average $\sim$15\% lower in the Type~I PNe in our sample than in non-Type~I objects.\footnote{Note that the comparison between O abundances of Type~I and non-Type~I PNe is not strictly an indicator of the overall level of O destruction in Type~I objects.  First, Type~I PNe arise from a younger population of stars, and should have larger initial O abundances than non-Type~I PNe.  Secondly, the proximity of Type~I PNe to the Galactic plane makes them difficult to detect at large distances.  Since non-Type~I PNe have a larger Galactic scale height (e.g., Corradi \& Schwarz 1995; Torres-Peimbert \& Peimbert 1997), they are observable at larger distances, and will exhibit a wider range of initial O abundances due to the chemical gradient in the Galaxy (e.g., Maciel \& Quireza 1999; Pottasch \& Bernard-Salas 2006).  Comparing the average abundances of these two classes of PNe therefore does not take into account the spread in O abundances due to the Galactic chemical gradient, nor the differences in the primordial abundances of stars from different populations.}  We note that these findings are not predicated upon a small number of objects.  Indeed, the only Type~I PNe in our sample that exhibit [Ar/O]~$<0.3$ are M~3-25, Me~2-2, NGC~1501, and NGC~6833.  Furthermore, these results are reproduced in other abundance studies of the same objects, which also find elevated [Ar/O] (see \S3.2.3).  We cannot exclude the possibility that Ar abundances are systematically overestimated in Type~I PNe.  However, this does not appear to be an excitation effect, since high [Ar/O] values are found over the entire range of O$^{++}$/O$^+$ of the Type~I PNe in our sample, from 1.25 (NGC~6778; Aller \& Czyzak 1983) to 13.7 (Me~1-1; Shen et al.\ 2004).  Also, if the high abundances relative to O are caused by errors in the atomic data, one would not expect a discrepancy between Type~I and non-Type~I objects.  Finally, although S and Cl abundances are generally not as well-determined as Ar, S/O and Cl/O are also larger in Type~I PNe than in non-Type~I objects by comparable amounts to Ar/O (a factor of 2.5 and 1.7, respectively), which provides further evidence that the Ar abundances are not in error.

Theoretical studies predict that O depletion can occur in IMS via ON-processing during HBB (e.g., Karakas 2003; Ventura \& D'Antona 2005a,b; Karakas et al.\ 2006), although this process is more efficient in low metallicity stars.  The 5 and 6~M$_{\odot}$ solar metallicity models computed by Karakas (2003) and Karakas et al.\ (2006) predict O depletions of only 0.05--0.1~dex, assuming solar compositions from Anders \& Grevesse (1989).  Adopting the more recent CNO abundances from Asplund et al.\ (2005) reduces the solar metallicity, and HBB is more efficient, producing O depletions of $\sim$0.15~dex in a 6~M$_{\odot}$, 1~$Z_{\odot}$ model (Karakas 2003; Karakas 2006, private communication).  Model uncertainties, including the treatment of mass loss and convection, affect the lifetime of the AGB phase and the temperature at the bottom of the convective envelope.  These in turn affect the HBB lifetime and efficiency (Ventura \& D'Antona 2005a,b; Karakas 2006, private communication), and thus it is possible that O depletions of a factor of two can occur in solar metallicity AGB stars.  However, efficient ON-cycling during HBB can lead to very large N enrichments, often larger than observed in Type~I PNe (Marigo et al.\ 2003).  Therefore, caution must be used in concluding whether efficient HBB is the cause of the observed O depletions in Type~I PNe.  This problem deserves further theoretical attention.

From the above arguments, we conclude that Ar is a more dependable metallicity tracer than O for Type~I PNe.  On the other hand, we have elected to use O as a reference element in non-Type~I PNe,\footnote{Four exceptions should be noted, where we use Ar as the reference element for non-Type~I PNe.  M~1-31, NGC~6881, and Vy~2-2 show signs of N enrichments, but their N/O falls just below the cutoff of 0.8 that would qualify them as Type~I PNe (Kingsburgh \& Barlow 1994).  [Ar/O] is also high in these objects, indicating that O depletion from ON-cycling may have occurred.  For IC~4997, Hyung et al.\ (1994b) found a high O abundance, which is inconsistent with the subsolar S, Cl, and Ar abundances for this object, and hence is likely to be uncertain.} since O abundances are generally better determined than Ar in PNe.  This is illustrated by the fact that some non-Type~I objects display [Ar/O] abundances more than 0.2~dex above or below the solar ratio.  The Ar abundances in many of these discrepant objects were determined from a single Ar ion, or disagree with determinations from other studies of the same object.  Therefore, the discordant [Ar/O] ratios in non-Type~I PNe are likely to be due to uncertainties in the Ar abundances.

It should be noted that the derived values of the Se and Kr enrichments depend on the source of solar abundances we adopt.  Throughout this paper, we utilize the solar composition reported by Asplund et al.\ (2005).  However, the solar Ar abundance from Asplund et al.\ is more than a factor of two (0.37~dex) smaller than that reported by Lodders (2003), although the solar O abundances of these two studies are identical within the cited errors.  If we adopted the Lodders (2003) solar abundances, her higher derived solar Ar abundance would lead us to classify more Type~I PNe as \emph{s}-process enriched than we do using the Asplund et al.\ solar abundances.

However, we believe that the solar Ar abundance of Asplund et al.\ (2005) is more reliable than that of Lodders (2003).  The abundances of Ar and other noble gases are difficult to determine, since these species cannot be directly observed in the solar photosphere.  Instead, the Ar abundance must be estimated from observations of the solar corona and measurements of solar energetic particles.  Asplund et al.\ used these techniques to obtain the Ar/O ratio, and then scaled the Ar abundance to the photospheric O value.  On the other hand, Lodders (2003) utilized local nuclear statistical equilibrium arguments to interpolate the $^{36}$Ar isotopic abundance from those of $^{28}$Si and $^{40}$Ca.  She determined the elemental Ar abundance from the $^{36}$Ar abundance and the isotope ratios measured in the solar wind by Wieler (2002).  This alternate method of determining the Ar abundance led to an Ar/O ratio that is 0.2--0.3~dex larger than other recent determinations of the solar Ar/O ratio (e.g., Anders \& Grevesse 1989; Grevesse \& Sauval 1998; Asplund et al.\ 2005).

The small Ar/O ratio determined by Lodders (2003) is inconsistent with our data.  Specifically, we find that the average logarithmic Ar/O ratio in non-Type~I PNe is $-2.74$, which is significantly closer to the solar value of $-2.48$ determined by Asplund et al.\ (2005) than that of Lodders ($-2.14$).  In addition, using the Lodders (2003) solar Ar value, we find that [Se/Ar] and [Kr/Ar] are systematically larger by a factor of two in non-Type~I PNe than [Se/O] and [Kr/O].  This inconsistency does not arise when the Asplund et al.\ (2005) solar values are used (see \S5.2.2).  Based on these arguments, we have chosen to use the solar abundances compiled by Asplund et al.\ (2005) in our analysis, and do not further consider the solar composition reported by Lodders (2003).

In Table~10 we list the Se and Kr abundances relative to O and Ar.  Since we find evidence for O depletion in Type~I PNe, we enclose the [Kr/O] and [Se/O] values for these objects in parentheses to indicate that they may be unreliable.  We use the [Se/Ar] and [Kr/Ar] abundances to determine \emph{s}-process enrichment factors in Type~I PNe, and [Se/O] and [Kr/O] for non-Type~I objects.

It is interesting to compare our derived [Kr/Ar] abundances with those of Sharpee et al.\ (2007) for the two objects that are common to our sample, IC~418 and NGC~7027.  Sharpee et al.\ derived Kr abundances using optical emission lines, and ICFs based on the similar ionization potential ranges of Kr and Ar ions with the same charge.  While our [Kr/Ar] values are systematically lower for IC~418 and NGC~7027 compared to Sharpee et al.\ (by 0.49 and 0.19 dex, respectively), this is primarily due to their use of the solar abundances of Lodders (2003) instead of Asplund et al.\ (2005).  If the 0.37~dex offset between the Lodders and Asplund et al.\ solar Ar value is subtracted from the [Kr/Ar] abundances of Sharpee et al., then our derived [Kr/Ar] agree within the errors with their determinations.

\subsubsection{Dependence of Se and Kr Abundances on the Choice of Abundance Reference}

Tables~11--14 (available only in the electronic version of this article) compare our Se and Kr abundance determinations using values of $T_{\rm e}$, $n_{\rm e}$, and abundances reported in different studies.  These tables display the same information as Tables~7--10, except that data from all of the abundance studies listed in Table~6 are given.  To avoid redundancy, in these tables we do not provide data for objects whose abundances have been determined in only one of the references listed in Table~6.  In many cases the derived Kr and (especially) Se abundances are in good agreement when temperatures and abundances from different sources are used.  However, some discrepancies are found (see Table~14).  For example, the [Kr/O] abundance of NGC~6629 ranges from 0.32 to 1.30~dex, depending on which abundance reference is used.  Similarly, [Se/O] in M~1-17 ranges from $-0.11$ to 0.51 dex when abundances and temperatures from different sources are utilized.

These discrepancies most often arise when the Ar$^{++}$, S$^{++}$, or O$^{++}$ ionic fractions used in our ICFs are small and uncertain.  However, this is also an effect of the inhomogeneous collection of abundance determinations we have utilized.  The O, S, and Ar abundances have been derived using various methods (e.g., photoionization modeling or empirical ICF methods, where the ICFs used by one group may differ from others).  Furthermore, some objects have been observed only in the optical spectral region, and hence the abundances of some elements (e.g., S) may be quite uncertain.

These effects underscore the dependence of our Se and Kr abundance determinations on those of lighter elements.  We have attempted to use the most reliable O, S, and Ar abundances available to compute the Se and Kr ICFs, preferentially using abundances determined from multiple wavelength regimes and high resolution data.  However, some of the objects in our sample have only been observed with low-dispersion instruments in the optical, and in these cases the ionic fractions used in our ICFs may be uncertain.

\section{SELF-ENRICHMENT OF SE AND KR AND COMPARISON TO MODEL PREDICTIONS}

\subsection{Criterion for Classifying Nebulae as Self-Enriched}

Our derived abundances (Tables~9--10) indicate that Se and Kr are enriched in several objects in our sample.  We find a range of abundances, from $-0.05$ to 1.89 in [Kr/(O,~Ar)]\footnote{We use the notation [Kr/(O,~Ar)] to remind the reader that we use different metallicity indicators for non-Type~I and Type~I PNe, as discussed in \S3.2.2.}, and $-0.56$ to 0.90 in [Se/(O,~Ar)]\footnote{We ignore a few exceptional objects whose Se abundances are highly uncertain.  The [Se/O] value of M~1-11 is very uncertain due to the large and uncertain ICF, and while [\ion{Se}{4}] was marginally detected in Hb~5, the derived [Se/Ar]~=~$-0.89$ is so far below the derived [Kr/Ar] that it should be regarded as uncertain.}.

In order to determine whether Se and Kr are self-enriched in a PN by the \emph{s}-process and TDU, it is necessary to consider the amount of primordial scatter in the initial abundances of these elements at the time of the progenitor star's formation.  Travaglio et al.\ (2004) found that the dispersion in the abundances of other light \emph{n}-capture elements (Sr, Y, and Zr; $Z=38$--40) is roughly 0.2~dex in unevolved stars of near-solar metallicity.  This confirmed the findings of Burris et al.\ (2000) for Y and Zr, and is similar to the star-to-star scatter of heavier \emph{n}-capture elements at near-solar metallicities (Simmerer et al.\ 2004).  Since the PNe in our sample are primarily Galactic disk objects with approximately solar metallicities, Se and Kr enrichments in excess of 0.2--0.3~dex relative to solar can generally be attributed to \emph{s}-process nucleosynthesis in their progenitor stars.  Although some individual objects may have larger initial Se and Kr abundances, leading us to incorrectly label them as self-enriched, others may have sufficiently low primordial abundances that they could have experienced \emph{s}-process nucleosynthesis and still exhibit relatively low [Se/(O,~Ar)] and [Kr/(O,~Ar)].

With these caveats in mind, we conservatively take [Se/(O,~Ar)] and [Kr/(O,~Ar)] values in excess of 0.3~dex to indicate that a PN is self-enriched by \emph{s}-process nucleosynthesis in its progenitor star.  According to this criterion, we find that 41 of the 79 PNe with determined Se and/or Kr abundances are self-enriched.  Most of the objects exhibiting [\ion{Kr}{3}] emission are enriched (28 out of 33 objects for which [Kr/(O,~Ar)] could be determined, or 85\%), and the average [Kr/(O,~Ar)] is 0.98$\pm$0.31~dex in these PNe.  In contrast, Se is enriched above 0.3~dex in only 24 of 68 objects with determined [Se/(O,~Ar)] (35\%).  The average Se enrichment is much smaller than that of Kr: [Se/(O,~Ar)]~=~0.31$\pm$0.27.

\subsection{Comparison to Predictions of \emph{s}-process Nucleosynthesis Models}

It is interesting to compare the observed Se and Kr enrichments to theoretical predictions.  Models of \emph{s}-process nucleosynthesis in AGB stars have been presented by Gallino et al.\ (1998; hereafter G98), GM00, and Busso et al.\ (2001; hereafter B01).  These studies all considered \emph{s}-process nucleosynthesis under radiative $^{13}$C burning conditions, as was shown to characterize the \emph{s}-process in low- and intermediate-mass stars by Straniero et al.\ (1995).  We consider results from these papers for metallicities in the range 0.1--2.0~$Z_{\odot}$, the metallicity range of the PNe in our sample, according to their O and Ar abundances.

In the models of G98 and B01, the mass of the layer in which the \emph{s}-process occurs (the $^{13}$C pocket) was treated as a free parameter, while GM00 reported enrichment factors only for a single choice of $^{13}$C pocket mass.  The $^{13}$C pocket mass, along with the metallicity, governs the ratio of free neutrons to Fe-peak seed nuclei, and hence controls the element-by-element pattern of \emph{s}-process enrichments (G98; GM00; B01).  Otherwise similar AGB stars are observed to exhibit a large scatter in \emph{s}-process enrichments, indicating a range of $^{13}$C pocket masses in stars with comparable masses and metallicities.  This is to be expected, given the stochastic nature of the processes that are presumed to form the $^{13}$C pocket (Herwig 2000; Denissenkov \& Tout 2003; Herwig et al.\ 2003; Siess et al.\ 2004).

According to the models, Se and Kr are more strongly enriched in stars with larger $^{13}$C pocket masses, a general feature seen for \emph{n}-capture elements in AGB stars of near-solar metallicity (B01).  Other factors, such as the metallicity (in the range 0.1--2.0~$Z_{\odot}$) and initial mass, play minor roles in the Se and Kr enrichment factors.  In general, the observed Se and Kr overabundances are in agreement with the theoretically predicted enrichments, given the uncertainties in our abundance determinations and the likelihood of star-to-star variations in the dredge-up efficiency and $^{13}$C pocket mass (G98; B01).

These models also predict that Kr should exhibit larger enrichment factors than Se.  This is likely to be the reason that so many more PNe exhibit Kr enrichments larger than 0.3~dex than Se; Se may well be enriched by the \emph{s}-process in several of the observed objects, but its enrichment factor more often falls below our criterion of 0.3~dex used to discern \emph{s}-process enrichments in the progenitor stars from primordial scatter than is the case for Kr.  The Kr enrichment relative to Se depends on the stellar parameters, $^{13}$C pocket mass, and which models are used (GM00 predict smaller [Kr/Se] than G98 and B01).  Nevertheless, all of the models predict that [Kr/Se]~=~0.0 to 0.5.

We have detected emission from both Se and Kr in 25 PNe, and [Se/(O,~Ar)] and [Kr/(O,~Ar)] were determined in 22 of these objects.  Excluding objects with uncertain ICFs or marginal Se and/or Kr detections, we find [Kr/Se]~=~0.5$\pm$0.2 for 18 PNe, which is in agreement with the model predictions.  The [Kr/Se] values of these PNe span a wide range (from $-0.01$~dex in NGC~6445 to 0.79~dex in Hb~12), which is at least partially due to uncertainties in the abundance determinations (a factor of 2--3 for most objects).

Some objects in our sample\footnote{Hb~5, Hb~12, Hu~2-1, IC~418, K~3-60, M~1-5, M~1-11, M~1-17, NGC~6629, and NGC~6644.  Note that we include objects in which only an upper limit is available for the Se or Kr abundance.} exhibit discrepant [Kr/Se] values (e.g., $>0.7$~dex or $<0.0$~dex) compared to the theoretical models.  In almost all of these cases, either [\ion{Kr}{3}] or [\ion{Se}{4}] were marginally detected, or one of the ICFs is uncertain.  Aside from the cases where the Se and Kr abundances are uncertain, we conclude that our abundance determinations largely agree with theoretical predictions.

\section{ABUNDANCE CORRELATIONS}

We have detected Se and/or Kr emission in 81 of 120 objects, which is a sufficiently large sample to search for correlations between \emph{s}-process enrichments and other nebular properties.  In this section, we compare the derived Se and Kr abundances with those of other elements, and with progenitor mass, dust chemistry, and central star type.  To quantify the robustness of correlations discussed below, we compute the correlation coefficient $r$ between each pair of quantities, as well as the probability $p_{r=0}$ that no correlation exists (i.e., the significance of the correlation), computed from $r$ and the number of data points.  To determine whether two distributions are statistically different, we use Kolmogorov-Smirnov (KS) tests to compute the probability $p_{\rm ks}$ that the distributions are drawn from the same cumulative distribution function (Press et al.\ 1992).

We begin by discussing the detection rates of Se and Kr in different morphological and population subclasses of PNe, and then examine correlations between Se and Kr enrichments and other nebular properties.

\subsection{Se and Kr Detection Rates Vs.\ Nebular Properties}

Before investigating the Se and Kr abundances in PNe with different nebular properties, it is illustrative to inspect the detection rates of [\ion{Se}{4}] and [\ion{Kr}{3}] in various subclasses of PNe.  In general, it is expected that Se and Kr are more easily detected when enriched by the \emph{s}-process, and thus PN subclasses with high detection rates of Se and Kr may be more highly \emph{s}-process enriched.  This is not an exact correspondence, since our abundance determinations indicate that we are sometimes able to detect Kr and especially Se even when they are not enriched.  However, this allows us to consider all PNe with Se and Kr detections, even when it was not possible to determine their abundances.

We have divided our sample into subclasses based on progenitor mass, morphology, central star type, and dust composition.  Type~I and bipolar PNe are believed to have intermediate-mass progenitor stars ($>3$--4~M$_{\odot}$), based on their He and N enrichments, Galactic distribution, and stellar and nebular masses (Peimbert 1978; Kingsburgh \& Barlow 1994; Corradi \& Schwarz 1995; G\'{o}rny et al.\ 1997; Torres-Peimbert \& Peimbert 1997; Stanghellini et al.\ 2002; see \S5.2).  Some PNe have H-deficient, C-rich central stars that exhibit emission features similar to massive Wolf-Rayet stars (Tylenda et al.\ 1993; Acker \& Neiner 2003).  These are classified as [WC] or [WO] stars, or weak emission line stars (WELS) if they exhibit weak and narrow stellar emission lines.

Table~15 lists the detection rates of [\ion{Kr}{3}] and [\ion{Se}{4}] for our full sample of PNe, as well as for Type~I and non-Type~I PNe, different morphological classes, central star types, H$_2$-emitting PNe, and objects with various dust chemistries.  We have detected [\ion{Kr}{3}] in 36 of 120 objects, for a detection rate of 30.0\%, and [\ion{Se}{4}] in 70 objects for a detection rate of 58.3\%.  We also note the detection rate of Se \textit{or} Kr emission.

The first correlation to be noted in Table~15 is that the detection rates of Se and/or Kr in Type~I PNe (41.4\%)  are lower than in non-Type~I objects (75.8\%).  This result suggests that Type~I PNe are on average less enriched in \emph{s}-process nuclei than other PNe.

On the other hand, while the Se detection rate is lower in bipolar PNe compared to elliptical nebulae (which have less massive progenitor stars; Stanghellini et al.\ 2002), the Kr detection rate is similar for these morphological classes.  However, the high detection rate of Kr in bipolar PNe can be largely explained by the ease of its detection in H$_2$-emitting PNe (Table~15).  As discussed in \S2.3, bipolar PNe have a stronger tendency to exhibit H$_2$ emission than other morphological types (Zuckerman \& Gatley 1988; Kastner et al.\ 1996; Guerrero et al.\ 2000).  [\ion{Kr}{3}] is much more easily detected in PNe exhibiting H$_2$ emission (59\%) than those which do not (16.0\%).  In contrast, [\ion{Se}{4}] has a comparable detection rate in PNe with and without detected H$_2$ emission.  Our high resolution spectra of H$_2$-emitting PNe indicate that the high detection rate of [\ion{Kr}{3}] in these objects is not due to confusion with the H$_2$~3-2~$S$(3)~2.201~$\mu$m line.  Instead, we believe that [\ion{Kr}{3}] is more easily detected in H$_2$-emitting PNe because these objects have a substantial amount of neutral and low-excitation material, and the low ionization potential range of Kr$^{++}$ (24.4--37.0~eV) causes it to have a larger fractional abundance in these objects.  [\ion{Se}{4}] does not show this tendency, because Se$^{3+}$ resides in higher excitation regions than Kr$^{++}$.

There is not a strong correlation between the detection rates in PNe with different central star types.  Kr tends to be more easily detected in [WC] PNe than in objects with WELS or H-rich stars, although the low detection rate in WELS PNe is likely an ionization balance effect (few WELS PNe exhibit H$_2$ emission, a tracer of neutral and low-excitation material, relative to those with [WC] or H-rich nuclei).  Se is detected more often in nebulae with either [WC] or WELS nuclei than those without.

While few objects in our sample currently have known dust chemistries, the detection rates in objects with different dust compositions are striking.  Se and/or Kr are detected in 90\% of PNe with C-rich or mixed (C-rich and O-rich) dust, but only 50--60\% of PNe with O-rich or unknown dust chemistry.  The objects exhibiting C-rich dust all likely experienced TDU and \emph{s}-process nucleosynthesis.  Furthermore, Se or Kr are detected in each of the two PNe exhibiting the 21~$\mu$m dust emission feature, which is associated with post-AGB stars that are strongly enriched in C and \emph{s}-process products (Kwok et al.\ 1989; Van~Winckel 2003 and references therein).  The high detection rate of Se and Kr in objects with C-rich dust composition provides evidence that C-rich PNe have larger \emph{s}-process enrichments than other objects, as is theoretically expected.

We now investigate correlations between the Se and Kr abundances and other nebular properties.

\subsection{Correlations With Progenitor Star Mass}

\subsubsection{Indicators of Progenitor Mass}

Type~I PNe are believed to be descendants of IMS ($M>3$--4~M$_{\odot}$), and are enriched in He and N, as would be expected if they experienced second dredge-up and HBB (Becker \& Iben 1979; Boothroyd et al.\ 1993).  Peimbert (1978, hereafter P78) originally classified objects with He/H~$>0.125$ and N/O~$>0.5$ as Type~I PNe.  Kingsburgh \& Barlow (1994, hereafter KB94) revised these criteria so that only objects with N/O ratios larger than their progenitor's initial (C~+~N)/O  are classified as Type~I PNe (corresponding to N/O~$>0.8$ in the solar neighborhood).  Such high N/O values can only be achieved if \textit{primary} (dredged-up) C undergoes CN-processing during HBB.  Some objects classified as Type~I PNe in the P78 scheme do not qualify as Type~I PNe according to the criteria of KB94, and vice versa.  To improve the statistics, we denote a PN as Type~I if it meets \textit{either} the P78 or KB94 criteria for He/H and/or N/O.  The results discussed below remain valid if we classify Type~I PNe by the P78 or KB94 scheme separately.

Two important caveats should be kept in mind regarding the classification of PNe into Type~I and non-Type~I objects.  First, many objects in our sample have been observed only in the optical, and thus the only N ion detected was N$^+$.  This is a trace ion in PNe, and hence the ICF N/O~=~N$^+$/O$^+$ (Torres-Peimbert \& Peimbert 1977) can be large and uncertain.  Furthermore, this ICF has been shown to break down under certain conditions (Alexander \& Balick 1997; Gon\c{c}alves et al.\ 2006), and uncertainties in the N abundance may be exacerbated by the high stellar temperatures and bipolar geometries that are typical of Type~I PNe (Gruenwald \& Viegas 1998).

The second caveat is that non-standard mixing in low-mass stars (LMS; $M<3$~M$_{\odot}$), called ``cool bottom processing'' (CBP), can enhance the N abundance in PN progenitors which are not massive enough to undergo second dredge-up or HBB.  CBP mixes material from the bottom of the convective envelope down into regions experiencing CNO-processing, and then back into the envelope.  This ``extra'' mixing can occur during both first and third dredge-up, and enriches the stellar surface with $^{13}$C, $^{14}$N, and $^{17}$O (Wasserburg et al.\ 1995; Boothroyd \& Sackmann 1999; Nollett et al.\ 2003).  The overall enrichment of N from CBP in AGB stars depends on the efficiency of this mixing process relative to TDU, and the temperature at which the CNO-processing occurs; N/O values as high as 8 in LMS are possible from this mechanism (Nollett et al.\ 2003).

Given the uncertainties in the N abundance determinations, one must expect some contamination of our Type~I subsample with objects that are not as enriched in N as required by the classification criterion, and vice versa for non-Type~I PNe.  The possibility of CBP also may contaminate our Type~I sample with LMS, although studies of the Galactic scale heights and stellar and nebular masses of Type~I PNe (e.g., Torres-Peimbert \& Peimbert 1997) indicate that (statistically) most of these objects are descendants of IMS.

PN morphologies are also potential probes of progenitor star mass.  Bipolar PNe are thought to have intermediate-mass progenitors, due to their small Galactic scale height, large central star masses, and their tendency to exhibit Type~I compositions (Corradi \& Schwarz 1995; G\'{o}rny et al.\ 1997; Torres-Peimbert \& Peimbert 1997; Stanghellini et al.\ 2002).  Moreover, the scale heights of elliptical and round PNe indicate that they arise from less massive progenitors (Manchado et al.\ 2000; Stanghellini et al.\ 2002).  It should be noted, however, that some bipolar PNe may be produced by binary systems (Soker 1997; Balick \& Frank 2002 and references therein), in which case these objects may not necessarily be related to IMS.

The central star temperature $T_{\rm eff}$ is also a possible probe of progenitor mass, as the evolutionary time of more massive PN progenitors is shorter, and they attain high temperatures more quickly than lower mass stars.  Furthermore, according to the evolutionary tracks of Bl\"{o}cker (1995), more massive PN nuclei can reach higher temperatures than is possible for lower mass objects.  Indeed, Corradi \& Schwarz (1995) found that bipolar PNe tend to have the hottest central stars of any morphological class of PNe.  However, the central star temperature is also dependent on the time elapsed since leaving the AGB, and low mass objects may have high $T_{\rm eff}$ if they are sufficiently evolved.

\subsubsection{Se and Kr Enrichments in PNe with Intermediate-Mass Progenitors}

In a preliminary analysis of our survey, Sterling \& Dinerstein (2006) found that Se and Kr tend to be more enriched in Type~I PNe than in non-Type~I objects.  However, they computed the Se and Kr enrichments relative to oxygen.  We have shown that O can be depleted in Type~I PNe relative to Ar (\S3.2.2), presumably as a result of ON-processing during HBB.  Therefore, Ar is a better metallicity indicator than O for Type~I PNe; if O is depleted, [Se/O] and [Kr/O] will overestimate the actual \emph{s}-process enrichments in Type~I PNe.

In Table~16, we show the mean [Se/(O,~Ar)] and [Kr/(O,~Ar)]\footnote{See footnote 13.} abundances and their mean absolute deviations for Type~I and non-Type~I PNe, different morphological classes, and the full sample.  In this table, only PNe with Se and/or Kr detections and known Ar (for Type~I PNe) or O abundances (non-Type~I PNe) are considered.

We find that when Ar is used as the reference element for Type~I PNe, the Se and Kr enrichments are actually smaller than in non-Type~I PNe.  Furthermore, the upper limits we have derived for Type~I PNe that do not exhibit Se or Kr emission often allow for only mild or no \emph{s}-process enrichments.  KS tests confirm that Type~I and non-Type~I PNe have different \emph{s}-process enrichment histories: the probability that the Se and Kr abundances in these two subclasses of PNe are drawn from the same distribution function is $p_{\rm ks}=0.02$ and 0.01, respectively.  Our analysis is not dependent on the use of O as a reference element for non-Type~I objects; indeed, the mean values of [Se/Ar] and [Kr/Ar] for non-Type~I PNe are 0.26 and 1.00, respectively, which are quite similar to the mean [Se/O] and [Kr/O] for these objects (Table~16).  These results indicate that PNe with IMS progenitors do not exhibit strong \emph{s}-process enrichments.

Bipolar PNe also tend to have smaller Kr enrichments than elliptical nebulae, although no significant difference is seen in the mean Se abundances.  However, the mean [Se/(O,~Ar)] for bipolar PNe is driven higher primarily by NGC~6445 ([Se/O]~=~0.90$\pm$0.44).  If this object is excluded, then the mean [Se/(O,~Ar)] drops to 0.15 for bipolar PNe, smaller than the value of 0.28 for elliptical PNe.  KS tests suggest that bipolar and elliptical PNe have different enrichment distributions ($p_{\rm ks}\mathrm{(Se)}=0.42$ and $p_{\rm ks}\mathrm{(Kr)}=0.21$), although the difference is not as robust as for Type~I and non-Type~I PNe.  Due to the small number of PNe in our sample with round and irregular morphologies, robust conclusions cannot be drawn regarding \emph{s}-process enrichments in morphological classes other than bipolar and elliptical.

To illustrate the range of Se and Kr enrichments in Type~I and bipolar PNe compared to the full sample, in Figure~4 we show histograms of the derived [Se/(O,~Ar)] and [Kr/(O,~Ar)] divided into 0.1~dex bins.  The top two panels exhibit enrichments in the full sample, with and without Type~I and bipolar PNe.  Abundances for Type~I and bipolar PNe are shown in the middle and bottom panels, respectively.  The distribution of Se and Kr abundances in bipolar and especially Type~I PNe is skewed toward smaller values than for the full sample of objects.  Most of these objects are marginally enriched, if at all, although exceptions do exist (e.g., the bipolar PNe Hu~2-1, IC~5117, J~900, and NGC~6445).

In Figure~5a, we plot the Se and Kr enrichments against three potential indicators of progenitor mass: He/H, N/O, and central star effective temperature $T_{\rm eff}$ (typical error bars are displayed in the left-hand panels).  Correlation coefficients $r$ and their significance $p_{r=0}$ (probability of no correlation) are indicated within each panel.  Note that in all cases, the values of $r$ are low, indicating weak correlations (although the small $p_{r=0}$ for the correlations with He/H and especially N/O suggest a trend may be present).  This is primarily due to the large scatter of Se and Kr enrichments in non-Type~I PNe at $12+\mathrm{log(He/H)}<11.1$ and $\mathrm{log(N/O)}<-0.3$.  The negative correlation coefficients are largely induced by the tendency of PNe with large He/H and N/O to have low Se and Kr enrichments.

There does not appear to be a correlation between Se and Kr enrichments and central star temperature, in the sense that PNe with the highest $T_{\rm eff}$ do not consistently display low Se and Kr enrichments.  While $T_{\rm eff}$ by itself may not be a robust tracer of progenitor mass (\S5.2.1), the lack of a correlation may also be due to the use of indirect $T_{\rm eff}$ determinations for many objects.  The Zanstra and energy balance methods for determining $T_{\rm eff}$ are predicated on the assumptions of blackbody ionizing flux distributions, optical thickness of the nebulae to H- or He$^+$-ionizing photons, and/or analytical corrections to unobserved cooling lines (Stanghellini et al.\ 1993; Preite-Martinez \& Pottasch 1983).  The majority of the PNe in our sample do not have robust $T_{\rm eff}$ determinations from an NLTE stellar atmosphere analysis.

In Figure~5b, we show the same correlations as Figure~5a, except that we include Se and Kr upper limits.  This increases the number of PNe that are strongly enriched in He and N.  Note that most objects with $12+\mathrm{Log(He/H)}~>11.15$ and Log(N/O)~$>-0.1$ do not exhibit Se and Kr emission lines, and the upper limits to \emph{s}-process enrichments in these objects are often small.

Figures~5a and 5b further illustrate that Type~I PNe show at most small enhancements of \emph{s}-processed material.  With this correlation and the Se and Kr abundances of bipolar PNe relative to elliptical nebulae, \textit{we find strong evidence that PNe with intermediate-mass progenitor stars have small (if any) \emph{s}-process enrichments compared to other PNe.}

\subsubsection{Implications for \emph{s}-process Nucleosynthesis in Intermediate-Mass AGB Stars}

Theoretically, it is not obvious \textit{a priori} that intermediate-mass AGB stars should have smaller \emph{s}-process enrichments than lower mass objects, particularly in the case of light \emph{n}-capture elements ($Z=30$--40) such as Se and Kr.  On one hand, it is conceivable that the $^{13}$C neutron source is not as important in an IMS as in stars of lower initial mass, since the intershell region is less massive than that of a LMS by about one order of magnitude (G98; Travaglio et al.\ 1999; Karakas 2003; Lattanzio \& Lugaro 2005).  This implies that the $^{13}$C pocket formed in an IMS is less massive than for LMS, and hence is not capable of producing as many \emph{s}-nuclei.  In addition, the interpulse time decreases with increasing mass (Paczynski 1974; Karakas 2003), allowing less time for free neutrons to be produced in IMS.  During dredge-up, the \emph{s}-process-enriched material that is produced is severely diluted into the massive envelope of an IMS, which leads to smaller enrichments at the stellar surface compared to LMS (GM00; Lattanzio \& Lugaro 2005).  Finally, Goriely \& Siess (2004, 2005) have shown that $^{13}$C burning can occur while protons are still diffusing into the intershell layers (i.e., during the formation of the $^{13}$C pocket) in massive AGB stars.  Depending on the strength of convective overshoot, this can lead to the neutron poison $^{14}$N exceeding the $^{13}$C abundance throughout the intershell region, so that the free neutrons are captured by $^{14}$N rather than Fe-peak nuclei.  In this case, the \emph{s}-process can be completely suppressed.

On the other hand, IMS attain intershell temperatures that are sufficiently high to activate the $^{22}$Ne neutron source, which plays only a minor role in less massive AGB stars (e.g., BGW99; GM00; Lugaro et al.\ 2003).  The $^{22}$Ne source produces an element-by-element enrichment pattern distinct from that of the $^{13}$C source.  In particular, the \emph{s}-process enrichments are expected to be much larger for light \emph{n}-capture elements ($Z=30$--40) than for heavier elements when $^{22}$Ne is the neutron source (Busso et al.\ 1988; GM00; Goriely \& Siess 2005).  This is primarily due to the fact that the neutrons released in this reaction are captured under convective conditions (i.e., during thermal pulses) rather than radiative conditions during the interpulse phase.  While the density of free neutrons is higher than that produced by $^{13}$C($\alpha,n$)$^{16}$O (by 3--4 orders of magnitude), they are available for a much shorter period of time (BGW99; GM00), leading to fewer \emph{n}-captures per seed nucleus.  However, the $^{22}$Ne-driven \emph{s}-process in IMS is subject to some of the same conditions that can limit enrichments by the $^{13}$C source: a small intershell mass and dilution into a massive stellar envelope (Lattanzio \& Lugaro 2005).

Given the number of theoretical uncertainties, it is not surprising that current models of \emph{s}-process nucleosynthesis in IMS are uncertain (BGW99; Travaglio et al.\ 1999; Lattanzio \& Lugaro 2005).  The lack of observational studies of massive AGB stars further limits the accuracy of the theoretical models.  These stars are extremely difficult to study in the optical during the AGB and post-AGB phases because of their high mass-loss rates and significant dust shielding (e.g., Habing 1996; Garc\'ia-Lario 2006).

It was not until recently that \emph{s}-process enrichments were studied in intermediate-mass AGB stars (Garc\'{i}a-Hern\'{a}ndez et al.\ 2007).  These authors searched for the \emph{n}-capture element Zr ($Z=40$) in the form of ZrO in a sample of OH/IR stars, luminous O-rich AGB stars that are bright in the infrared and exhibit OH maser emission.  They determined that the initial masses of these stars are $M>3$~M$_{\odot}$, based on their long pulsation periods, large expansion velocities, and Li enrichments (from HBB).  Little to no Zr enrichment was found in these objects, in agreement with the small Se and Kr enrichments we find in Galactic Type~I PNe.

The small \emph{s}-process enrichments of Type~I and bipolar PNe are likely due to the small intershell masses and efficient dilution of processed material into the massive envelopes of their IMS progenitors, as discussed above.  These factors can be significant regardless of the neutron source.  The possible quenching of the \emph{s}-process from the $^{13}$C neutron source by $^{13}$C-burning during proton diffusion (Goriely \& Siess 2004, 2005) may also reduce \emph{s}-process enrichments in IMS.

\subsection{Correlations With Central Star Type}

\subsubsection{H-Deficient Central Stars}

Based on Ge abundance determinations in four PNe, Sterling et al.\ (2002) suggested that PNe with [WC] central stars may tend to exhibit larger \emph{s}-process enrichments than objects with H-rich central stars.  Large \emph{s}-process enrichments in [WC] PNe would not be surprising, given the deep mixing and heavy mass loss these objects must have experienced in their transition from H-rich to H-deficient objects (Bl\"{o}cker 2001; Herwig 2001; De~Marco \& Soker 2002).  In fact, [WC] central stars exhibit surface abundances similar to that of intershell material (Werner \& Herwig 2006 and references therein), and therefore their nebulae could be enriched in C and \emph{s}-process nuclei.  Indeed, Pe\~{n}a et al.\ (1997) found that PNe with [WC] central stars in the Magellanic Clouds exhibit extreme C enrichments.  On the other hand, studies of Galactic [WC] PNe indicate that their nebular compositions are not significantly different from other PNe (G\'{o}rny \& Stasi\'{n}ska 1995; Pe\~{n}a et al.\ 2001; Girard et al.\ 2007), even for C (De~Marco \& Barlow 2001; G\'{o}rny 2001).

The mean Se and Kr enrichments in PNe with [WC], WELS, and H-rich central stars are reported in Table~17.  We find no significant difference between the Se and Kr abundances in objects with different central star types.  Indeed, when Ar is used as a metallicity indicator for Type~I PNe, the high Kr enrichments found by Sterling \& Dinerstein (2006) in [WC] PNe relative to other objects vanishes (in fact, the mean Kr abundances in [WC] and WELS PNe are formally smaller than those with H-rich central stars; however, KS tests indicate that this difference is not significant).

The similarity between the Se and Kr abundances of PNe with H-rich and H-deficient central stars is illustrated by the distribution of enrichments in these objects.  Figure~6 shows histograms of the Se and Kr enrichments, separated into 0.1~dex bins, in the full sample of objects and in [WC] and WELS PNe.  The distribution of Se and Kr enrichments in [WC] and WELS PNe is very similar to that for the full sample of objects.  KS tests confirm this similarity: $p_{\rm ks}\mathrm{(Se)}=0.84$ and $p_{\rm ks}\mathrm{(Kr)}=0.99$ for PNe with [WC] and H-rich central stars, and $p_{\rm ks}\mathrm{(Se)}=0.62$ and $p_{\rm ks}\mathrm{(Kr)}=0.99$ when all PNe with H-deficient stars are compared to those with H-rich nuclei.  Note that the inclusion of Type~I PNe does not affect the Se and Kr enrichment distributions in [WC] or WELS PNe relative to objects with H-rich stars; the $p_{\rm ks}$ values differ by $\lesssim 0.05$ when Type~I and bipolar PNe are excluded from the samples of objects with different central star types.

We conclude that \emph{s}-process enrichments in PNe with [WC] and WELS central stars are \textit{not} significantly different from other PNe.  This adds to the evidence that the compositions of Galactic PNe around H-deficient central stars are not distinguishable from those with H-rich central stars, even for elements enhanced in the central stars themselves.

\subsubsection{Binary Central Stars}

To this point, our analysis of \emph{s}-process enrichments has been predicated on single star evolution.  However, some PNe are known to have binary central stars (e.g., Bond 2000; De~Marco 2006), and recent surveys of radial velocity variations in PN central stars indicate that a large fraction may be members of binary systems (De~Marco et al.\ 2004; Sorensen \& Pollaco 2004; Af\c{s}ar \& Bond 2005).  In fact, some authors have suggested that most PNe originate in binary star systems (Yungelson et al.\ 1993; Soker 1997; Moe \& De~Marco 2006).

Nucleosynthesis in binary systems has not been well-studied theoretically, due to the number of additional free parameters introduced by a close companion (e.g., Izzard et al.\ 2006).  However, Izzard (2004) used synthetic stellar evolution and nucleosynthesis algorithms to investigate this problem, and found that a companion star can enhance the mass-loss rate during AGB evolution, thereby truncating this phase.  If binary interactions occur during the TP-AGB, then (depending on the orbital separation) C and \emph{s}-process enrichments can be reduced by $\sim$60\% in binary systems compared to single stars with similar initial mass (Izzard 2004).

Few PNe in our sample are known to have a binary central star system.  Two objects exhibit evidence of cooler main sequence stars in their spectra (Me~1-1 and NGC~6302; Shen et al.\ 2004; Feibelman 2001).  The radial velocity surveys of De~Marco et al.\ (2004), Af\c{s}ar \& Bond (2005), and Sorensen \& Pollaco (2004) have found variations in the central star radial velocities of some objects in our sample.  These variations do not furnish proof of stellar companions, since inhomogeneous stellar winds or pulsations may result in similar effects.  However, the radial velocity of the central star of IC~4593 has been found to vary periodically (De~Marco et al.\ 2004), indicating that it has a binary companion.  The remainder of the objects in these studies have not been observed sufficiently to determine whether or not the variations are periodic (De~Marco et al.\ 2004; Sorensen \& Pollaco 2004).  Other evidence for binary progenitors is indirect, such as point-symmetric outflows which are suggestive of precessing jets (e.g., Sahai \& Trauger 1998).

In Table~18, we list the PNe in our sample known to have binary central stars or that exhibit properties suggestive of a multiple star system.  The [Se/(O,~Ar)] and [Kr/(O,~Ar)] values for each object are listed, and the mean abundances for objects with possible or known binary central stars are given in Table~17.  It is interesting that most of these objects either do not exhibit [\ion{Se}{4}] or [\ion{Kr}{3}] emission lines, or have small enrichment factors.  The mean [Se/(O,~Ar)] of objects with possible binary central stars is subsolar, while that of [Kr/(O,~Ar)] is larger due to enrichments in Hb~12 and Hu~2-1.  Overall, the small enrichments of Se and Kr in PNe with known or possible binary central stars is consistent with the prediction that binary companions can truncate the AGB phase, reducing the amount of \emph{s}-process enrichment.  However, other properties (particularly stellar mass) can also lead to small enrichments.  Given the small number of objects in our sample with observational evidence of binary central stars, the low \emph{s}-process enrichments in these PNe cannot definitively be attributed to the presence of binary companions.  A more conclusive study of the effects of binary companions on \emph{s}-process nucleosynthesis awaits the discovery of additional PNe with multiple central stars and more detailed nucleosynthetic  predictions for these systems.

\subsection{Correlations With C/O}

Neutron-capture element abundances are expected to correlate with the C/O ratio, as carbon is brought to the surface of AGB stars along with \emph{s}-processed material during TDU.  There is strong empirical evidence for this correlation: the abundances of \emph{n}-capture elements have been found to scale with the C/O ratio in AGB (Smith \& Lambert 1990; Abia et al.\ 2002) and post-AGB stars (Van~Winckel 2003).

In Figure~7, we plot [Se/(O,~Ar)] and [Kr/(O,~Ar)] against C/O in the objects with known C abundances (determined from UV collisionally-excited lines).  There is significant scatter in the plots, but a trend of increasing Se and Kr enrichments with increasing C/O may be present.  For [Se/(O,~Ar)] vs.\ log(C/O), $r=0.45$ and $p_{\rm r=0}=0.01$, indicating a marginal but significant correlation.  On the other hand, $r=0.29$ and $p_{\rm r=0}=0.34$ for [Kr/(O,~Ar)] against log(C/O), with the discrepant object Hb~12 leading to the poor correlation; if Hb~12 is excluded, the correlation becomes much stronger ($r=0.64$ and $p_{\rm r=0}=0.03$).

The large amount of scatter in Figure~7 is partially due to the uncertainties in our Se and Kr abundance determinations (0.3--0.5~dex).  However, it should be emphasized that the C abundances are also quite uncertain in PNe.  As discussed in \S1.2, the C/O ratios of PNe derived by different authors are often dissimilar by a factor of two or more.  It is unclear whether the C/O value derived by Hyung \& Aller (1996) for Hb~12 (the discrepant object in the right-hand panel of Figure~7) may be in error, since no other C abundance determinations have been performed for this object.

We computed the best linear fits to the correlations between the Se and Kr enrichments and C/O (excluding Hb~12 from the Kr fit), using a least-squares fitting routine written in IDL.  These fits are plotted as solid lines in Figure~7, and correspond to:
\begin{equation}
\mathrm{[Se/(O,~Ar)]} = (0.16\pm0.04) + (0.43\pm0.14)\mathrm{log(C/O)},
\end{equation}
and
\begin{equation}
\mathrm{[Kr/(O,~Ar)]} = (0.38\pm0.10) + (0.79\pm0.29)\mathrm{log(C/O)},
\end{equation}
where the uncertainties include only the dispersion in the fits.  Note that the correlation between [Kr/(O,~Ar)] and C/O (Equation~(5)) has a steeper slope than that for [Se/(O,~Ar)].  This is likely due to the tendency of Kr to be more highly enriched by the \emph{s}-process than Se (\S4.2), as predicted by theoretical studies (G98; GM00; B01).  The higher Kr concentration in dredged-up material causes its abundance to increase more rapidly with C/O than the increase in Se with C/O.

For comparison, we display as dashed lines the correlation between the averaged abundances of the ``light-\emph{s}'' (ls) elements Sr, Y, and Zr and log(C/O) in the M, MS, and S AGB stars of Smith \& Lambert (1985, 1990), the CH subgiants of Smith et al.\ (1993), and the 21~$\mu$m-emitting post-AGB stars of Van~Winckel \& Reyniers (2000).  The best fit to this correlation is:
\begin{equation}
\mathrm{[ls/Fe]} = (0.89\pm0.06) + (1.47\pm0.18)\mathrm{log(C/O)}.
\end{equation}
The slope of the [ls/Fe] curve is much steeper than those of [Se/(O,~Ar)] and [Kr/(O,~Ar)], as previously noted by Gustafsson \& Wahlin (2006).  This can be explained by the higher \emph{s}-process yields that theoretical models predict for Sr, Y, and Zr relative to Se and Kr.  At solar metallicity, GM00 and B01 predict that these three elements are more enriched than Se and Kr in the \emph{s}-processed intershell material by a factor of $\sim$0.5~dex (depending on the mass of the $^{13}$C pocket).  Therefore, the incremental enrichments of Sr, Y, and Zr during dredge-up are larger, and their abundances should increase more rapidly with C/O than do Se and Kr.

The dust chemistry of PNe is an indirect tracer of the C/O ratio.  Because of the stability of the CO molecule, the minority species of C and O is assumed to be locked up in molecular form, leaving the majority species to be incorporated into dust (e.g., Treffers \& Cohen 1974; Barlow 1983; Lodders \& Fegley 1999).  Therefore, if a PN progenitor is C-rich, it will exhibit C-rich dust features; if the progenitor star is O-rich (either because it did not experience enough TDU episodes to become C-rich, or HBB prevented the formation of a C star), then it will exhibit O-rich (silicate) dust emission.

In Table~17, we display the average [Se/(O,~Ar)] and [Kr/(O,~Ar)] for PNe with different dust chemistries, including objects with 21~$\mu$m dust emission and those with mixed (C- and O-rich) dust chemistry.  The number of PNe in our sample with known dust chemistries is quite small, but the objects with O-rich dust (IC~4997, NGC~6302, NGC~6537, and Vy~2-2) show no \emph{s}-process enrichment.  On the other hand, Se and Kr tend to be strongly enriched in objects with mixed or C-rich dust.  While KS tests indicate that the distributions of Se enrichments in PNe with O-rich and C-rich dust ($p_{\rm ks}=0.10$) are different, the distribution of Se and Kr enrichments in PNe with mixed and C-rich dust are similar ($p_{\rm ks}=0.54$ and 0.84, respectively).

NGC~40 and NGC~6369, which exhibit the 21~$\mu$m dust emission feature (Hony et al.\ 2001), show large Se and Kr enrichments, and are among the most enriched objects in our sample.  These objects have H-deficient central stars with significantly different temperatures (NGC~40 has a [WC8] nucleus, while NGC~6369 has a much hotter [WO3] central star; Acker \& Neiner 2003), but do not exhibit obvious differences from other [WC] PNe aside from this dust feature.  The 21~$\mu$m feature lacks a clear identification at this time, but has been associated with post-AGB stars that have strong C and \emph{s}-process enrichments (Kwok et al.\ 1989; Van~Winckel 2003).

These results indicate that there is a trend of increasing Se and Kr enrichments as the dust emission features change from O-rich to C-rich.  This implies that Se and Kr enrichments increase with the C/O ratio, and supports the correlations we find between [Se/(O,~Ar)] and [Kr/(O,~Ar)] and the gaseous C/O ratio.

In principle, \emph{n}-capture elements are potential indicators of C enrichments in PNe, since both are conveyed to AGB star envelopes via TDU.  This is important due to the difficulty in accurately determining the C abundance in ionized nebulae (Kaler 1983; Rola \& Stasi\'{n}ska 1994).  Furthermore, there is only limited spectroscopic access to the UV with existing space observatories, and consequently C abundances can be determined only for PNe that have already been observed at these wavelengths.  In contrast, \emph{n}-capture elements are detectable in a large number of PNe (as we have shown) with ground-based observatories.  The low initial abundances of \emph{n}-capture elements make them more sensitive tracers of moderate enrichments than elements such as C and He, where the incremental enrichments from TDU can be small compared to their initial values.  Improvements to \emph{n}-capture element abundance determinations (\S7) are needed to more accurately constrain the correlation between C and \emph{s}-process enrichments in PNe (Figure~7, and Equations~(4)--(5)).

\section{WHAT FRACTION OF GALACTIC PLANETARY NEBULAE ARE SELF-ENRICHED IN \emph{S}-PROCESS PRODUCTS?}

Models of AGB star evolution predict that TDU only occurs in stars with masses $\gtrsim$1.5~M$_{\odot}$ at solar metallicity (BGW99; Straniero et al.\ 2006).  Because the Galactic initial mass function favors lower stellar masses, this leads to the prediction that the majority of AGB stars and PNe should not be enriched in \emph{n}-capture elements or C.\footnote{This statement depends on the value of the minimum initial stellar mass required to form a PN, commonly assumed to be $\sim$1.0~M$_{\odot}$.  If the minimum mass is much higher than this value, then this statement may not be correct.  Another uncertainty is that it is possible for dredge-up to occur without significant C or \emph{s}-process enrichments (as in the case of IMS; see \S5.2).  This may cause us to slightly underestimate the fraction of PN progenitors that experience TDU, although the effect should be small, given the small number of IMS relative to LMS.}  We test this prediction by using the results of our survey to estimate the fraction of Galactic PNe self-enriched in \emph{s}-processed material.

We have found that 41 of the 79 PNe (51.9\%) in our sample with measured [Se/(O,~Ar)] and [Kr/(O,~Ar)] are \emph{s}-process enriched (\S4.1).  Including meaningful non-detections, where [Se/(O,~Ar)] and/or [Kr/(O,~Ar)]~$<0.3$~dex, the enrichment rate is 41/94, or 43.6\%.  Note that this is quite similar to the fraction of C-rich PNe (35\%) computed by Rola \& Stasi\'{n}ska (1994).  However, our sample is flux-limited, and hence the fraction of \emph{s}-process enriched PNe in our sample may not be indicative of the fraction of \emph{all} Galactic PNe that are enriched.

We construct a PN luminosity function (PNLF) to correct our sample for incompleteness. PNLFs are derived from [\ion{O}{3}]~$\lambda5007$ magnitudes, and are commonly used as extragalactic standard candles (e.g., Jacoby 1989; Ciardullo et al.\ 1989; Ciardullo 2005), based on the similarity of the [\ion{O}{3}] luminosities of the brightest PNe in different galaxies.  However, generating a PNLF for our sample requires distances, which in general are poorly known; most Galactic PNe have only statistical distance determinations, which can be uncertain by more than a factor of two (Terzian 1993).  In the following subsections, we describe the construction of a PNLF for our sample, and investigate \emph{s}-process enrichments as a function of absolute [\ion{O}{3}] luminosity.  We use these results to estimate the fraction of Galactic PNe whose progenitors experienced \emph{s}-process nucleosynthesis and TDU.

\subsection{PN Distances and [\ion{O}{3}]~$\lambda5007$ Magnitudes}

In order to derive the absolute [\ion{O}{3}]~$\lambda5007$ magnitude $M_{5007}$ of a PN, the distance to the object and the global $\lambda$5007 flux are needed.  Whenever possible, we utilize direct distance measurements from nebular expansion or stellar trigonometric parallaxes.  For expansion parallaxes, we employ the distances of Mellema (2004), which he corrected for the differences between pattern and material velocities.  However, direct distance measurements are available for only 14 PNe in our sample.

For the other objects, we are forced to use statistical distances.  We consider four statistical scales, based on the assumption that all PNe have the same ionized mass (Cahn et al.\ 1992, hereafter CKS92) or on empirical correlations between radio continuum temperature brightness ($T_{\rm b}$) and radius (Van~de~Steene \& Zijlstra 1994; Zhang 1995), ionized mass and radius (Zhang 1995), or $T_{\rm b}$ and 5~GHz luminosity (Phillips 2004).  For each object in our sample without a direct distance measurement, we average the distances from each of these different statistical scales (when available) to derive an adopted distance.  The standard deviations of these estimates are used as error bars, although the actual uncertainties may be much greater in some cases.  The distance to each PN is given in Table~19 (available in the electronic version of this article), where we list the direct distance determination $d_{\rm meas}$ (if available), followed by the statistical distances from the four scales mentioned above, and the adopted distances.

Most spectroscopic studies of the PNe in our sample (Table~6) have determined $F(\lambda5007)$ for only a small portion of the nebulae.  To compute the global $F(\lambda5007)$, we use global H$\beta$ fluxes from CKS92 or the Vizier Strasbourg-ESO Catalog database\footnote{http://vizier.u-strasbg.fr/cgi-bin/VizieR?-source=V/84/main.  While these are not necessarily global H$\beta$ fluxes, we have used this database primarily for compact PNe whose emission is largely included in the entrance aperture of Acker et al.'s survey.} (Acker et al.\ 1992), when not listed by CKS92.  We deredden these global H$\beta$ fluxes and the $F(\lambda5007)$ from the primary abundance references using the extinction coefficients listed in Table~19, and assume the measured intensity ratio $I(\lambda5007)$/$I$(H$\beta$) is typical of the global ratio in each nebula.  Occasionally, [\ion{O}{3}]~$\lambda5007$ was not observed or was saturated in the optical spectra.  In these cases, we compute $I(\lambda5007)$ from the [\ion{O}{3}]~$\lambda4959$ intensity, which is related to $I(\lambda5007)$ by the ratio of their transition probabilities, since both lines arise from the same upper level.

The apparent $\lambda5007$ magnitude of each PN was determined using the relation
\begin{equation}
m_{5007} = -2.5~\mathrm{log}(I_{5007}) - 13.74
\end{equation}
(Jacoby 1989).  We convert these to absolute magnitudes $M_{5007}$ using the adopted distances in Table~19.  The global $F$(H$\beta$), $c_{\rm H\beta}$, and $M_{5007}$ of each PN in our sample are given in Table~19.  Five objects were excluded from this analysis, due to the lack of a distance determination (M~1-71 and Vy~1-2), $F$(H$\beta$) (K~3-17 and K~3-55), or $F(\lambda5007)$ measurement (K~3-62).  The main uncertainty in the derived $M_{5007}$ stems from the statistical distances that we use, which lead to an average uncertainty of 0.67~mag.

Interestingly, we find that the most luminous PN in our sample, NGC~6543, has $M_{5007}=-2.7$, about 1.8 magnitudes fainter than the bright-end cutoff of extragalactic PNLFs ($-4.48$; Ciardullo et al.\ 1989).  This implies that the statistical distance scales we have utilized systematically underestimate the actual distances to PNe by about a factor of 2.3 (it is unlikely that our sample does not include some of the most luminous PNe in the Galaxy).  Therefore, we have shifted the derived $M_{5007}$ by $-1.78$ magnitudes in order for the bright end of the PNLF to match the expected bright limit of $-4.48$~mag.  This adjustment has no effect on our subsequent analysis.

The faint limit of the Galactic PNLF is expected to be about 8 mag below the bright limit, or at $M_{5007}=+3.5$~mag, based largely on detectability arguments (Jacoby 1980; 2006).  We derive fainter $M_{5007}$ for four objects (He~2-459, M~1-11, M~3-41, and M~4-18), which indicates either that their (statistical) distances are overestimated, or that they are not PNe.  However, the [WC] central stars of He~2-459 and M~4-18 and the large \emph{s}-process enrichment of M~1-11 (the Kr$^{++}$/H$^+$ abundance \textit{alone} is enriched relative to solar) indicate that the PN status of three of these objects are secure.  These four PNe are all very low-excitation objects, and the assumption of a standard ionized mass (e.g., CKS92; Zhang 1995) and other empirical correlations used for distance determinations may not be valid if the ionization front in these objects has not advanced through the bulk of their circumstellar envelopes.  In addition, their low ionization balance causes much of their O to be singly ionized, and hence their [\ion{O}{3}]~$\lambda$5007 fluxes are correspondingly low.  Therefore, it is possible that these PNe have inherently faint $\lambda$5007 luminosities.  For the remainder of this discussion, we ignore these four objects.

\subsection{PN Luminosity Function and \emph{s}-process Enrichments}

We are now able to examine \emph{s}-process enrichments as a function of luminosity, and correct for incompleteness in our sample in order to estimate the fraction of Galactic PNe whose progenitors experienced \emph{s}-process nucleosynthesis and TDU.  We categorize the \emph{s}-process enrichment of each PN in our sample in the following manner (Table~19; see also \S4.1):
\begin{itemize}
\item \textit{Enriched} -- [Se/(O,~Ar)] and/or [Kr/(O,~Ar)]~$\geq0.3$~dex (i.e., larger than the level of primordial scatter of light \emph{n}-capture element abundances in unevolved stars with near-solar metallicity; Travaglio et al.\ 2004).  We also require that the abundance uncertainties do not allow for [Se/(O,~Ar)] or [Kr/(O,~Ar)]~$<0.0$~dex.\footnote{This requirement reduces the number of PNe we previously defined as enriched (\S4.1) from 41 to 38.}  When both Se and Kr abundances have been determined, we preferentially use Kr since it is more enriched than Se by the \emph{s}-process (G98; GM00; B01), and since the Kr abundances are likely to be more accurately determined (Paper~I).

\item \textit{Not Enriched} -- [Se/(O,~Ar)] and/or [Kr/(O,~Ar)] upper limits are below 0.3~dex, or the abundance uncertainties do not allow for enrichments larger than 0.3~dex.

\item \textit{Indeterminate} -- It could not be determined whether the PN is \emph{s}-process enriched or not.  [Se/(O,~Ar)] and [Kr/(O,~Ar)] upper limits are larger than 0.3~dex, or cannot be classified as enriched or not enriched due to abundance uncertainties.
\end{itemize}

Lower and upper limits to the fraction of \emph{s}-process enriched PNe can be obtained by ignoring and including objects with indeterminate enrichments, respectively.  Figure~8 shows the cumulative number of PNe in our sample brighter than a given $M_{5007}$, along with the minimum and maximum fraction of enriched PNe.  This plot shows that most objects in our sample have absolute $\lambda5007$ magnitudes of +2.5 or smaller.  Considering PNe up to this limit, we find that the number of \emph{s}-process enriched PNe in our sample is between 30 and 75\%.

However, our sample is not complete at faint $\lambda$5007 magnitudes.  Figure~9 (upper panel) shows the PNLF of our sample up to the expected faint limit of +3.5~mag (Jacoby 1980, 2006), along with the fraction of enriched and possibly enriched (indeterminate) PNe as a function of $M_{5007}$.  Our sample can be corrected for incompleteness by using a theoretical PNLF (Figure~9, lower panel), calculated from Equation~(2) of Ciardullo et al.\ (1989):
\begin{equation}
N(M) \propto e^{0.307M}(1 - e^{3(M^* - M)}),
\end{equation}
where $N(M)$ is the number of PNe, $M=M_{5007}$, and $M^*$ is the bright end cutoff of $-4.48$ mag.  We normalize the theoretical PNLF by assuming our sample is complete up to $M_{5007}=-3$~mag.  It can be seen that our sample quickly becomes incomplete at fainter magnitudes.

We have not detected any PNe with $+2.5\leq M_{5007} \leq +3.5$~mag.  It is tempting to assume that most PNe at such faint $M_{5007}$ arise from the least massive PN progenitors (1--1.5~M$_{\odot}$), which do not experience TDU.  However, Jacoby \& De~Marco (2002) found a high incidence of Type~I objects among the faintest PNe in the Small Magellanic Cloud (SMC).  While the SMC has a significantly lower metallicity than the Galaxy and hence a different stellar population, this result indicates that the faint end of the PNLF is likely to be occupied by PNe with a range of progenitor masses.  Therefore, it cannot be assumed that intrinsically faint PNe do not exhibit \emph{s}-process enrichments.

To derive a lower limit to the fraction of \emph{s}-process enriched PNe, we assume that all objects with indeterminate enrichments, as well as unobserved objects with $+2.5\leq M_{5007} \leq +3.5$, are not enriched.  Assuming that the fraction of enriched PNe in each magnitude bin of our sample (Figure~9, upper panel) is representative of the actual fraction of enriched objects at that luminosity, we find that at least 20\% of Galactic PNe are self-enriched in \emph{s}-process nuclei.  This limit is rather uncertain, due to uncertainties in the statistical distances used for most of the objects in our sample, the small number of intrinsically faint PNe observed, and uncertainties in our derived Se and Kr abundances.  Furthermore, our supposition that the PNLF cuts off at 8~mag below the bright limit may be questionable, in light of the recent discovery of a number of large Galactic PNe with very low surface brightnesses (Parker et al.\ 2006).  We emphasize that our lower limit is valid only for PNe within 8~mag of the bright cutoff of the Galactic PNLF.  Another uncertainty is our assumption that the fraction of enriched objects in each magnitude bin of our sample is representative of the fraction of all enriched Galactic PNe at that luminosity; if our sample is biased toward enriched objects at faint luminosities, this could also reduce the lower limit of \emph{s}-process enriched PNe.

We determine the upper limit to the fraction of \emph{s}-process enriched PNe by assuming that \emph{all} objects with indeterminate enrichments or $+2.5\leq M_{5007} \leq +3.5$~mag are enriched.  This leads to an upper limit of \emph{s}-process enrichments in 80\% of Galactic PNe.  Considering our conservative assumptions in this estimate, the actual fraction of \emph{s}-process enriched PNe is likely to be much smaller than this upper limit.

The lower limit we derive is in qualitative agreement with the prediction that most PN progenitors did not experience \emph{s}-process nucleosynthesis or TDU (BGW99; Straniero et al.\ 2006).  However, given the uncertainties in our analysis, we cannot definitively rule out that a higher fraction of PNe are enriched.  While we are not currently able to conclusively constrain the fraction of Galactic PN progenitors that experienced TDU and the \emph{s}-process, we have demonstrated the utility of PNe for empirically testing models of AGB mixing and nucleosynthesis in this manner.

To more accurately constrain the fraction of PN progenitors that experience TDU, improvement on three fronts is necessary.  First, individual (as opposed to statistical) distances to more PNe are needed.  This is a difficult and long-standing problem in the field of PNe, but progress has been made in recent years (see references in footnote \textit{a} of Table~19).  Secondly, the shape of the PNLF is not well-constrained at this time (e.g., Jacoby \& De~Marco 2002), particularly at faint luminosities.  However, results from the MASH survey (Parker et al.\ 2006) have the potential to considerably improve our understanding of the faint end of the PNLF.  Finally, the accuracies of our derived Se and Kr abundances (0.3--0.5~dex) are not sufficient to determine whether many PNe in our sample are self-enriched in \emph{s}-processed material.  We discuss prospects for improving the accuracy of \emph{n}-capture element abundance determinations in the following section.

\section{CONCLUSIONS AND FUTURE WORK}

We have presented results from the first large-scale survey of \emph{n}-capture elements in PNe.  Over 100 Galactic PNe have been observed in the $K$ band to search for emission lines of [\ion{Kr}{3}] and [\ion{Se}{4}], and we expanded our sample to 120 objects by using data from the literature.  We derived elemental Se and Kr abundances to investigate \emph{s}-process enrichments in PNe and their relation to other nucleosynthetic and nebular properties.  The primary conclusions of our study are now highlighted:

\begin{enumerate}

\item We have detected Se and/or Kr emission in 81 of 120 objects, for a detection rate of 67.5\%.  [\ion{Se}{4}]~2.287~$\mu$m is detected in 70 objects (58\%), while [\ion{Kr}{3}]~2.199~$\mu$m is detected in about half as many objects (36, or 30\%).  This is likely to be an excitation effect.  In H$_2$-emitting PNe, we have removed contamination of the [\ion{Kr}{3}] and [\ion{Se}{4}] fluxes by H$_2$~3-2~$S$(3)~2.201 and H$_2$~3-2~$S$(2)~2.287~$\mu$m, using high resolution observations and the measured ratios of other observed H$_2$ lines.  These lines are the only important contaminants to the observed [\ion{Kr}{3}] and [\ion{Se}{4}] features (Paper~I).

\item We determined the ionic abundances (or upper limits) of Kr$^{++}$ and Se$^{3+}$ in all PNe of our sample, using electron temperatures and densities from the literature.  Employing formulae derived from photoionization models (Paper~I), we computed ionization correction factors (ICFs) for each object in order to determine the elemental Se and Kr abundances from the ionic abundances.  The ICFs require ionic and elemental abundances of O, Ar, and S for each PN.  We have conducted an extensive search of PN composition studies in the literature, and utilize the most reliable abundance determinations to compute the Se and Kr ICFs.  The Se and Kr abundances are determined to an accuracy of 0.3--0.5~dex for most objects in our sample, taking into account uncertainties in the [\ion{Se}{4}] and [\ion{Kr}{3}] line fluxes, electron temperatures and densities, and the ionic and elemental abundances used in the ICFs.

\item Se and Kr enrichment factors have been determined for each PN by using O and Ar as reference elements.  Notably, we find that Ar/O, S/O, and Cl/O are systematically larger in Type~I PNe than in non-Type~I objects by about a factor of two, which we attribute to O depletion via ON-cycling during hot bottom burning in Type~I progenitor stars.  We have therefore used Ar as a reference element in Type~I PNe, and O for all other objects.

\item We find a range of Se and Kr abundances, from $-0.05$ to 1.89~dex for [Kr/(O,~Ar)] and $-0.56$ to 0.90~dex for [Se/(O,~Ar)].  We consider PNe to be self-enriched by \emph{s}-process nucleosynthesis and TDU in their progenitor stars if [Se/(O,~Ar)] and/or [Kr/(O,~Ar)]~$>0.3$~dex, which is the level of scatter of light \emph{n}-capture element abundances in unevolved stars with near-solar metallicities (Travaglio et al.\ 2004).  Using this criterion, we find that 41 of the 94 objects with [Se/(O,~Ar)] and/or [Kr/(O,~Ar)] determinations or meaningful upper limits exhibit \emph{s}-process enrichments.

\item Kr tends to be more highly enriched than Se ([Kr/Se]~=~0.5$\pm$0.2 in 18 objects exhibiting both Se and Kr emission), as predicted by theoretical models of \emph{s}-process nucleosynthesis.  The enrichment factors of Se and Kr vary widely in our sample, which can be attributed to a range of $^{13}$C pocket masses and TDU efficiencies in PN progenitor stars.

\item We find strong evidence that Se and Kr are only marginally (if at all) enriched in Type~I and (to a lesser extent) bipolar PNe.  Type~I and bipolar PNe are believed to be descendants of intermediate-mass stars (IMS; $>3$--4~M$_{\odot}$), based on their chemical (CNO and He) compositions, Galactic distribution, and estimated nebular and central star masses.  This result implies that IMS experience smaller \emph{s}-process enrichments than low-mass AGB stars ($<3$~M$_{\odot}$).  This is likely due to the small intershell masses of IMS, which limits the amount of material that undergoes \emph{s}-process nucleosynthesis; and to the large envelope masses of these objects, which significantly dilute the processed material dredged up to the surface.  Similar results have been found recently for intermediate-mass, O-rich Galactic AGB stars (Garc\'{i}a-Hern\'{a}ndez et al.\ 2007).

\item In contrast to previous suggestions, we do not find systematically larger \emph{s}-process enrichments in PNe with H-deficient, C-rich [WC] or WELS central stars compared to objects with H-rich nuclei.  In fact, the distribution of enrichments among [WC] and WELS PNe is quite similar to those with H-rich central stars.  This is somewhat surprising, in that [WC] and WELS central stars are enriched in C and probably \emph{s}-processed material, indicating that these stars experienced TDU.  Nevertheless, this result is consistent with previous studies that found no significant differences in the compositions of Galactic [WC] and non-[WC] PNe, even for C (G\'{o}rny \& Stasi\'{n}ska 1995; De~Marco \& Barlow 2001; Girard et al.\ 2007).

\item We find evidence that \emph{s}-process enrichments correlate with the gaseous C/O ratio, as predicted theoretically and observed in AGB and post-AGB stars.  The Se and Kr abundances do not increase as rapidly with increasing C/O as do Sr, Y, and Zr in AGB and post-AGB stars, due to their smaller yields from \emph{s}-process nucleosynthesis.  The correlation between Se and Kr abundances with the gaseous C/O ratio is strengthened by the \emph{s}-process enrichments of PNe with different dust compositions.  We find that PNe exhibiting C-rich dust emission display markedly larger Se and Kr enrichments than objects with only O-rich dust features.

\item Theoretical models of AGB evolution (e.g., BGW99; Straniero et al.\ 2006) predict that TDU and the \emph{s}-process do not operate in solar-metallicity stars with initial masses less than $\sim$1.5~M$_{\odot}$.  Since the initial mass function favors low-mass star formation, a consequence of this prediction is that most AGB stars and PNe should not exhibit \emph{s}-process or C enrichments.  We have estimated the fraction of \emph{s}-process enriched Galactic PNe by dividing our sample into enriched, non-enriched, and indeterminate enrichment objects.  We constructed a PN luminosity function (PNLF) for our sample, and corrected it for incompleteness using a theoretical PNLF.  We find that at least 20\% of Galactic PN progenitors experienced \emph{s}-process nucleosynthesis and TDU, considering PNe within 8~mag of the bright limit of the PNLF.  By assuming that all objects with indeterminate enrichments or at the faint end of the PNLF are enriched, we conservatively estimate that at most 80\% of Galactic PNe are \emph{s}-process enriched.  The lower limit is in general agreement with the prediction that TDU and the \emph{s}-process do not operate in stars less massive than 1.5~M$_{\odot}$.

\end{enumerate}

Further improvements to the accuracy of \emph{n}-capture element abundance determinations in PNe require reducing the uncertainties in the ICFs, which arise from two major sources.  First, the atomic data governing the ionization equilibrium of Se and Kr (photoionization cross-sections and rate coefficients for various recombination processes) are poorly known, and the ICF formulae derived in Paper~I rely on approximations to these data.  Unfortunately, Se and Kr are not alone in this regard; most \emph{n}-capture elements have poorly (if at all) determined photoionization cross-sections and recombination rate coefficients.  One of us (NCS) has begun a laboratory astrophysics investigation to determine these atomic data for Se, Kr, and Xe ions.  With more accurate atomic data, it will be possible to derive more reliable ICFs, using the methods introduced in Paper~I.

Second, we have detected only one ion each of Se and Kr.  The ICFs can be large if the ionic fractions of Kr$^{++}$ or Se$^{3+}$ are small, and also depend on the accuracy of the fractional ionic abundances of Ar$^{++}$, S$^{++}$, and O$^{++}$ derived from optical spectra.  Observing multiple ionization stages of Se and Kr can reduce the magnitude and importance of uncertainties in the ICFs.  For example, [\ion{Kr}{4}] and [\ion{Kr}{5}] have transitions in the optical, and we have used these lines to derive Kr abundances in ten objects from our sample (Paper~I).  While it is difficult to observe multiple ionization stages of \emph{n}-capture elements due to their low abundances and the consequent weakness of their emission lines, the detection of at least two ionization stages of Br, Kr, Rb, and Xe in the optical spectra of some PNe (Liu et al.\ 2004a; Zhang et al.\ 2005; Sharpee et al.\ 2007) shows that this difficulty is surmountable.

\acknowledgements

We are grateful to K.\ Butler for calculating the [\ion{Se}{4}]~2.287~$\mu$m effective collision strength, G.\ Jacoby for many helpful conversations and his careful reading of this manuscript, M.\ Busso and R.\ Gallino for enlightening discussions of AGB nucleosynthesis, A.\ Karakas for discussions of O destruction in IMS, and D.\ Lester for assistance with CoolSpec operations.  We also thank the staff at McDonald Observatory, whose tireless support made these observations possible.  This work has been supported by NSF grants AST~97-31156 and AST~04-06809.

\clearpage

\begin{deluxetable}{lccccccccccc}
\tablecolumns{12}
\tablewidth{0pc} 
\tabletypesize{\scriptsize}
\tablecaption{Observing Log and Nebular Properties}
\tablehead{
\colhead{Object} & \colhead{Obs.} & \colhead{Exp.} & \colhead{} & \colhead{Peimbert} & \colhead{CS} & \colhead{$T_{\rm eff}$} & \colhead{$T_{\rm eff}$} & \colhead{} & \colhead{Morphology} & \colhead{Dust} & \colhead{Dust} \\
\colhead{Name} & \colhead{Date} & \colhead{Time (s)} & \colhead{Res.\tablenotemark{a}} & \colhead{Type} & \colhead{Type\tablenotemark{b}} & \colhead{(10$^4$ K)} & \colhead{Ref.\tablenotemark{c}} & \colhead{Morphology\tablenotemark{d}} & \colhead{Ref.\tablenotemark{c}} & \colhead{Type\tablenotemark{e}} & \colhead{Ref.\tablenotemark{c}}}
\startdata
Cn 3-1 & 21 June 2003 &  1080 & L & Non-I & WELS &   6.44 & CS21 & E & M11 & \nodata & \nodata \\
DdDm 1 & 28 June 2005 &  960 & L & Non-I & H-rich &   3.70 & CS13 & \nodata & \nodata & \nodata & \nodata \\
Hb 4 & 22 July 2003 &  600 & L & I & $[$WO3$]$ &   8.50 & 4 & E & M5 & \nodata & \nodata \\
  & 28 June 2005 &  1800 & L &   &   &   &   &   &   &   &   \\
Hb 5 & 02 Aug.\ 2004 &  600 & L & I & H-rich &  15.00 & 4 & B & M4 & C & D1 \\
  & 23 July 2005 &  1320 & L &   &   &   &   &   &   &   &   \\
  & 25 July 2005 &  840 & L &   &   &   &   &   &   &   &   \\
  & 05 Sept.\ 2004 &  720 & H &   &   &   &   &   &   &   &   \\
\enddata
\scriptsize
\tablenotetext{a}{H = high resolution ($R=4400$), L = low (survey) resolution ($R=500$)}
\tablenotetext{b}{Taken from the compilation of Acker \& Neiner (2003).  The various central star types are Wolf-Rayet ([WC] or [WO]), weak emission line star (WELS), and H-rich.}
\tablenotetext{c}{See Table 6 for $T_{\rm eff}$, morphology, and dust-type references}
\tablenotetext{d}{B = bipolar, E = elliptical, R = round, I = irregular (includes point-symmetric nebulae)}
\tablenotetext{e}{C = C-rich dust, O = O-rich dust, M = mixture of C-rich and O-rich dust, 21$\mu$m = exhibits 21~$\mu$m dust emission feature}
\end{deluxetable}
 
\clearpage

\begin{deluxetable}{lcccccccc}
\tablecolumns{9}
\tablewidth{0pc} 
\tabletypesize{\scriptsize}
\tablecaption{Observed Fluxes -- Survey Resolution Data}
\tablehead{
\colhead{} & \multicolumn{8}{c}{$F$(Br$\gamma$) or $F$($\lambda$)/$F$(Br$\gamma$)$\times$100} \\
\cline{2-9}
\colhead{} & \colhead{} & \colhead{} & \colhead{H$_2$ 3-2~$S$(3)+} & \colhead{} & \colhead{} & \colhead{} & \colhead{} & \colhead{H$_2$ 3-2~$S$(2)+} \\
\colhead{Object} & \colhead{Br$\gamma$} & \colhead{He II} & \colhead{[Kr III]} & \colhead{[Fe III]} & \colhead{H$_2$ 1-0~$S$(0)} & \colhead{[Fe III]} & \colhead{H$_2$ 2-1~$S$(1)} & \colhead{[Se IV]} \\
\colhead{Name} & \colhead{2.166\tablenotemark{a}} & \colhead{2.189} & \colhead{2.199\tablenotemark{b}} & \colhead{2.218} & \colhead{2.224} & \colhead{2.243} & \colhead{2.248} & \colhead{2.287\tablenotemark{b}}}
\startdata
Cn 3-1 & 5.32$\pm$0.15 & \nodata & $<$1.84 & \nodata & \nodata & \nodata & \nodata & $<$1.70 \\
DdDm 1 & 5.45$\pm$0.24 & \nodata & $<$9.50 & \nodata & \nodata & \nodata & \nodata & $<$10.8 \\
Hb 4 & 6.31$\pm$0.14 & 6.31$\pm$1.07 & $<$1.54 & \nodata & \nodata & \nodata & \nodata & 3.77$\pm$0.65 \\
Hb 5 & 16.8$\pm$0.4 & 18.3$\pm$0.9 & 1.28$\pm$0.32 & 1.33$\pm$0.49\tablenotemark{d} & 6.49$\pm$0.44 & 0.52$\pm$0.24\tablenotemark{d} & 1.83$\pm$0.47 & 1.23$\pm$0.40\tablenotemark{d} \\
Hb 6 & 11.4$\pm$0.3 & 6.54$\pm$0.91 & $<$1.52 & 0.97$\pm$0.71\tablenotemark{d} & \nodata & \nodata & \nodata & 3.80$\pm$0.40 \\
Hb 7 & 1.81$\pm$0.09 & \nodata & $<$6.30 & \nodata & \nodata & \nodata & \nodata & $<$6.85 \\
He 2-459 & 7.27$\pm$0.15 & 2.48$\pm$1.09\tablenotemark{d} & $<$2.61 & 1.28$\pm$0.33 & 6.64$\pm$0.99 & \nodata & 2.79$\pm$0.54 & $<$1.62 \\
Hu 1-1 & 0.765$\pm$0.060 & \nodata & $<$10.5 & \nodata & \nodata & \nodata & \nodata & $<$12.4 \\
\enddata
\scriptsize
\tablenotetext{a}{Units of 10$^{-13}$ erg cm$^{-2}$ s$^{-1}$}
\tablenotetext{b}{Not corrected for contamination from H$_2$ 3-2~$S$(3) 2.201 or H$_2$ 3-2~$S$(2) 2.287~$\mu$m (see \S2.2)}
\tablenotetext{c}{Flux uncertain due to thin clouds}
\tablenotetext{d}{Weak or uncertain detection}
\tablenotetext{e}{\protect Low resolution data taken with a non-standard setting ($1\farcs 8$ slit width, 85~l/mm grating) under a separate program}
\end{deluxetable}

\clearpage

\begin{deluxetable}{lccccccccccc}
\rotate
\tablecolumns{12}
\tablewidth{0pc} 
\tabletypesize{\scriptsize}
\tablecaption{Literature Fluxes and Nebular Properties}
\tablehead{
\colhead{Object} & \colhead{} & \colhead{$F$([Kr III])/} & \colhead{$F$([Se~IV])/} & \colhead{Peimbert} & \colhead{CS} & \colhead{$T_{\rm eff}$} & \colhead{$T_{\rm eff}$} & \colhead{} & \colhead{Morph.} & \colhead{Dust} & \colhead{Dust} \\
\colhead{Name} & \colhead{Ref.\tablenotemark{a}} & \colhead{$F$(Br$\gamma$)$\times$100\tablenotemark{b}} & \colhead{$F$(Br$\gamma$)$\times$100\tablenotemark{b}} & \colhead{Type} & \colhead{Type\tablenotemark{c}} & \colhead{(10$^4$ K)} & \colhead{Ref.\tablenotemark{a}} & \colhead{Morphology\tablenotemark{d}} & \colhead{Ref.\tablenotemark{a}} & \colhead{Type\tablenotemark{e}} & \colhead{Ref.\tablenotemark{a}}}
\startdata
BD+30$^{\rm o}$3639 & IR2 & $<$5.00 & $<$1.00 & Non-I & $[$WC9$]$ &   4.70 & CS10 & E & CS21 & M & D3 \\
  & IR4 & $<$4.21 & \nodata &   &   &   &   &   &   &   &   \\
  & IR6 & 0.90$\pm$0.22 & \nodata &   &   &   &   &   &   &   &   \\
Hb 12 & IR5 & 1.50$\pm$0.20 & 2.40$\pm$0.26 & Non-I & H-rich &   3.50 & 42 & B & M13 & M & D1 \\
IC 418 & IR3 & 2.10$\pm$1.40\tablenotemark{f} & \nodata & Non-I & H-rich &   3.60 & CS14 & E & M7 & C & D1 \\
IC 2003 & IR2 & $<$5.00 & 15.00$\pm$3.00 & Non-I & $[$WC3$]$? &   8.99 & CS21 & E & M10 & \nodata & \nodata \\
  & IR3 & \nodata & 13.00$\pm$6.00\tablenotemark{f} &   &   &   &   &   &   &   &   \\
IC 2165 & IR2 & $<$2.00 & 7.00$\pm$2.00 & Non-I & WELS &  14.00 & 39 & B & M7 & C & D1 \\
\enddata
\scriptsize
\tablenotetext{a}{See Table 6 for NIR flux, $T_{\rm eff}$, morphology, and dust-type references}
\tablenotetext{b}{Not corrected for contamination from H$_2$ 3-2~$S$(3) 2.201 or H$_2$ 3-2~$S$(2) 2.287~$\mu$m}
\tablenotetext{c}{Taken from the compilation of Acker \& Neiner 2003}
\tablenotetext{d}{B = bipolar, E = elliptical, R = round, I = irregular (includes point-symmetric nebulae)}
\tablenotetext{e}{C = C-rich dust, O = O-rich dust, M = mixture of C-rich and O-rich dust, 21$\mu$m = exhibits 21~$\mu$m dust emission feature}
\tablenotetext{f}{Weak or uncertain detection}
\end{deluxetable}

\clearpage

\begin{deluxetable}{lcccc}
\tablecolumns{5}
\tablewidth{0pc} 
\tablecaption{Observed Fluxes -- High Resolution Data}
\tablehead{
\colhead{} & \multicolumn{4}{c}{$F$(Br$\gamma$) or $F$($\lambda$)/$F$(Br$\gamma$)$\times$100} \\
\cline{2-5}
\colhead{Object} & \colhead{Br$\gamma$} & \colhead{He II} & \colhead{[Kr III]} & \colhead{H$_2$ 3-2~$S$(3)} \\
\colhead{Name} & \colhead{2.166\tablenotemark{a}} & \colhead{2.189} & \colhead{2.199} & \colhead{2.201}}
\startdata
Hb 5 & 12.5$\pm$0.3 & 16.3$\pm$0.6 & 2.22$\pm$0.47 & \nodata \\
He 2-459 & 4.73$\pm$0.11 & \nodata & $<$2.41 & \nodata \\
J 900 & 5.71$\pm$0.20 & 13.4$\pm$2.8 & $<$3.57 & \nodata \\
K 3-17 & 8.55$\pm$0.27\tablenotemark{b} & 14.3$\pm$0.9 & 4.23$\pm$1.23 & \nodata \\
K 3-60 & 1.76$\pm$0.05 & 12.3$\pm$1.0 & 3.32$\pm$0.86 & \nodata \\
M 1-11 & 30.3$\pm$0.8 & \nodata & 3.47$\pm$0.63 & \nodata \\
M 1-17 & 1.94$\pm$0.12 & \nodata & $<$9.18 & \nodata \\
M 1-32 & 2.76$\pm$0.08\tablenotemark{b} & \nodata & 6.49$\pm$1.53 & \nodata \\
M 1-40 & 6.03$\pm$0.12\tablenotemark{b} & 12.5$\pm$1.1 & 2.27$\pm$1.21\tablenotemark{c} & 2.19$\pm$1.10\tablenotemark{c} \\
M 1-51 & 7.49$\pm$0.20 & \nodata & 4.59$\pm$1.03 & \nodata \\
M 1-57 & 5.83$\pm$0.21 & 13.6$\pm$0.9 & 2.42$\pm$0.69 & 2.26$\pm$0.37 \\
M 2-43 & 24.2$\pm$0.5 & \nodata & 5.74$\pm$0.83 & \nodata \\
M 3-28 & 1.30$\pm$0.13 & 6.39$\pm$2.84\tablenotemark{c} & $<$10.0 & \nodata \\
NGC 6445 & 2.14$\pm$0.58 & \nodata & $<$29.4 & \nodata \\
NGC 6537 & 12.1$\pm$0.3 & 24.1$\pm$1.3 & $<$1.41 & \nodata \\
NGC 6881 & 4.40$\pm$0.10 & 11.0$\pm$0.1 & 2.18$\pm$0.60 & \nodata \\
NGC 6886 & 3.83$\pm$0.08 & 8.49$\pm$1.73 & $<$1.94 & \nodata \\
Vy 2-2 & 24.0$\pm$0.4\tablenotemark{b} & \nodata & $<$0.73 & \nodata \\
\tableline
\enddata
\scriptsize
\tablenotetext{a}{Units of 10$^{-13}$ erg cm$^{-2}$ s$^{-1}$}
\tablenotetext{b}{Flux uncertain due to thin clouds}
\tablenotetext{c}{Weak or uncertain detection}
\end{deluxetable}
 
\clearpage

\begin{deluxetable}{lccccccccc}
\tablecolumns{10}
\tablewidth{0pc} 
\tabletypesize{\scriptsize}
\tablecaption{H$_2$ Line Ratios and Contribution to 2.199, 2.287 $\mu$m Features}
\tablehead{
\colhead{} & \colhead{} & \colhead{} & \colhead{} & \colhead{} & \colhead{} & \colhead{} & \colhead{} & \colhead{Adopted} & \colhead{Adopted} \\
\colhead{Object} & \colhead{} & \colhead{$F$(2.199)/} & \colhead{$F$(2.248)/} & \colhead{$F$(2.287)/} & \colhead{H$_2$} & \colhead{$F$(H$_2$~2.201)/} & \colhead{$F$(H$_2$~2.287)/} & \colhead{$F$([Kr~III])/}  & \colhead{$F$([Se~IV])/}\\
\colhead{Name} & \colhead{Ref.} & \colhead{$F$(2.224)} & \colhead{$F$(2.224)} & \colhead{$F$(2.224)} & \colhead{Model} & \colhead{$F$(Br$\gamma$)$\times$100}  & \colhead{$F$(Br$\gamma$)$\times$100}   & \colhead{$F$(Br$\gamma$)$\times$100}  & \colhead{$F$(Br$\gamma$)$\times$100}}
\startdata
Model 14\tablenotemark{a} & \nodata & 0.39 & 1.22 & 0.30 & \nodata & \nodata & \nodata & \nodata & \nodata \\
Model S2\tablenotemark{b} & \nodata & 0.03 & 0.38 & 0.00 & \nodata & \nodata & \nodata & \nodata & \nodata \\
BD+30$^{\rm o}$3639 & IR2 & $<$2.78 & $<$0.56 & $<$0.56 & S2 & \nodata & \nodata & $<$5.00 & $<$1.00 \\
 & IR4 & 0.41$\pm$0.10\tablenotemark{c} & 0.86$\pm$0.14 & \nodata & 14/S2 & \nodata & \nodata & $<$4.21\tablenotemark{d} & \nodata \\
 & IR6 & 0.49$\pm$0.17 & 0.56$\pm$0.20 & \nodata & S2 & 0.00 & \nodata & 0.90$\pm$0.22 & \nodata \\
Hb 5 & IR1 & 0.20$\pm$0.05 & 0.28$\pm$0.07 & 0.19$\pm$0.06 & S2 & 0.00 & 0.00 & 1.28$\pm$0.32\tablenotemark{d} & 1.23$\pm$0.40\tablenotemark{d} \\
Hb 12 & IR5 & 0.48$\pm$0.08\tablenotemark{c} & 1.18$\pm$0.14 & 1.32$\pm$0.15 & 14 & \nodata & 0.54$\pm$0.06 & 1.50$\pm$0.20\tablenotemark{d} & 1.86$\pm$0.16 \\
He 2-459 & IR1 & $<$0.39 & 0.42$\pm$0.10 & $<$0.24 & S2 & \nodata & \nodata & $<$2.61 & $<$1.62 \\
\enddata
\scriptsize
\tablenotetext{a}{Model 14 of Black \& van~Dishoeck 1987; purely fluorescent H$_2$ excitation with $n_{\rm H}=3.0\times10^3$~cm$^{-3}$}
\tablenotetext{b}{Model S2 of Black \& van~Dishoeck 1987; thermally excited H$_2$ at $T=2000$~K}
\tablenotetext{c}{Flux due to H$_2$ 3-2~$S$(3) only (resolved from [Kr III])}
\tablenotetext{d}{Based on high resolution spectrum}
\tablenotetext{e}{Due to the large uncertainty of the [Kr III] and H$_2$ 3-2~$S$(3) detections in the high resolution spectrum, we chose not to correct these line ratios, since the low measured $F$(2.248/2.224) indicates that the H$_2$ is collisionally excited.}
\end{deluxetable}
 
\clearpage

\begin{deluxetable}{lll}
\tablecolumns{3}
\tablewidth{0pt} 
\tablecaption{References for PN Nebular Properties}
\tabletypesize{\footnotesize}
\tablehead{
\multicolumn{1}{l}{Index} & \multicolumn{1}{l}{Reference} & \multicolumn{1}{l}{Comments}}
\startdata
\cline{1-3}
\multicolumn{3}{c}{\textit{Abundance References\tablenotemark{a}}} \\
\cline{1-3}
1 & Aller \& Czyzak 1983 & UV/Optical \\
2 & Aller \& Hyung 1995 & Optical \\
3 & Aller et al.\ 1996 & UV/Optical \\
4 & Aller \& Keyes 1987 & Optical \\
5 & Aller et al.\ 1985 & UV/Optical \\
6 & Aller et al.\ 1986 & UV/Optical \\
7 & Aller et al.\ 1988 & UV/Optical \\
\enddata
\tablenotetext{a}{Wavelength regimes used for abundance analysis are noted under comments.  Here, we use ``optical'' to refer to measurements within the range 3000--11000~\AA.}
\tablenotetext{b}{Method of $T_{\rm eff}$ determination is noted.  Whenever possible, we use $T_{\rm eff}$ determined from NLTE model atmosphere analyses.  Otherwise, $T_{\rm eff}$ determinations from photoionization modeling are preferred.  \ion{He}{2} Zanstra and energy balance temperature estimates are based on assumptions of blackbody ionizing flux distributions, optical thickness of the nebulae to He$^+$ ionizing photons, and/or analytical corrections to unobserved cooling lines.  These temperature estimates are not expected to be as robust as NLTE or photoionization model analyses.}
\end{deluxetable}

\clearpage

\begin{deluxetable}{lccccccc}
\tablecolumns{8}
\tablewidth{0pc} 
\tabletypesize{\scriptsize}
\tablecaption{PN $T_{\rm e}$, $n_{\rm e}$, and Ionic Abundances used for ICF(Kr) and ICF(Se)}
\tablehead{
\colhead{Object} & \colhead{$T_{\rm e}$(O III)} & \colhead{$T_{\rm e}$(N II)} & \colhead{$n_{\rm e}$} & \colhead{10$^{7}\times$} & \colhead{10$^{6}\times$} & \colhead{10$^{4}\times$} & \colhead{} \\
\colhead{Name} & \colhead{(10$^3$ K)} & \colhead{(10$^3$ K)} & \colhead{(10$^3$ cm$^{-3}$)} & \colhead{Ar$^{++}$/H$^+$} & \colhead{S$^{++}$/H$^+$} & \colhead{O$^{++}$/H$^+$} & \colhead{Ref.}}
\startdata
BD+30$^{\rm o}$3639 & \nodata & 8.40$\pm$1.00 & 11.00$\pm$1.10 & 2.76$\pm$0.55 & 5.07$\pm$1.01 & 0.041$\pm$0.008 & 11 \\
Cn 3-1 & 7.67$\pm$1.00 & 7.84$\pm$1.00 & 6.83$\pm$2.68 & 6.58$\pm$1.32 & 5.33$\pm$1.07 & 0.206$\pm$0.041 & 93 \\
DdDm 1 & 12.30$\pm$1.00 & 12.98$\pm$1.00 & 4.50$\pm$0.50 & 0.980$\pm$0.196 & 1.68$\pm$0.34 & 0.842$\pm$0.168 & 93 \\
Hb 4 & 10.50$\pm$1.00 & \nodata & 5.60$\pm$1.12 & 12.5$\pm$0.2\tablenotemark{a} & 2.93$\pm$1.15\tablenotemark{a} & 2.22$\pm$0.68\tablenotemark{a} &  4 \\
Hb 5 & 13.00$\pm$1.00 & 9.00$\pm$1.00 & 12.00$\pm$2.40 & 25.7$\pm$5.1 & 2.48$\pm$0.50 & 2.42$\pm$0.48 & 80 \\
Hb 6 & 11.00$\pm$1.00 & \nodata & 6.00$\pm$1.20 & 13.6$\pm$2.6\tablenotemark{a} & 2.68$\pm$0.95\tablenotemark{a} & 2.39$\pm$0.67\tablenotemark{a} &  4 \\
Hb 7 & 9.29$\pm$1.00\tablenotemark{d} & \nodata & 6.00$\pm$3.00 & \nodata & \nodata & 2.27$\pm$0.90\tablenotemark{a} & 86 \\
Hb 12 & 13.50$\pm$1.00 & 13.50$\pm$1.00 & 500.0$\pm$300.0 & 9.84$\pm$1.97 & 2.52$\pm$0.50 & 1.19$\pm$0.24 & 42 \\
\enddata
\scriptsize
\tablecomments{The complete version of this table is in the electronic edition of the Journal.  The printed edition contains only a sample.  Values are from the primary abundance reference for each object.  If several positions of a nebula were observed, we list averaged values.}
\tablenotetext{a}{Ionic abundances calculated from observed fluxes and the IRAF \textit{nebular.ionic} task (Shaw \& Dufour 1995).}
\tablenotetext{b}{$n_{\rm e}$ for the low-density region of the nebula assumed (see reference 48).}
\tablenotetext{c}{S$^{++}$/H$^+$ from near-IR lines used for Ref.\ 38, 65, 66, and 69.}
\tablenotetext{d}{$T_{\rm e}$ adopted from Ref.\ 70 since not given in Ref.\ 86 (see Table~11).}
\end{deluxetable}
 
\clearpage

\begin{deluxetable}{lccccccccc}
\tablecolumns{10}
\tablewidth{0pc}
\rotate
\tabletypesize{\scriptsize}
\tablecaption{Nebular Abundances From the Literature}
\tablehead{
\colhead{Object} & \colhead{} & \colhead{} & \colhead{10$^4\times$} & \colhead{10$^4\times$} & \colhead{10$^4\times$} & \colhead{10$^4\times$} & \colhead{10$^5\times$} & \colhead{10$^6\times$} & \colhead{10$^7\times$}\\
\colhead{Name} & \colhead{Ref.} & \colhead{He/H}  & \colhead{O/H} & \colhead{C/H\tablenotemark{a}} & \colhead{N/H} & \colhead{Ne/H} & \colhead{S/H} & \colhead{Ar/H} & \colhead{Cl/H}}
\startdata
BD+30$^{\rm o}$3639 & 11 & \nodata & 4.60$\pm$0.92 & 7.30$\pm$1.46 & 1.10$\pm$0.22 & 1.90$\pm$0.38 & 0.640$\pm$0.128 & 5.20$\pm$1.04 & 1.40$\pm$0.28 \\
Cn 3-1 & 93 & \nodata & 4.22$\pm$0.84 & \nodata & 0.749$\pm$0.150 & \nodata & \nodata & 1.23$\pm$0.25 & \nodata \\
DdDm 1 & 93 & 0.089$\pm$0.018 & 1.12$\pm$0.22 & 0.081$\pm$0.016 & 0.206$\pm$0.041 & 0.174$\pm$0.035 & 0.223$\pm$0.045 & 0.144$\pm$0.029 & 0.486$\pm$0.097 \\
Hb 4 &  4 & 0.126$\pm$0.029 & 4.79$\pm$1.10 & \nodata & 2.82$\pm$0.65 & 0.912$\pm$0.210 & 1.95$\pm$0.45 & 3.16$\pm$1.10 & 5.01$\pm$1.15 \\
Hb 5 & 80 & 0.123$\pm$0.025 & 5.00$\pm$1.00 & \nodata & 8.40$\pm$1.68 & 2.20$\pm$0.44 & 1.10$\pm$0.22 & 6.00$\pm$1.20 & 2.80$\pm$0.56 \\
Hb 6 &  4 & 0.110$\pm$0.039 & 5.13$\pm$1.80 & \nodata & 4.47$\pm$1.56 & 1.00$\pm$0.35 & 2.00$\pm$0.70 & 5.01$\pm$1.75 & 2.57$\pm$0.90 \\
Hb 7 & 86 & 0.060$\pm$0.012 & 4.47$\pm$1.34 & \nodata & 0.269$\pm$0.081 & \nodata & 0.295$\pm$0.089 & 1.05$\pm$0.32 & \nodata \\
Hb 12 & 42 & 0.109$\pm$0.016 & 2.20$\pm$0.44 & 1.15$\pm$0.23 & 0.600$\pm$0.120 & 0.360$\pm$0.072 & 0.420$\pm$0.084 & 2.00$\pm$0.40 & 0.450$\pm$0.090 \\
\enddata
\scriptsize
\tablecomments{The complete version of this table is in the electronic edition of the Journal.  The printed edition contains only a sample.  Values are from the primary abundance reference for each object.  If several positions of a nebula were observed, we list the average abundances.}
\tablenotetext{a}{C abundances from UV collisionally-excited lines only.}
\end{deluxetable}
 
\clearpage

\begin{deluxetable}{lccccccc}
\tablecolumns{8}
\tablewidth{0pc} 
\tabletypesize{\scriptsize}
\tablecaption{Ionic and Elemental Se and Kr Abundances}
\tablehead{
\colhead{PN Name} & \colhead{$\frac{n(\mathrm{Kr}^{++})}{n(\mathrm{H}^+)}$} & \colhead{$\frac{n(\mathrm{Se}^{3+})}{n(\mathrm{H}^+)}$} & \colhead{ICF(Kr)\tablenotemark{a}} & \colhead{ICF(Se)} & \colhead{Ref.\tablenotemark{b}} & \colhead{[Kr/H]} & \colhead{[Se/H]}}
\startdata
BD+30$^{\rm o}$3639\tablenotemark{c} & $<$6.54E$-9$ & $<$2.83E$-10$ & 1.95$\pm$1.29\tablenotemark{e} & \nodata & IR2, 11 & $<$0.82 & \nodata \\
  & $<$5.51E$-9$ & \nodata & 1.95$\pm$1.29\tablenotemark{e} & \nodata & IR4, 11 & $<$0.75 & \nodata \\
  & (1.18$\pm$0.35)E$-9$ & \nodata & 1.95$\pm$1.29\tablenotemark{e} & \nodata & IR6, 11 & 0.08$\pm$0.33 & \nodata \\
Cn 3-1 & $<$2.71E$-9$ & $<$5.39E$-10$ & 5.57$\pm$2.53 & 114.03$\pm$50.96 & IR1, 93 & $<$0.90 & $<$1.46 \\
DdDm 1 & $<$7.42E$-9$ & $<$1.88E$-9$ & 3.85$\pm$2.28 & 2.67$\pm$1.58 & IR1, 93 & $<$1.18 & $<$0.37 \\
Hb 4 & $<$1.47E$-9$ & (7.94$\pm$1.42)E$-10$ & 8.13$\pm$4.02 & 5.75$\pm$4.00 & IR1,  4 & $<$0.80 & 0.33$\pm$0.39 \\
Hb 5\tablenotemark{c} & (9.39$\pm$2.51)E$-10$ & (2.01$\pm$0.68)E$-10$\tablenotemark{d} & 7.39$\pm$2.95 & 5.39$\pm$2.71 & IR1, 80 & 0.56$\pm$0.23 & $-0.30\pm$0.31 \\
Hb 6 & $<$1.37E$-9$ & (7.59$\pm$0.86)E$-10$ & 12.61$\pm$7.28 & 5.70$\pm$4.96 & IR1,  4 & $<$0.96 & 0.31$\pm$0.55 \\
\enddata
\scriptsize
\tablecomments{The ICFs are derived from the primary abundance references.  Solar abundance values are taken from Asplund et al.\ (2005).}
\tablenotetext{a}{ICF(Kr) from Equation~(1) except where noted}
\tablenotetext{b}{See Table~6}
\tablenotetext{c}{Contains H$_2$}
\tablenotetext{d}{Based on a weak or uncertain detection}
\tablenotetext{e}{ICF(Kr) from Equation~(2)}
\end{deluxetable}
 
\clearpage

\begin{deluxetable}{lccccccc}
\tablecolumns{8}
\tablewidth{0pc} 
\tabletypesize{\scriptsize}
\tablecaption{Se and Kr Abundances Relative to O and Ar}
\tablehead{
\colhead{PN Name} & \colhead{[Kr/H]} & \colhead{[Kr/O]\tablenotemark{a}} & \colhead{[Kr/Ar]} & \colhead{[Se/H]} & \colhead{[Se/O]\tablenotemark{a}} & \colhead{[Se/Ar]} & \colhead{Ref.}}
\startdata
BD+30$^{\rm o}$3639 & $<$0.82 & $<$   0.82 & $<$   0.29 & \nodata & \nodata & \nodata & IR2, 11 \\
  & $<$0.75 & $<$   0.75 & $<$   0.21 & \nodata & \nodata & \nodata & IR4, 11 \\
  & 0.08$\pm$0.33 &    0.08$\pm$0.36 &   $-0.46\pm$0.36 & \nodata & \nodata & \nodata & IR6, 11 \\
Cn 3-1 & $<$0.90 & $<$   0.93 & $<$   0.99 & $<$1.46 & $<$   1.49 & $<$   1.55 & IR1, 93 \\
DdDm 1 & $<$1.18 & $<$   1.78 & $<$   2.20 & $<$0.37 & $<$   0.98 & $<$   1.39 & IR1, 93 \\
Hb 4 & $<$0.80 & ($<$   0.78) & $<$   0.48 & 0.33$\pm$0.39 &    (0.31$\pm$0.43) &   0.01$\pm$0.48 & IR1,  4 \\
Hb 5 & 0.56$\pm$0.23 & (0.52$\pm$0.25) & $-0.04\pm$0.25 & $-0.30\pm$0.31\tablenotemark{d} & ($-0.33\pm$0.33)\tablenotemark{d} & $-0.89\pm$0.33\tablenotemark{d} & IR1, 80 \\
Hb 6 & $<$0.96 & ($<$   0.91) & $<$   0.44 & 0.31$\pm$0.55 &    (0.26$\pm$0.70) &   $-0.21\pm$0.70 & IR1,  4 \\
\enddata
\scriptsize
\tablecomments{Solar abundance values are taken from Asplund et al.\ (2005).}
\tablenotetext{a}{The [Kr/O] and [Se/O] values of Type~I PNe are placed in parentheses.  O may be depleted via the ON-cycle in the progenitor stars of Type~I PNe, and therefore the Kr and Se abundances relative to O may not be reliable indicators of \emph{s}-process enrichments in these objects.}
\tablenotetext{b}{IC~4997 has an uncertain O abundance, and Ar is a better metallicity indicator for this object.}
\tablenotetext{c}{Although not classified as Type~I PNe, M~1-31, NGC~6881 and Vy~2-2 exhibit N enrichments indicative of second dredge-up and hot bottom burning.  For these objects, we use Ar as a reference element, since O may have been depleted by ON-cycling.}
\tablenotetext{d}{Based on a weak or uncertain detection.}
\end{deluxetable}
 
\clearpage

\begin{deluxetable}{lccccccc}
\tablecolumns{8}
\tablewidth{0pc} 
\tabletypesize{\footnotesize}
\tablecaption{Comparison of $T_{\rm e}$, $n_{\rm e}$, and Ionic Abundances from Different References}
\tablehead{
\colhead{Object} & \colhead{$T_{\rm e}$(O III)} & \colhead{$T_{\rm e}$(N II)} & \colhead{$n_{\rm e}$} & \colhead{10$^{-7}$} & \colhead{10$^{-6}$} & \colhead{10$^{-4}$} & \colhead{} \\
\colhead{Name} & \colhead{(10$^3$ K)} & \colhead{(10$^3$ K)} & \colhead{(10$^3$ cm$^{-3}$)} & \colhead{Ar$^{++}$/H$^+$} & \colhead{S$^{++}$/H$^+$} & \colhead{O$^{++}$/H$^+$} & \colhead{Ref.}}
\startdata
BD+30$^{\rm o}$3639 & \nodata & 8.40$\pm$1.00 & 11.00$\pm$1.10 & 2.76$\pm$0.55 & 5.07$\pm$1.01 & 0.041$\pm$0.008 & 11 \\
  & \nodata & 8.80$\pm$0.20 & 13.00$\pm$2.60 & 2.46$\pm$0.49 & 3.85$\pm$0.77 & 0.040$\pm$0.008 &  2 \\
Cn 3-1 & 7.67$\pm$1.00 & 7.84$\pm$1.00 & 6.83$\pm$2.68 & 6.58$\pm$1.32 & 5.33$\pm$1.07 & 0.206$\pm$0.041 & 93 \\
  & 10.10$\pm$1.00 & 6.50$\pm$1.00 & 7.28$\pm$1.46 & \nodata & 1.05$\pm$0.21 & 0.057$\pm$0.011 &  1 \\
  & \nodata & 7.50$\pm$1.00 & 6.90$\pm$1.38 & 5.89$\pm$2.41 & 3.45$\pm$2.94 & 0.126$\pm$0.079 & 31 \\
DdDm 1 & 12.30$\pm$1.00 & 12.98$\pm$1.00 & 4.50$\pm$0.50 & 0.980$\pm$0.196 & 1.68$\pm$0.34 & 0.842$\pm$0.168 & 93 \\
  & 11.80$\pm$0.80 & 11.00$\pm$1.20 & 4.40$\pm$0.88 & 4.90$\pm$0.98 & 2.56$\pm$0.51 & 0.896$\pm$0.179 & 14 \\
  & 11.70$\pm$1.00 & 11.40$\pm$1.00 & 4.00$\pm$0.80 & 5.16$\pm$1.03 & 1.87$\pm$0.37\tablenotemark{c} & 1.00$\pm$0.20 & 38 \\
\enddata
\scriptsize
\tablecomments{The complete version of this table is in the electronic edition of the Journal.  The printed edition contains only a sample.  Values from primary references listed first.  If several positions of a nebula were observed, averaged values are listed except where noted.    If a PN has only one abundance reference from Table~6, it is not listed here (see Table~7).}
\tablenotetext{a}{Ionic abundances calculated from observed fluxes and the IRAF \textit{nebular.ionic} task (Shaw \& Dufour 1995).}
\tablenotetext{b}{$T_{\rm e}$(\ion{N}{2}) used for ionic and Se and Kr abundance determinations.}
\tablenotetext{c}{S$^{++}$/H$^+$ from near-IR lines used for Ref.\ 38, 65, 66, and 69.}
\end{deluxetable}
 
\clearpage

\begin{deluxetable}{lccccccccc}
\tablecolumns{10}
\rotate
\tablewidth{0pc}
\tabletypesize{\scriptsize}
\tablecaption{Comparison of Nebular Abundances From Different References}
\tablehead{
\colhead{Object} & \colhead{} & \colhead{} & \colhead{10$^4$} & \colhead{10$^4$} & \colhead{10$^4$} & \colhead{10$^4$} & \colhead{10$^5$} & \colhead{10$^6$} & \colhead{10$^7$}\\
\colhead{Name} & \colhead{Ref.} & \colhead{He/H}  & \colhead{O/H} & \colhead{C/H\tablenotemark{a}} & \colhead{N/H} & \colhead{Ne/H} & \colhead{S/H} & \colhead{Ar/H} & \colhead{Cl/H}}
\startdata
BD+30$^{\rm o}$3639 & 11 & \nodata & 4.60$\pm$0.92 & 7.30$\pm$1.46 & 1.10$\pm$0.22 & 1.90$\pm$0.38 & 0.640$\pm$0.128 & 5.20$\pm$1.04 & 1.40$\pm$0.28 \\
  &  2 & \nodata & 3.72$\pm$0.74 & 3.98$\pm$2.39 & 1.12$\pm$0.22 & \nodata & 0.537$\pm$0.107 & \nodata & 1.45$\pm$0.29 \\
Cn 3-1 & 93 & \nodata & 4.22$\pm$0.84 & \nodata & 0.749$\pm$0.150 & \nodata & \nodata & 1.23$\pm$0.25 & \nodata \\
  &  1 & \nodata & 1.68$\pm$1.51 & \nodata & 1.85$\pm$1.11 & 0.605$\pm$0.545 & 0.250$\pm$0.150 & \nodata & 2.50$\pm$2.30 \\
  & 31 & 0.048$\pm$0.010 & 6.17$\pm$1.85 & \nodata & 0.891$\pm$0.267 & \nodata & 0.977$\pm$0.293 & 1.41$\pm$0.42 & 0.891$\pm$0.668 \\
DdDm 1 & 93 & 0.089$\pm$0.018 & 1.12$\pm$0.22 & 0.081$\pm$0.016 & 0.206$\pm$0.041 & 0.174$\pm$0.035 & 0.223$\pm$0.045 & 0.144$\pm$0.029 & 0.486$\pm$0.097 \\
  & 14 & 0.100$\pm$0.020 & 1.40$\pm$0.42 & \nodata & 0.250$\pm$0.080 & 0.200$\pm$0.060 & 0.340$\pm$0.100 & 0.730$\pm$0.220 & \nodata \\
  & 38 & 0.100$\pm$0.020 & 1.38$\pm$0.41 & 0.059$\pm$0.024 & 0.290$\pm$0.087 & 0.230$\pm$0.070 & 0.240$\pm$0.070 & 0.520$\pm$0.160 & 0.210$\pm$0.060 \\
\enddata
\scriptsize
\tablecomments{The complete version of this table is in the electronic edition of the Journal.  The printed edition contains only a sample.  Values from primary references listed first.  If several positions of a nebula were observed, averaged abundances are listed except where noted.    If a PN has only one abundance reference from Table~6, it is not listed here (see Table~8).}
\tablenotetext{a}{C abundances from UV collisionally-excited lines only.}
\end{deluxetable}
 
\clearpage

\begin{deluxetable}{lccccccc}
\tablecolumns{8}
\tablewidth{0pc} 
\tabletypesize{\scriptsize}
\tablecaption{Comparison of Ionic and Total Se and Kr Abundances Derived Using Different Abundance References}
\tablehead{
\colhead{PN Name} & \colhead{$\frac{n(\mathrm{Kr}^{++})}{n(\mathrm{H}^+)}$} & \colhead{$\frac{n(\mathrm{Se}^{3+})}{n(\mathrm{H}^+)}$} & \colhead{ICF(Kr)\tablenotemark{a}} & \colhead{ICF(Se)} & \colhead{Ref.\tablenotemark{b}} & \colhead{[Kr/H]} & \colhead{[Se/H]}}
\startdata
BD+30$^{\rm o}$3639$^{\rm o}$3639\tablenotemark{c} & (1.18$\pm$0.35)E$-9$ & \nodata & 1.95$\pm$1.29\tablenotemark{e} & \nodata & IR6, 11 & 0.08$\pm$0.33 & \nodata \\
  & (1.11$\pm$0.27)E$-9$ & \nodata & 2.26$\pm$1.40\tablenotemark{e} & \nodata & IR6,  2 & 0.12$\pm$0.33 & \nodata \\
Cn 3-1 & $<$2.71E$-9$ & $<$5.39E$-10$ & 5.57$\pm$2.53 & 114.03$\pm$50.96 & IR1, 93 & $<$0.90 & $<$1.46 \\
  & $<$1.84E$-9$ & $<$3.73E$-10$ & 4.62$\pm$6.31\tablenotemark{e} & 212.88$\pm$116.50 & IR1,  1 & $<$0.65 & $<$1.57 \\
  & $<$2.79E$-9$ & $<$5.54E$-10$ & 7.62$\pm$6.56 & 923.29$\pm$478.35 & IR1, 31 & $<$1.05 & $<$2.37 \\
DdDm 1 & $<$7.42E$-9$ & $<$1.88E$-9$ & 3.85$\pm$2.28 & 2.67$\pm$1.58 & IR1, 93 & $<$1.18 & $<$0.37 \\
  & $<$7.85E$-9$ & $<$1.98E$-9$ & 3.95$\pm$2.94 & 3.48$\pm$2.49 & IR1, 14 & $<$1.21 & $<$0.51 \\
  & $<$7.94E$-9$ & $<$2.00E$-9$ & 1.41$\pm$1.67 & 2.83$\pm$1.93 & IR1, 38 & $<$0.77 & $<$0.42 \\
\enddata
\scriptsize
\tablecomments{The complete version of this table is in the electronic edition of the Journal.  The printed edition contains only a sample.  Abundances derived with primary abundance references are listed first.    If a PN has only one abundance reference from Table~6, it is not listed here (see Table~9).  Solar abundance values are taken from Asplund et al.\ (2005).}
\tablenotetext{a}{ICF(Kr) from Equation~(1) except where noted}
\tablenotetext{b}{See Table~6}
\tablenotetext{c}{Contains H$_2$}
\tablenotetext{d}{Based on a weak or uncertain detection}
\tablenotetext{e}{ICF(Kr) from Equation~(2)}
\end{deluxetable}
 
\clearpage

\begin{deluxetable}{lccccccc}
\tablecolumns{8}
\tablewidth{0pc} 
\tabletypesize{\scriptsize}
\tablecaption{Comparison of Se and Kr Abundances Derived from Different Abundance References}
\tablehead{
\colhead{PN Name} & \colhead{[Kr/H]} & \colhead{[Kr/O]} & \colhead{[Kr/Ar]} & \colhead{[Se/H]} & \colhead{[Se/O]} & \colhead{[Se/Ar]}& \colhead{Ref.}}
\startdata
BD+30$^{\rm o}$3639$^{\rm o}$3639 & 0.08$\pm$0.33 &    0.08$\pm$0.36 &   $-0.46\pm$0.36 & \nodata & \nodata & \nodata & IR6, 11 \\
  & 0.12$\pm$0.33 &    0.20$\pm$0.35 & \nodata & \nodata & \nodata & \nodata & IR6,  2 \\
Cn 3-1 & $<$0.90 & $<$   0.93 & $<$   0.99 & $<$1.46 & $<$   1.49 & $<$   1.55 & IR1, 93 \\
  & $<$0.65 & $<$   1.08 & \nodata & $<$1.57 & $<$   2.00 & \nodata & IR1,  1 \\
  & $<$1.05 & $<$   0.92 & $<$   1.08 & $<$2.37 & $<$   2.24 & $<$   2.40 & IR1, 31 \\
DdDm 1 & $<$1.18 & $<$   1.78 & $<$   2.20 & $<$0.37 & $<$   0.98 & $<$   1.39 & IR1, 93 \\
  & $<$1.21 & $<$   1.72 & $<$   1.53 & $<$0.51 & $<$   1.02 & $<$   0.82 & IR1, 14 \\
  & $<$0.77 & $<$   1.29 & $<$   1.23 & $<$0.42 & $<$   0.94 & $<$   0.89 & IR1, 38 \\
\enddata
\scriptsize
\tablecomments{The complete version of this table is in the electronic edition of the Journal.  The printed edition contains only a sample.  Abundances derived using the primary abundance references are listed first.  If a PN has only one abundance reference from Table~6, it is not listed here (see Table~10).  The [Kr/O] and [Se/O] values of Type~I PNe are placed in parentheses.  O may be depleted via the ON-cycle in the progenitor stars of Type~I PNe, and therefore the Kr and Se abundances relative to O may not be reliable indicators of \emph{s}-process enrichments in these objects.  Solar abundance values are taken from Asplund et al.\ (2005).}
\end{deluxetable}

\clearpage

\begin{deluxetable}{lccccccc}
\tablecolumns{8}
\tablewidth{0pc}
\tabletypesize{\scriptsize}
\tablecaption{Kr and Se Detection Rates Vs.\ Nebular Properties}
\tablehead{
\multicolumn{1}{l}{} & \colhead{Number} & \colhead{Kr} & \colhead{Kr Det.} & \colhead{Se} & \colhead{Se Det.} & \colhead{Se \textit{or} Kr} & \colhead{Se \textit{or} Kr} \\
\multicolumn{1}{l}{Property} & \colhead{of PNe} & \colhead{Detections} & \colhead{Rate (\%)} & \colhead{Detections} & \colhead{Rate (\%)} & \colhead{Detections} & \colhead{Det.\ Rate (\%)}}
\startdata
Full Sample & 120 & 36 & 30.0 & 70 & 58.3 & 81 & 67.5 \\
\tableline
\multicolumn{8}{c}{\textit{Progenitor Mass}} \\
\tableline
Type I & 29 & 3 & 10.3 & 12 & 41.4 & 12 & 41.4 \\
Non-Type I & 91 & 31 & 34.1 & 58 & 63.7 & 69 & 75.8 \\
\tableline
\multicolumn{8}{c}{\textit{Morphology}} \\
\tableline
Bipolar & 28 & 9 & 32.1 & 14 & 50.0 & 15 & 53.4 \\
Elliptical & 47 & 15 & 31.9 & 31 & 66.0 & 37 & 78.7 \\
Round & 4 & 1 & 25.0 & 3 & 75.0 & 3 & 75.0 \\
Irregular\tablenotemark{a} & 7 & 2 & 28.6 & 4 & 57.1 & 4 & 57.1 \\
Unknown Morphology & 34 & 9 & 26.5 & 18 & 52.9 & 22 & 64.7 \\
\tableline
\multicolumn{8}{c}{\textit{Central Star Type}} \\
\tableline
$[$WC$]$ & 26 & 10 & 38.5 & 16 & 61.5 & 20 & 76.9 \\
WELS & 17 & 3 & 17.6 & 12 & 70.6 & 12 & 70.6 \\
Non-[WC]/WELS & 77 & 23 & 29.9 & 42 & 54.5 & 49 & 63.6 \\
\tableline
\multicolumn{8}{c}{\textit{H$_2$ Emission}} \\
\tableline
H$_2$ & 39 & 23 & 59.0 & 25 & 64.1 & 30 & 76.9 \\
Non-H$_2$ & 81 & 13 & 16.0 & 45 & 54.5 & 51 & 63.0 \\
\tableline
\multicolumn{8}{c}{\textit{Dust Chemistry}} \\
\tableline
O-rich Dust & 4 & 0 & 0.0 & 2 & 50.0 & 2 & 50.0 \\
Mixed Dust & 9 & 7 & 77.8 & 5 & 55.6 & 8 & 88.9 \\
C-rich Dust & 21 & 15 & 71.4 & 14 & 66.7 & 19 & 90.5 \\
21$\mu$m Dust & 2 & 1 & 50.0 & 1 & 50.0 & 2 & 100.0 \\
Unknown Dust Chemistry & 84 & 13 & 15.5 & 48 & 57.1 & 50 & 59.5 \\
\tableline
\enddata
\tablenotetext{a}{Includes point-symmetric nebulae} 
\end{deluxetable}

\clearpage

\begin{deluxetable}{lcccccc}
\tablecolumns{7}
\tablewidth{0pc}
\tabletypesize{\footnotesize}
\tablecaption{Kr and Se Abundances Vs. Nebular Properties Tracing Progenitor Mass}
\tablehead{
\colhead{} & \colhead{Mean} & \colhead{} & \colhead{Number of} & \colhead{Mean} & \colhead{} & \colhead{Number of} \\
\colhead{Property} & \colhead{[Se/(O,~Ar)]\tablenotemark{a}} & \colhead{$< \sigma >$\tablenotemark{b}} & \colhead{Se Detections} & \colhead{[Kr/(O,~Ar)]\tablenotemark{a}} & \colhead{$< \sigma >$\tablenotemark{b}} & \colhead{Kr Detections}}
\startdata
Type I & $-0.03$ & 0.27 & 12 & 0.09 & 0.14 & 3 \\
Non-Type I\tablenotemark{c} & 0.36 & 0.26 & 55 & 1.02 & 0.27 & 30 \\
\tableline
Bipolar & 0.27 & 0.38 & 14 & 0.68 & 0.25 & 8 \\
Elliptical & 0.28 & 0.22 & 28 & 1.09 & 0.38 & 15 \\
Round & 0.59 & 0.17 & 2 & \nodata & \nodata & 0 \\
Irregular\tablenotemark{d} & 0.06 & 0.21 & 4 & 0.52 & 0.40 & 2 \\
\tableline
Full Sample & 0.31 & 0.27 & 67 & 0.98 & 0.31 & 33 \\
\enddata
\tablecomments{Only PNe exhibiting Se and/or Kr emission and with determined O and Ar abundances are considered.}
\tablenotetext{a}{Ar is used as the reference element for Type~I PNe, as well as M~1-31, NGC~6881, and Vy~2-2 (which have N/O ratios just below the Type~I cutoff); O is used for all other objects.} 
\tablenotetext{b}{The $< \sigma >$ are the mean absolute deviations in the Se and Kr abundances.} 
\tablenotetext{c}{For non-Type~I PNe, the mean [Se/Ar]~=~0.26 (for 53 objects) and [Kr/Ar]~=~1.00 (for 29 objects), which are very similar to their abundances relative to O.  This shows that using O as a reference element for non-Type~I PNe instead of Ar does not affect our results.} 
\tablenotetext{d}{Includes point-symmetric nebulae} 
\end{deluxetable}

\clearpage

\begin{deluxetable}{lcccccc}
\tablecolumns{7}
\tablewidth{0pc}
\tabletypesize{\footnotesize}
\tablecaption{Kr and Se Abundances Vs.\ Central Star Type and Dust Chemistry}
\tablehead{
\colhead{} & \colhead{Mean} & \colhead{} & \colhead{Number of} & \colhead{Mean} & \colhead{} & \colhead{Number of} \\
\colhead{Property} & \colhead{[Se/(O,~Ar)]\tablenotemark{a}} & \colhead{$< \sigma >$\tablenotemark{b}} & \colhead{Se Detections} & \colhead{[Kr/(O,~Ar)]\tablenotemark{a}} & \colhead{$< \sigma >$\tablenotemark{b}} & \colhead{Kr Detections}}
\startdata
$[$WC$]$ & 0.39 & 0.28 & 16 & 0.90 & 0.34 & 10 \\
WELS & 0.26 & 0.25 & 12 & 0.63 & 0.21 & 3 \\
Non-[WC]/WELS & 0.29 & 0.28 & 39 & 1.05 & 0.32 & 20 \\
Binary & $-0.07$ & 0.12 & 5 & 0.82 & 0.08 & 2 \\
\tableline
O-rich Dust & $-0.47$ & 0.08 & 2 & \nodata & \nodata & 0 \\
Mixed Dust & 0.34 & 0.14 & 3 & 1.19 & 0.36 & 7 \\
C-rich Dust & 0.28 & 0.25 & 13 & 0.83 & 0.28 & 14 \\
21$\mu$m Dust & 0.79 & \nodata & 1 & 1.03 & \nodata & 1 \\
\tableline
Full Sample & 0.31 & 0.27 & 67 & 0.98 & 0.31 & 33 \\
\enddata
\tablecomments{Only PNe exhibiting Se and/or Kr emission and with determined O and Ar abundances are considered.}
\tablenotetext{a}{Ar is used as the reference element for Type~I PNe, as well as M~1-31, NGC~6881, and Vy~2-2 (which have N/O ratios that leave them just short of the Type~I cutoff); O is used for all other objects.} 
\tablenotetext{b}{The $< \sigma >$ are the mean absolute deviations in the Se and Kr abundances for each group of PNe.} 
\end{deluxetable}

\clearpage

\begin{deluxetable}{lclcc}
\tablecolumns{7}
\tablewidth{0pc}
\tablecaption{PNe with Possible Binary Central Stars}
\tablehead{
\colhead{Object} & \colhead{Evidence for} & \colhead{} & \colhead{} & \colhead{} \\
\colhead{Name} & \colhead{Binarity\tablenotemark{a}} & \multicolumn{1}{c}{Reference} & \colhead{[Se/(O,~Ar)]\tablenotemark{b}} & \colhead{[Kr/(O,~Ar)]\tablenotemark{b}}}
\startdata
Cn 3-1 & RV & Sorensen \& Pollaco 2004 & $<1.49$ & $<$0.93 \\
Hb 12 & P & Hsia et al.\ 2006 & 0.11$\pm$0.27 & 0.89$\pm$0.24 \\
Hu 2-1 & M & Miranda et al.\ 2001b & $<-0.44$ & 0.74$\pm$0.46 \\
IC 4593 & RV & De~Marco et al.\ 2004\tablenotemark{c} & $<$0.02 & $<0.54$ \\
IC 4634 & M & G\'{o}rny et al.\ 1997 & $-0.15\pm$0.21 & $<0.35$ \\
IC 4846 & M & Miranda et al.\ 2001a & $<-0.04$ & $<0.80$ \\
J 320 & M & Harman et al.\ 2004 & $<0.36$ & $<1.22$ \\
Me 1-1 & C & Shen et al.\ 2004 & $<-0.08$ & $<0.74$ \\
NGC 2392 & RV & Af\c{s}ar \& Bond 2005 & $<0.44$ & $<1.18$ \\
NGC 6210 & RV & De~Marco et al.\ 2004 & 0.02$\pm$0.21 & $<0.29$ \\
NGC 6302 & C & Feibelman 2001 & $<-0.53$ & $<0.29$ \\
NGC 6543 & RV & Sorensen \& Pollaco 2004 & $-0.23\pm$0.24 & \nodata \\
NGC 6826 & RV & De Marco 2006 & $-0.19\pm$0.30 & $<0.61$ \\
NGC 6891 & RV & De Marco et al.\ 2004\tablenotemark{c} & $<0.02$ & $<0.64$ \\
\enddata
\tablenotetext{a}{C -- composite spectrum with cool stellar companion; M -- morphological properties (e.g., point-symmetric geometry); P -- photometric variations in light curve;  RV -- radial velocity variations.} 
\tablenotetext{b}{Ar is used as the reference element for Type~I PNe, while O is used for other objects.} 
\tablenotetext{c}{Confirmed by Sorensen \& Pollaco (2004) and/or Af\c{s}ar \& Bond (2005).}
\end{deluxetable}

\clearpage

\begin{deluxetable}{lcccccccccc}
\tablecolumns{11}
\tablewidth{0pc} 
\tabletypesize{\scriptsize}
\tablecaption{Distances and [\ion{O}{3}] $\lambda$5007 Magnitudes}
\tablehead{
\colhead{Object} & \colhead{$d_{\rm meas}$} & \colhead{$d_{\rm CKS}$} & \colhead{$d_{\rm Phil}$} & \colhead{$d_{\rm VdSZ}$} & \colhead{$d_{\rm Zh}$} & \colhead{Adopted} & \colhead{$-$Log($F$(H$\beta$))} & \colhead{} & \colhead{} & \colhead{}\\
\colhead{Name} & \colhead{(kpc)\tablenotemark{a}} & \colhead{(kpc)} & \colhead{(kpc)} & \colhead{(kpc)} & \colhead{(kpc)} & \colhead{$d$ (kpc)} & \colhead{(erg cm$^{-2}$ s$^{-1}$)} & \colhead{$c_{\rm H\beta}$} & \colhead{$M_{5007}$\tablenotemark{b}} & \colhead{Enriched?\tablenotemark{c}}}
\startdata
BD+30$^{\rm o}$3639 &  1.30$\pm$0.20 &  1.16 &  2.14 &  1.84 &  1.85 &  1.30$\pm$0.20 & 10.03$\pm$0.01 &  0.40 &   1.87$\pm$0.35 & I \\
Cn 3-1 & \nodata &  3.58 &  4.00 &  3.92 &  4.05 &  3.89$\pm$0.21 & 10.94$\pm$0.02 &  0.46 &   0.60$\pm$0.12 & I \\
DdDm 1 & \nodata & 11.03 & 15.85 & \nodata & 15.35 & 14.08$\pm$2.65 & 11.57$\pm$0.10 &  0.14 &  $-3.98\pm$0.43 & I \\
Hb 4 & \nodata &  2.08 & \nodata &  2.60 &  2.68 &  2.45$\pm$0.33 & 11.96$\pm$0.01 &  1.94 &  $-0.27\pm$0.30 & I \\
Hb 5 & \nodata &  1.24 & \nodata &  1.26 &  1.32 &  1.27$\pm$0.04 & 11.52$\pm$0.04 &  1.69 &  $-0.36\pm$0.07 & N \\
Hb 6 & \nodata &  1.66 & \nodata &  2.40 &  2.45 &  2.17$\pm$0.44 & 12.05$\pm$0.01 &  2.10 &   0.01$\pm$0.47 & I \\
Hb 7 & \nodata &  5.54 & \nodata & \nodata &  5.90 &  5.72$\pm$0.25 & 11.25$\pm$0.07 &  0.28 &  $-3.48\pm$0.10 & I \\
Hb 12 & \nodata &  2.24 & 10.46 & \nodata &  8.11 &  6.94$\pm$4.24 & 11.02$\pm$0.04 &  1.35 &  $-3.94\pm$1.00 & Y \\
\enddata
\scriptsize
\tablecomments{The complete version of this table is in the electronic edition of the Journal.  The printed edition contains only a sample.  Adopted distances, $F$(H$\beta$), extinction coefficients, $[$\ion{O}{3}$]$ $\lambda$5007 magnitudes, and \emph{s}-process enrichments are given for each PN.  The adopted distances are either measured distances or the averaged statistical distances of CKS (Cahn et al.\ 1992), Phillips (2004), VdSZ (Van~de~Steene \& Zijlstra 1994), and Zhang (1995).  The $F$(H$\beta$) values are all from CKS or the Vizier Strasbourg-ESO Catalog database, and $c_{\rm H\beta}$ are from the primary abundance references (Table~6).  The following PNe have been excluded from this analysis, due to lack of distance, $F$(H$\beta$), or optical data: K~3-17, K~3-55, K~3-62, M~1-71, and Vy~1-2.}
\tablenotetext{a}{References for trigonometric or expansion parallax distances ($d_{\rm meas}$): Acker et al.\ 1998 (NGC 2392), G\'{o}mez et al.\ 1993 (NGC 6302), Guzm\'{a}n et al.\ 2006 (M 2-43), Hajian et al.\ 1995 (NGC 6210, NGC 6572), Harris et al.\ 1997 (NGC 6720), and Mellema 2004 (BD+30$^{\rm o}$3639, NGC 3242, NGC 6543, NGC 6578, NGC 6884, NGC 7027, NGC 7662, Vy 2-2).}
\tablenotetext{b}{The derived $M_{5007}$ have been shifted by $-1.78$ magnitudes to match the expected bright limit of $-4.48$ mag (Ciardullo et al.\ 1989).}
\tablenotetext{c}{Determination of \emph{s}-process enrichment: Y -- yes (enriched), N -- not enriched, I -- indeterminate enrichment.  See \S6.2.}
\end{deluxetable}

\clearpage

\begin{figure}
\plotone{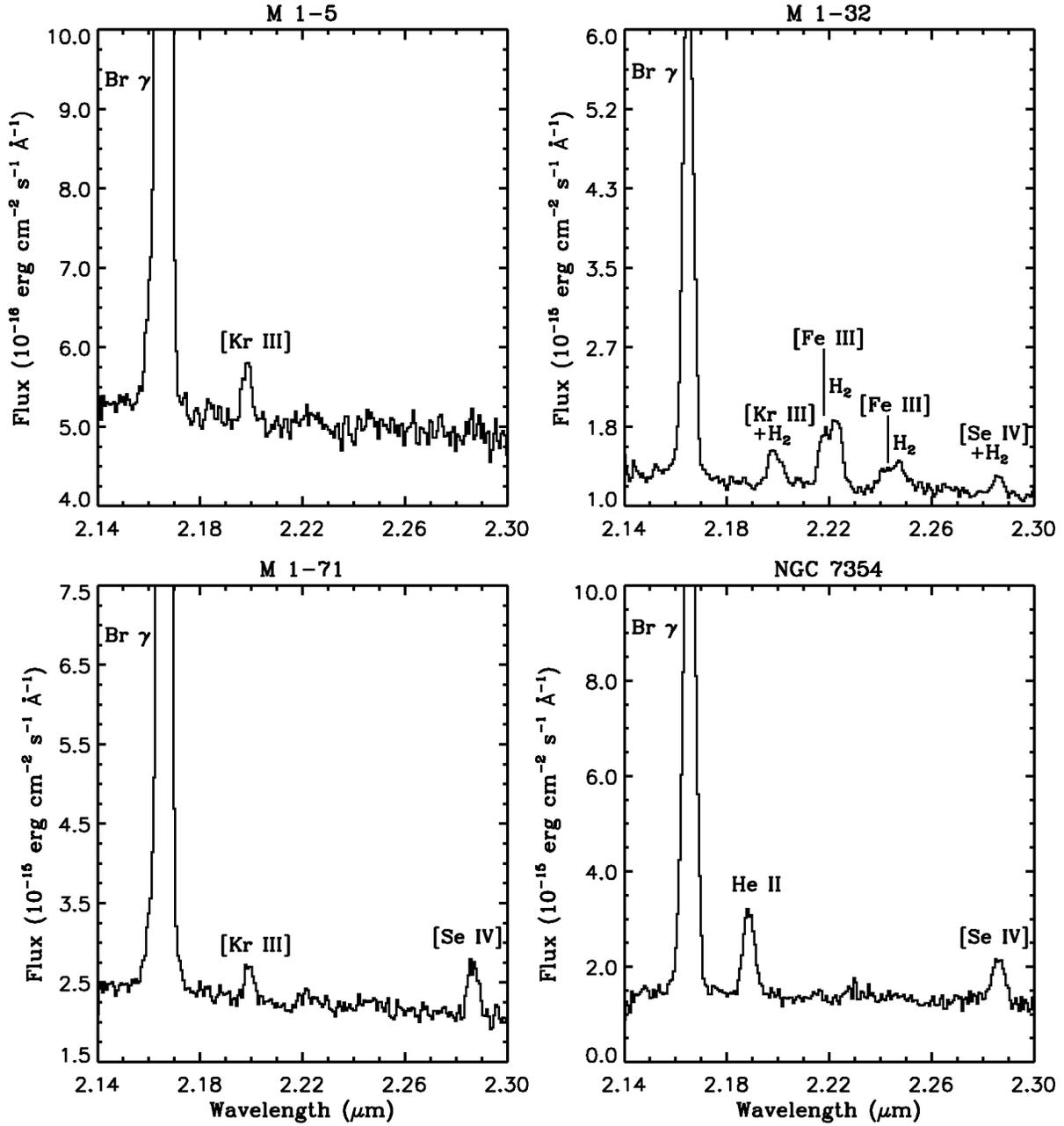}
\figcaption[f1.eps]{$K$ band spectra of four PNe from our sample.  The Br$\gamma$ line is truncated in all spectra except M~1-32 for display of weak nebular features.}
\end{figure}

\clearpage

\begin{figure}
\plotone{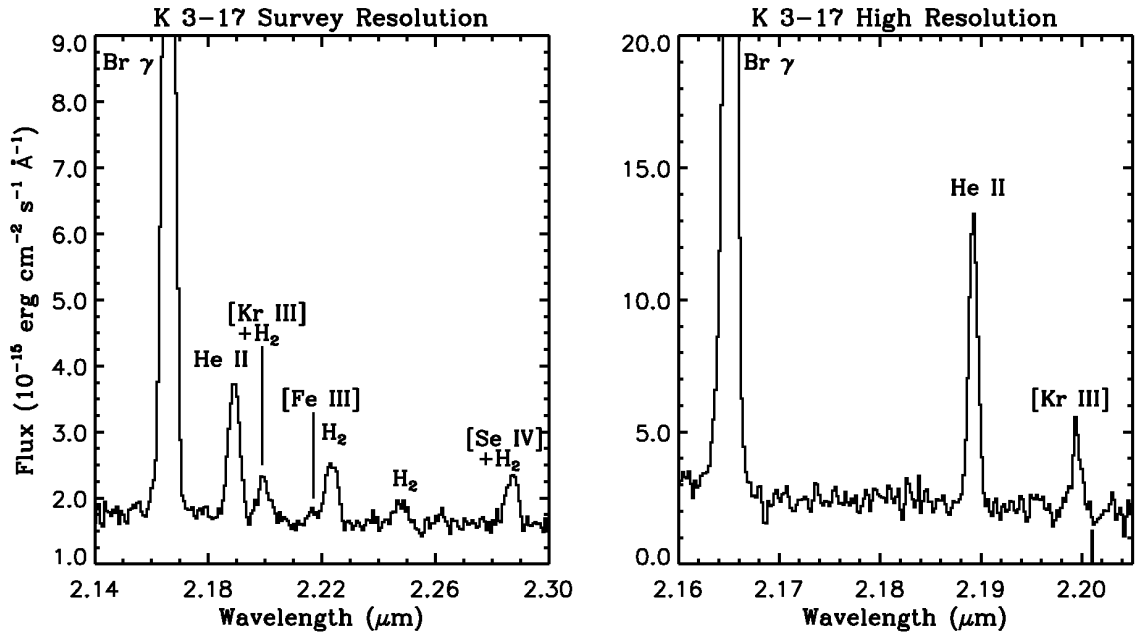}
\figcaption[f2.eps]{Survey and high resolution spectra of K~3-17, with emission features identified.  The Br$\gamma$ line is truncated for display of weak nebular features.  The high resolution spectrum clearly shows the 2.199~$\mu$m line to be due to [\ion{Kr}{3}] and not H$_2$~3-2~$S$(3)~2.201~$\mu$m (whose wavelength is indicated with a tick mark below the spectrum).}
\end{figure}

\clearpage

\begin{figure}
\plotone{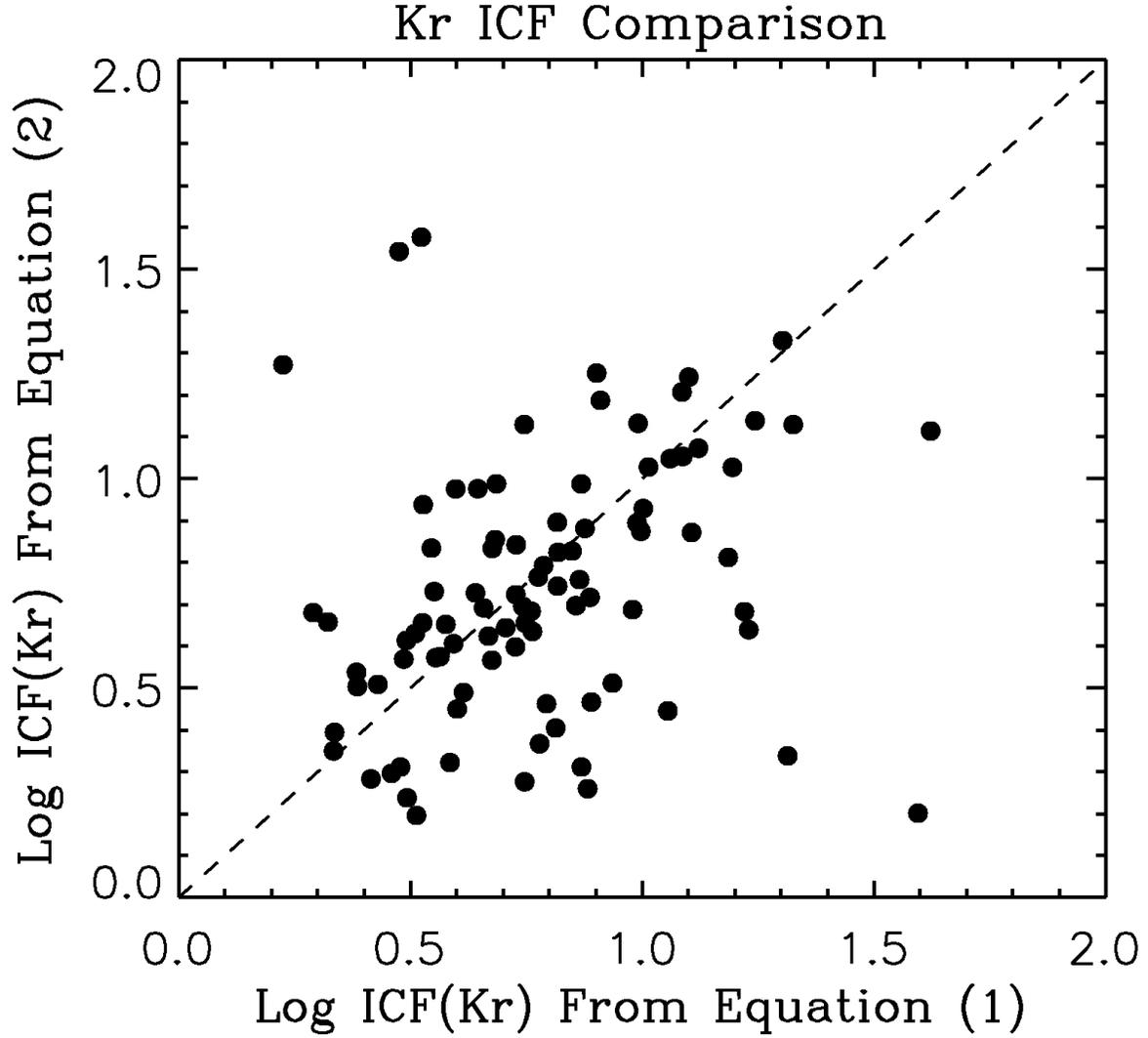}
\figcaption[f3.eps]{Comparison of Kr ICFs derived using Equations~(1) and (2).  The dashed line corresponds to perfect agreement.  The outliers in this plot correspond to PNe with very low Ar$^{++}$ or S$^{++}$ fractional abundances, which cause the ICFs to be highly uncertain.}
\end{figure}

\clearpage

\begin{figure}
\epsscale{0.85}
\plotone{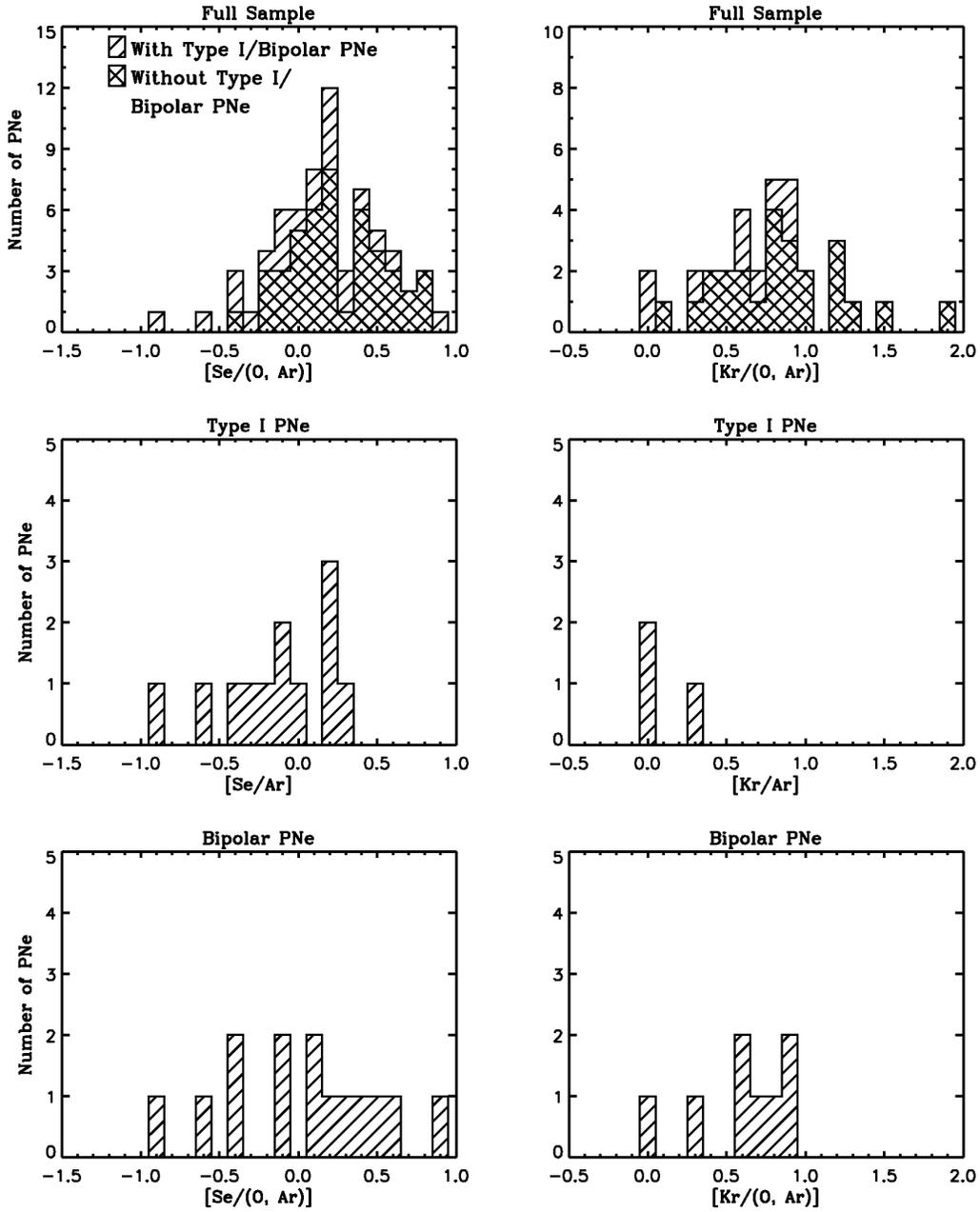}
\figcaption[f4.eps]{Histograms of Se and Kr abundances relative to O (non-Type~I PNe) or Ar (Type~I PNe), separated into 0.1~dex bins, are shown.  The top two panels display enrichments in the full sample, including and excluding bipolar/Type~I PNe.  Abundances for Type~I and bipolar PNe are shown in the middle and bottom panels, respectively.  In general, Type~I and bipolar PNe exhibit smaller \emph{s}-process enrichments than other objects in our sample (see \S5.2.2).}
\end{figure}

\clearpage

\begin{figure}
\epsscale{0.85}
\figurenum{5a}
\plotone{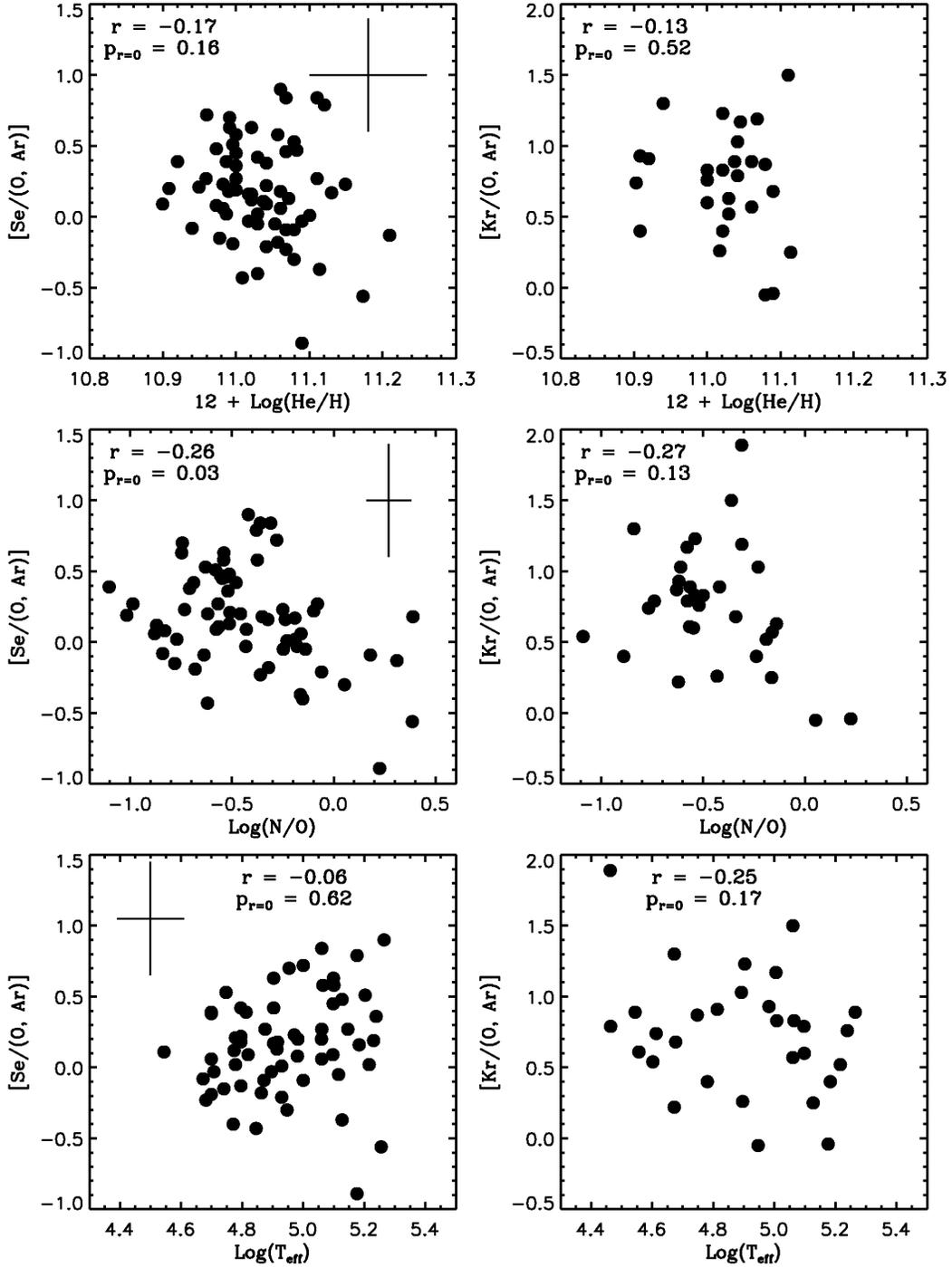}
\figcaption[f5a.eps]{(\textit{a}) [Se/(O,~Ar)] (left hand panels) and [Kr/(O,~Ar)] (right hand panels) are plotted against tracers of PN progenitor mass: He/H, N/O, and central star effective temperature $T_{\rm eff}$.  Typical error bars in the abundances or $T_{\rm eff}$ are shown in the panels on the left side of the figure.  Correlation coefficients $r$ and their significance $p_{\rm r=0}$ are indicated in each panel.  (\textit{b}) Same as (\textit{a}), except that [Se/(O,~Ar)] and [Kr/(O,~Ar)] upper limits are shown as empty circles with a line extending downward.  This increases the number of strongly He- and N-enriched PNe displayed.}
\end{figure}
\clearpage
\epsscale{0.85}
{\plotone{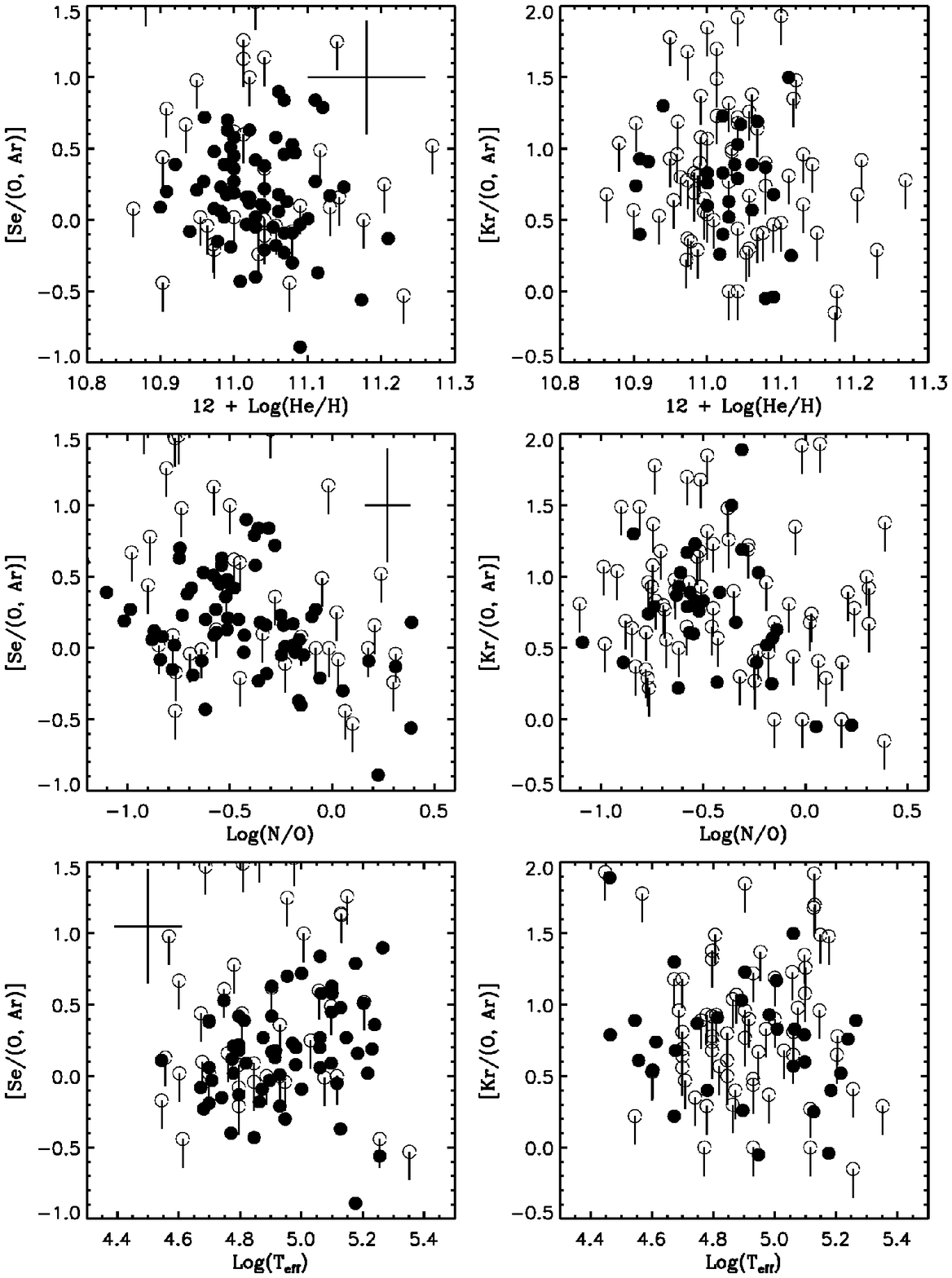}}\\
\centerline{Fig. 5b. ---}
\clearpage

\begin{figure}
\epsscale{0.85}
\figurenum{6}
\plotone{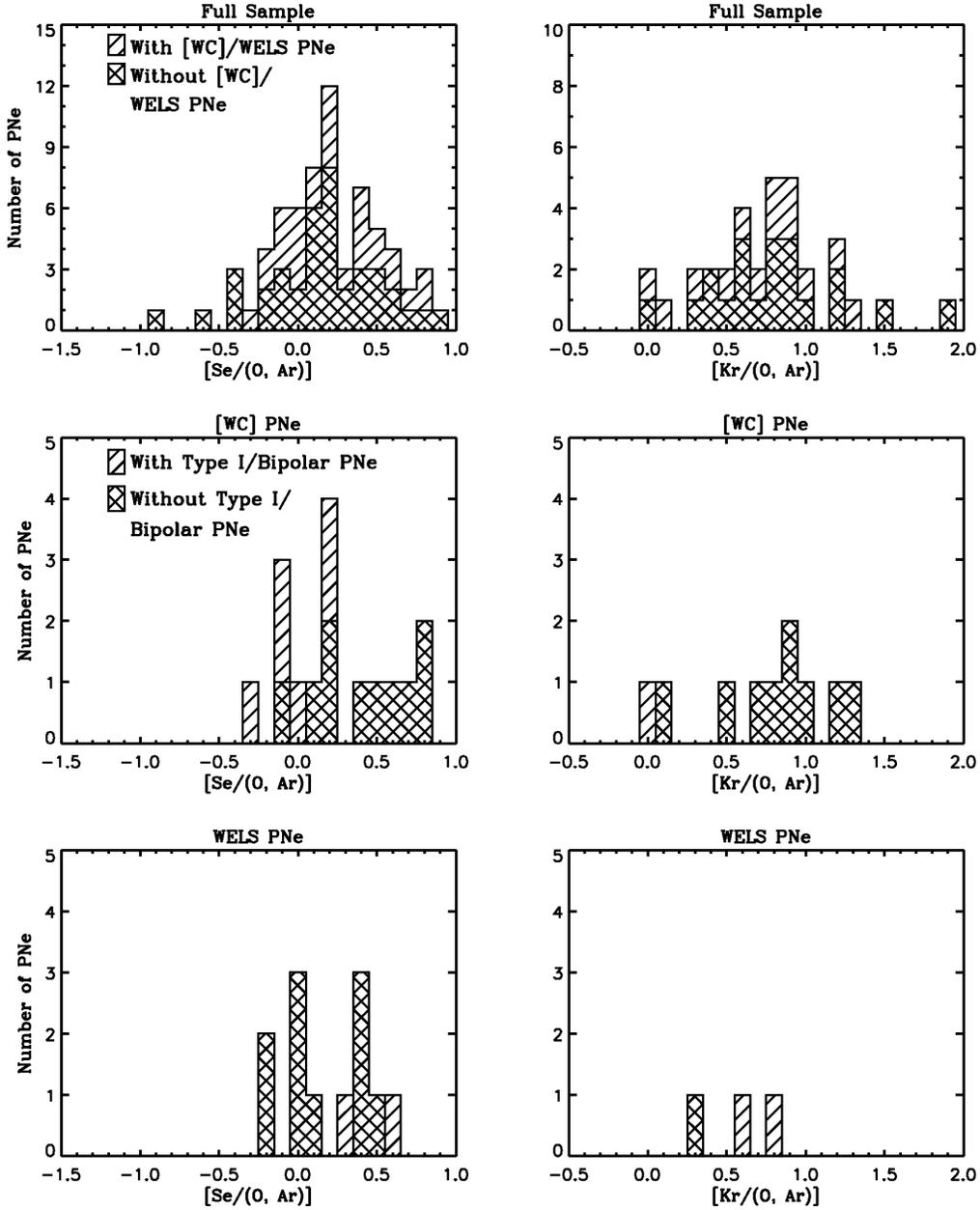}
\figcaption[f6.eps]{Histograms of Se and Kr abundances relative to O or Ar are shown.  The top two panels display enrichments in the full sample, including and excluding [WC] and WELS PNe.  Abundances for PNe with [WC] and WELS central stars are shown in the middle and bottom panels, respectively, both including and excluding Type~I and bipolar PNe.  No significant difference is seen between the distribution of Se and Kr enrichments in [WC] or WELS PNe compared to the full sample, as confirmed by KS tests (see \S5.3.1).}
\end{figure}

\clearpage

\begin{figure}
\figurenum{7}
\plotone{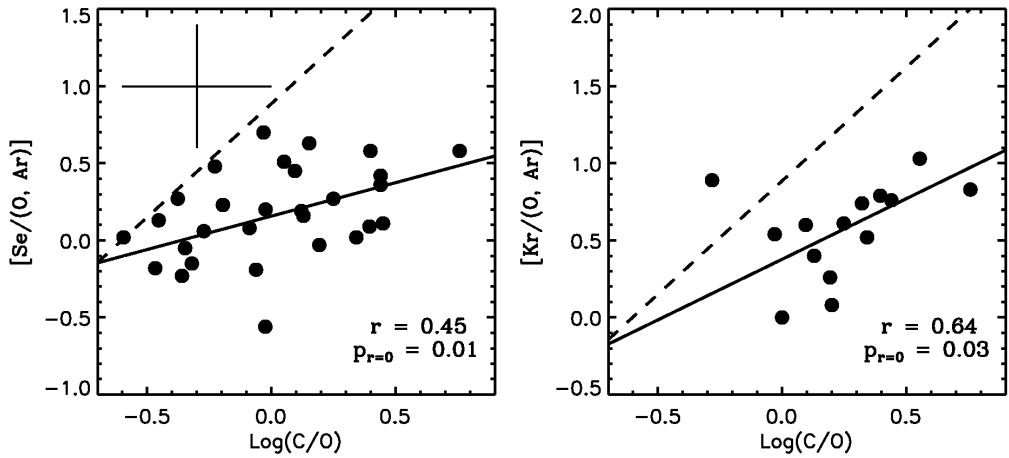}
\figcaption[f7.eps]{[Se/(O,~Ar)] (left) and [Kr/(O,~Ar)] (right) are plotted against the logarithmic C/O ratio of objects in our sample.  Typical error bars are indicated in the left panel.  The best linear fit to each correlation is plotted as a solid line (the discrepant object Hb~12 is excluded from the fit in the right-hand panel; see \S5.4).  Fits to the correlation between [$<$Sr, Y, Zr$>$/Fe] and log(C/O) in AGB and post-AGB stars are shown as dashed lines for comparison (stellar abundances taken from Smith \& Lambert 1985, 1990; Smith et al.\ 1993, and Van~Winckel \& Reyniers 2000).}
\end{figure}

\clearpage

\begin{figure}
\figurenum{8}
\plotone{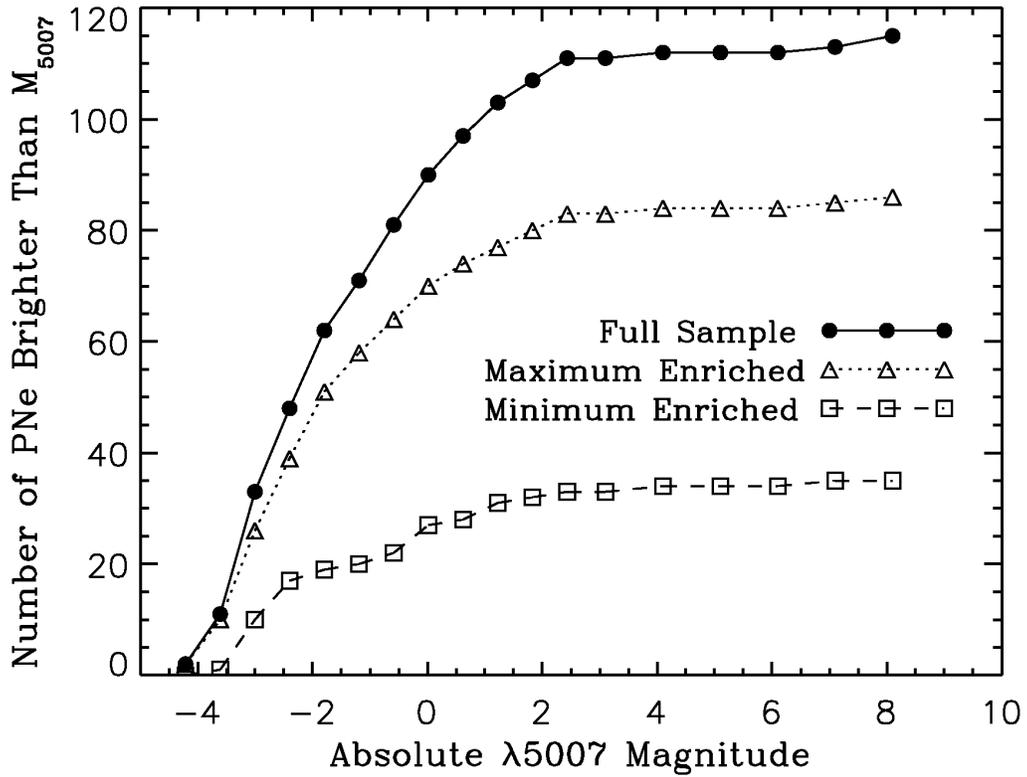}
\figcaption[f8.eps]{Cumulative number of PNe brighter than a given absolute [\ion{O}{3}]~$\lambda5007$ magnitude.  Also shown are the minimum (assuming all PNe with indeterminate \emph{s}-process enrichments are \emph{not} enriched) and maximum (assuming all PNe with indeterminate \emph{s}-process enrichments \emph{are} enriched) number of \emph{s}-process enriched PNe.}
\end{figure}

\clearpage

\begin{figure}
\figurenum{9}
\plotone{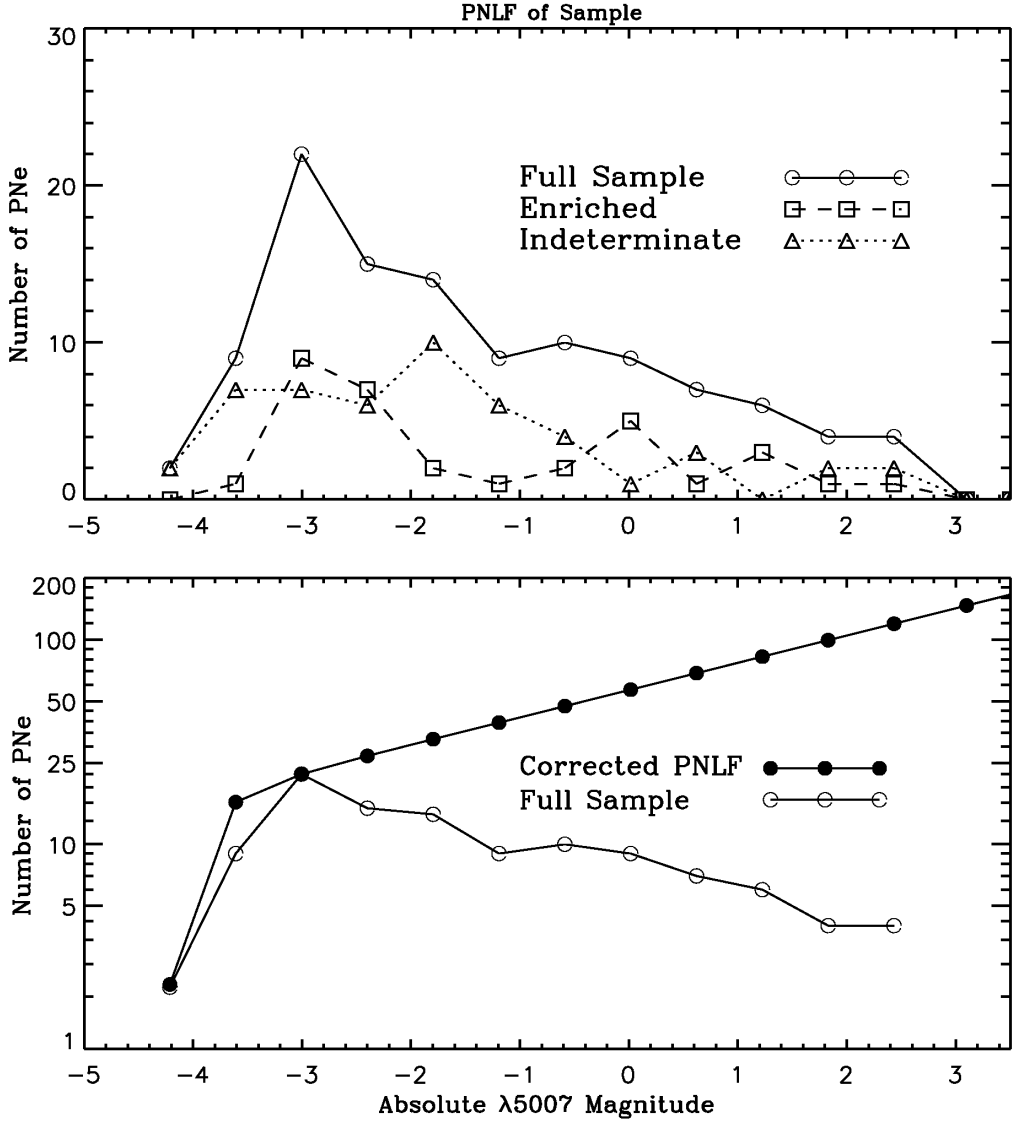}
\figcaption[f9.eps]{\textit{Upper Panel}: The derived PNLF of our full sample, along with the number of enriched and possibly enriched (indeterminate) objects in each magnitude bin.  \textit{Lower Panel}: The PNLF of our sample plotted up to the expected faint limit of 3.5~mag (Jacoby 1980, 2006).  A theoretical PNLF (Ciardullo et al.\ 1989), normalized to the number of objects in the $M_{5007}=-3.0$ bin, is shown for comparison.  The theoretical PNLF is used to correct for incompleteness in our sample at faint $M_{5007}$.}
\end{figure}

\end{document}